\documentclass[11pt,a4paper]{article}
\RequirePackage{amsmath,amssymb}
\RequirePackage[dvipsnames,usenames]{color}
\usepackage{cite}
\usepackage{fullpage}
\usepackage[british]{babel}
\usepackage[latin1]{inputenc}
\usepackage[T1]{fontenc}
\usepackage[final]{showkeys} 
\usepackage[bookmarks]{hyperref}
\usepackage{amsthm}
\usepackage{graphicx}
\usepackage{subfigure}
\usepackage{braket}
\usepackage{mathrsfs}
\def\theequation{\arabic{section}.\arabic{equation}}
\usepackage{psfrag}
\usepackage{subfigure}
\usepackage{color}
\usepackage{mathrsfs}
\usepackage{graphicx}
\usepackage{amssymb, bm}
\usepackage{amsmath, amsthm}
\usepackage{epstopdf}
\usepackage{hyperref}
\usepackage{enumerate}
\usepackage{longtable}
\usepackage{amsmath}

\usepackage{amsfonts}

\newcommand{\be}{\begin{equation}}
\newcommand{\ee}{\end{equation}}

\usepackage{bm}
\usepackage{amsfonts}
\usepackage{braket}

\title{\bf Multi-fluid cosmology in Einstein gravity: analytical solutions}

\author{
Valerio~Faraoni\thanks{E-mail: vfaraoni@ubishops.ca},
Sonia~Jose\thanks{E-mail: soniatjose96@gmail.com},
~and
Steve~Dussault\thanks{E-mail: sdussault19@ubishops.ca}     
$\,$
\\
\\
{\em Department of Physics \& Astronomy, Bishop's University}
\\
{\em 2600 College Street, Sherbrooke Qu\'ebec, Canada J1M 1Z7}
}

\begin{document}
\maketitle
\begin{abstract}

We review analytical solutions of the Einstein equations which are 
expressed in terms of elementary functions and describe 
Friedmann-Lema\^itre-Robertson-Walker universes sourced by multiple (real 
or effective) perfect fluids with constant equations of state. Effective 
fluids include spatial curvature, the cosmological constant, and scalar 
fields. We provide a description with unified notation, explicit and 
parametric forms of the solutions, and relations between different 
expressions present in the literature. Interesting solutions from a modern 
point of view include interacting fluids and scalar fields. Old solutions, 
integrability conditions, and solution methods keep being rediscovered, 
which motivates a review with modern eyes.

\end{abstract}

\newpage

\section{Introduction}
\setcounter{equation}{0}
\label{sec:1}

Analytical solutions of the Einstein-Friedmann equations describing 
cosmology in the context of Einstein's theory of gravity have been known 
since its early days, beginning with those of de Sitter 
\cite{deSitter:1917zz}, Friedmann \cite{Friedmann1922,Friedmann1924},  
Lema\^itre \cite{Lemaitre1927, Lemaitre1930, Lemaitre1931, 
Lemaitre1933}, and Einstein himself, including the famous and ill-fated 
Einstein static universe \cite{Einstein1917} (see the textbooks 
\cite{LandauLifschitz, Wald, Carroll, Inverno, Peeblesbook, 
EllisMaartensMacCallum, Liddle, KolbTurner, Slava} and see 
Ref.~\cite{BarrowBook} for a popular exposition).

The qualitative analysis of the phase spaces of solutions of the relevant 
differential equations provide much information about the cosmological 
dynamics with multiple fluids ({\em e.g.}, \cite{StabellRefsdal, 
WainwrightEllisBook, Coleybook, Felten:1986zz, Uzan01, UrenaLopez:2006ay, 
Sonego:2011rb, Faraoni:2012bf}). However, analytical solutions are often 
needed for the investigation of physics beyond the cosmic dynamics.  
Various physical motivations to look for exact solutions include the need 
for toy models in theoretical research; testing and calibrating numerical 
codes; 
describing the smooth transition between different eras (most notably, 
between radiation- and matter-dominated epochs, between inflation and 
radiation, or the onset of dark energy domination); obtaining solutions of 
the many inflationary scenarios of the early universe; modelling the 
acceleration of the cosmic expansion in the present era.  The latter 
requires the simultaneous analysis of dark energy and dark/baryonic 
matter, and two-component dark energy models have also been proposed 
\cite{Guo:2004fq, Zhang:2005kj, Gong:2007se, Vazquez:2012ag} (in particular, the 
quintom model consisting of two interacting scalar fields, one regular and 
one phantom, has received much attention \cite{Feng:2004ff, Xia:2004rw, 
Faraoni:2004bb, Zhao:2005vj, Wu:2005apa, Wei:2005nw, Cai:2006dm, 
Zhao:2006mp, Lazkoz:2006pa, Feng:2006ya, Guo:2006pc, Zhang:2006ck, 
Alimohammadi:2006tw, Setare:2006rf, MohseniSadjadi:2006hb, 
Cai:2007gs, Wei:2007rp, Cai:2007qw, Cai:2007zv, 
Lazkoz:2007mx, Alimohammadi:2007jj, Cai:2008ed, Setare:2008sf, 
Xiong:2008ic, Shi:2008df, Setare:2008ci, Zhang:2008ac, Setare:2008pz, 
Setare:2008si, Cai:2008gk, Setare:2008dw, Chimento:2008ws, Sadeghi:2008qp, 
Zhang:2009un, Saridakis:2009jq, Saridakis:2009ej, Wang:2009ae, 
Wang:2009ag, Qiu:2010ux, Leon:2012vt, Dutta:2016exd, Mishra:2018dzq, 
Leon:2018lnd}, see \cite{Cai:2009zp} for a review).

Most solutions of the Einstein-Friedmann equations found in the 1960s and 
1970s and the conditions for their integrability are spread in a 
body of literature 
that spans over half a century and are being forgotten, as 
demonstrated by 
the fact that two- and three- (effective) fluid solutions found in the 
1960s and 1970s have been re-investigated and rediscovered in recent 
years.  Moreover, the theoretical motivation for studing multiple fluids 
or effective fluids has changed: scalar fields have assumed a primary role 
in cosmology with the advent of the theory of inflation and with the 
introduction of the dark energy concept to explain the 1998 discovery that 
the present expansion of the universe is accelerated 
\cite{Perlmutter:1997zf, Riess:1998cb, Perlmutter:1998np}.  This situation 
motivates reviewing analytical multi-fluid solutions of the 
Einstein-Friedmann equations from a modern point of view, including scalar 
fields and interacting fluids in the picture, the study of which was much 
less motivated in the 1970s.

\section{Field equations, fluids, and scalar fields}
\setcounter{equation}{0}
\label{sec:2}

We restrict our discussion to cosmology and analytical solutions in 
the context of Einstein gravity, referring the reader to \cite{FujiiMaeda, 
Faraoni:2004pi, CapozzielloFaraoni, Sotiriou:2008rp, DeFelice:2010aj, 
Nojiri:2010wj} for scalar-tensor (including $f(R)$) cosmology and to 
Refs.~\cite{Clifton:2011jh, HarkoLobobook, Heisenberg:2018vsk} for more 
general 
theories of 
gravity. We follow the notation of Ref.~\cite{Wald}: the signature of the 
metric tensor $g_{ab}$ is ${-}{+}{+}{+}$ and units in which the speed of 
light $c$ and Newton's constant $G$ are unity are used.

The Einstein equations read 
\be 
R_{ab}-\frac{1}{2} \, g_{ab} R +\Lambda g_{ab}  =8\pi T_{ab} 
\,,\label{Einsteineqs}
\ee
where $R_{ab}$ and $T_{ab}$ are, respectively, the Ricci tensor and the 
matter stress-energy tensor, while $R \equiv g^{ab}R_{ab}$ is the Ricci 
scalar and $\Lambda$ is the cosmological constant. In order to discuss 
solution 
methods, one must necessarily 
restrict the scope. We begin with the simplest situation, and the 
one most common in the research literature on multi-fluid cosmology,  
consisting of two non-interacting\footnote{Sec.~\ref{sec:8} discusses 
interacting fluids.} perfect fluids with stress-energy 
tensors  
\begin{eqnarray}
T_{ab}^{(1)} & = & \left(P_1+\rho_1 \right) u_a u_b +P_1 g_{ab} \,,\\
&&\nonumber\\
T_{ab}^{(2)} & = & \left(P_2+\rho_2 \right) u_a u_b +P_2 g_{ab} \,,
\end{eqnarray}
where $u^c $ is the common four-velocity,\footnote{Again, the assumption 
that the two fluids are collinear is not mandatory: mutually tilted fluids 
are the subject of a considerable literature ({\em e.g.}, 
\cite{SmootGorensteinMuller77, ColeyTupper86, ColeyTupper86b, ColeyASS89, 
ColeyWainwright92, Nesteruk94, Nesteruk:1995uu, Coley:1995dpj, 
WainwrightEllisBook, Verma07}).} 
$\rho_{1,2}$ are the energy densities, and $P_{1,2}$ are the 
isotropic pressures. A 
constant linear barotropic equation of state
\be
P=w\rho \,, \;\;\;\;\;\;\;\;\;\;\;\;\; w=\, \mbox{const.} 
\label{barotropic}
\ee
is assumed for both fluids.  Indeed, when multiple gravitating 
fluids are present, we will assume that they all have the same 
four-velocity, according to the assumptions of spatial homogeneity and 
isotropy that require the existence of a single comoving frame associated 
with observers who see these symmetries in the surrounding universe.

By imposing spatial homogeneity and isotropy to satisfy the Copernican 
principle, the Einstein equations admit as a solution the 
Friedmann-Lema\^itre-Robertson-Walker (FLRW) geometries 
\cite{LandauLifschitz, Wald, Carroll, Inverno, Liddle, KolbTurner, Slava}. 
The FLRW line element in spherical comoving coordinates 
$\left(t,r,\vartheta, \varphi \right)$ is
\be
ds^2=-dt^2 +a^2(t)\left( \frac{dr^2}{1-Kr^2} + r^2 d\Omega_{(2)}^2 \right) 
\,,
\ee
where $d\Omega_{(2)}^2 \equiv d\vartheta^2 +\sin^2  \vartheta \, 
d\varphi^2$ is 
the line element on the unit 2-sphere, $K$ is the curvature index (usually  
normalized to $0, \pm 1$), and $a(t)$ is the scale factor 
encoding the expansion history of the universe. We will also use the 
conformal time $\eta$ defined by $dt\equiv a d\eta$.

Assuming the FLRW line element, the spatial symmetries 
reduce Eq.~(\ref{Einsteineqs}) to the Einstein-Friedmann equations 
\begin{eqnarray}
&& H^2= \frac{8\pi}{3} \, \rho +\frac{\Lambda}{3}-\frac{K}{a^2} 
\,,\label{Friedmann}\\
&&\nonumber\\
&& \frac{\ddot{a}}{a} = -\frac{4\pi}{3} \left( \rho +3P \right) 
+\frac{\Lambda}{3} \,,\label{acceleration}\\
&&\nonumber\\
&& \dot{\rho}+3H\left( P+\rho \right) =0 \,, \label{conservation}
\end{eqnarray}
where $H\equiv \dot{a}/a$ is the Hubble function and an overdot denotes 
differentiation with respect to the comoving time $t$. 
Eq.~(\ref{conservation}) follows from the covariant conservation  of the 
matter energy-momentum tensor $\nabla^b T_{ab}=0$.  
Only two of the three Einstein-Friedmann equations are independent. If two 
of them are given, the last one can be deduced from them.  Without loss of 
generality, we can regard the Friedmann equation~(\ref{Friedmann}) and the 
energy conservation equation~(\ref{conservation}) as independent and the 
acceleration equation~(\ref{acceleration}) as derived.

If  a fluid is not coupled explicitly to other matter or fields and 
has equation of state~(\ref{barotropic}),  the conservation 
equation~(\ref{conservation}) implies that its  energy density scales as 
\be
\rho(a)=\frac{\rho^{(0)}}{a^{3(w+1)} } \label{rhoscaling}
\ee
with $\rho^{(0)}$ a constant. Because fluids with different equation of 
state parameters $w$ scale differently with the scale factor $a$, in the 
presence of multiple fluids one of them will come to dominate the cosmic 
dynamics if the universe keeps expanding. For example, if there are only 
radiation ($w=1/3$) and dust 
($w=0$) in an expanding universe, the radiation fluid dominates early on 
for small $a$ but its energy density $\rho_\text{r}\sim a^{-4}$ decays 
faster  than the energy density of dust $\rho_\text{m} \sim a^{-3}$, which 
comes to dominate later on if $a(t)$ always increases. Therefore, it is 
customary to approximate the 
history of the universe with a radiation era followed by a matter era. 
This approximation is adequate for many problems, but there are 
situations in which one is interested in the detailed two-fluid solution.

In cosmology, common terminology refers to a {\em dark energy} fluid if 
$-1\leq w <-1/3 $ (which, according to the acceleration 
equation~(\ref{acceleration}) implies that the universe accelerates, 
$\ddot{a}>0$), and to {\em phantom} matter if $w<-1$.

If the relation~(\ref{barotropic}) between the cosmic fluid pressure $P$ 
and energy density $\rho$ is not 
constant, one can still define an effective equation of state parameter 
$w_\text{eff} \equiv P/\rho$. Non-linear barotropic equations of 
state $P=P(\rho)$ have been studied in the literature, especially in 
relation with phantom fluids and sudden future singularities 
\cite{BarrowGallowayTipler, ShtanovSahni02, Kofinas03, Calcagni:2004bh, 
Gorini04, Nojirietal05, Barrow04, Diego, Borowiec16, Nojiri:2004pf, 
NojiriOdintsov05, Stefancic:2004kb, Frampton:2011sp, Bouhmadi15, 
Ananda:2006gf, Ananda:2005xp, Capozzielloetal06,SilvaCosta09} but here we 
restrict ourselves to linear equations of state.

The energy density of a matter component $\rho$ can be expressed in terms 
of the dimensionless density parameter
\be
\Omega \equiv \frac{\rho}{\rho_c} \,,
\ee
where $\rho_c(t) \equiv \frac{3H^2}{8\pi}$ is the energy density of a 
spatially flat universe with $K=0$ and $\Omega=1$.

The standard model of cosmology, the so-called $\Lambda$--Cold Dark Matter 
or $\Lambda$CDM model consists of a spatially flat FLRW universe 
containing approximately 70\% dark energy (with a measured equation of 
state parameter consistent with the cosmological constant signature 
$w=-1$) and 
30\% matter, modelled as a dust, adding up to $\Omega_\text{tot}=1$. 
 
Due to their non-linearity, there are significant mathematical 
difficulties in solving analytically the Einstein-Friedmann equations of 
cosmology sourced by multiple fluids or effective fluids (the latter 
include the cosmological constant and spatial curvature, see below). It is 
even more difficult to solve analytically the more general Einstein 
equations (without spatial homogeneity and isotropy) with two fluids as 
the matter source, for example in stellar models ({\em e.g.}, 
\cite{Carter:1995mj, Carter:1998rn, Naidu:2021nwh}).  In cosmology, there 
are two approaches to the analytical solution: the first method attempts 
to integrate the Friedmann equation in comoving or in conformal time and 
deals with an integral that, in general, can only be expressed in terms 
or elliptic or hypergeometric functions \cite{AssadLima88}. The second 
method looks for 
solutions expressed in parametric form employing the conformal time as the 
parameter. Also in this second approach, there is no guarantee that the 
integration can be performed in closed form in terms of elementary 
functions, which is instead the main goal of the present paper. The two 
approaches are complementary: one approach may succeed in situations where 
the other fails, and {\em vice-versa}.

A large variety of situations, which is best described by qualitative 
methods and phase space analysis, can present themselves for general forms 
of matter which may include perfect or imperfect fluids, tilted fluids 
\cite{SmootGorensteinMuller77, ColeyTupper86, ColeyTupper86b, ColeyASS89, 
ColeyWainwright92, Coley:1995dpj, WainwrightEllisBook, Verma07}, 
fluids with non-linear and/or non-constant equation of state 
\cite{BarrowGallowayTipler, ShtanovSahni02, Kofinas03, Calcagni:2004bh, 
Gorini04, Barrow04, Nojiri:2004pf, NojiriOdintsov05, Stefancic:2004kb,  
Nojirietal05, Ananda:2006gf, Ananda:2005xp, Capozzielloetal06, 
SilvaCosta09, Diego, Borowiec16, Frampton:2011sp, Bouhmadi15}, scalar 
fields, {\em etc.} One also has to distinguish between non-interacting 
fluids and interacting (or explicitly coupled) fluids. There is now a 
large amount of literature devoted to the latter, because of the 
hypothesis that dark energy and dark matter may be coupled directly, which 
would in principle explain the coincidence between the orders of magnitude 
of the density parameters $\Omega_\text{DE}$ and $\Omega_\text{DM}$ of 
these two fluids.

Even under the simplifying assumptions of spatial homogeneity and 
isotropy, perfect fluids, and constant barotropic equations of state, the 
landscape of analytical FLRW solutions and their physics is rich and is 
covered by a considerable amount of literature spanning five decades. The  
older literature is sometimes forgotten, which results in analytical 
solutions and solution methods being rediscovered, or in the use of 
numerical solutions, which necessarily commit to specific values of the 
parameters and initial conditions, when analytical formulas are instead 
available. Our goal is to provide a bird's eye view of this area of 
multi-fluid cosmology.

As said, the integration of the Einstein-Friedmann equations usually leads 
to solutions expressed by elliptic integrals or hypergeometric functions 
which cannot be reduced to simpler forms \cite{AssadLima88}. These 
expressions are not useful 
in practice, making for a rather sterile catalogue of formal situations in 
which numerical solution of the Einstein-Friedmann 
equations~(\ref{Friedmann1})-(\ref{conservation1}) is more convenient for 
practical purposes. Here we focus on situations in which the field 
equations are integrable and the solution can be expressed explicitly in 
the form of a finite number of elementary functions.

\subsection{Single fluid and effective curvature~/ $\Lambda$-fluids}

If $K=0$ and $\Lambda=0$ and there is a single fluid with constant 
equation of state~(\ref{barotropic}), the well-known textbook solution 
for the scale factor is 
\be
a(t)=a_0 \left( t-t_0 \right)^{ \frac{2}{3(w+1)}} \,, \label{usual}
\ee
where $a_0$ and $t_0$ are constants and $ \rho(a)$ is given by 
Eq.~(\ref{rhoscaling}).  Models with a single real fluid, possibly with 
cosmological constant and spatial curvature, were reviewed and classified 
early on (\cite{Harrison1967,Vajk1969, Hughston69, HughstonShepley1970, 
HughstonJacobs70, McIntosh1972, McIntoshFoyster1972}, see also 
\cite{Inverno}).

For a fluid mixture, one defines the total effective density by
\be
\frac{8\pi}{3} \, \rho_\text{tot} = \frac{8\pi}{3} \, \rho 
+\frac{\Lambda}{3} 
-\frac{K}{a^2} \equiv \frac{8\pi}{3} \left( \rho+ 
\rho_{\Lambda}+ \rho_K \right) \,,
\ee
where
\be
\rho_{\Lambda} \equiv \frac{\Lambda}{8\pi} \,, \;\;\;\;\;\;\;\;
\rho_K \equiv  -\frac{3K}{8\pi a^2} 
\ee
can have any sign, as long as $\rho_\text{tot} \geq 0$ (otherwise 
the Friedmann equation~(\ref{Friedmann}) cannot be satisfied because the 
left hand side is always non-negative). 

To be consistent in treating the curvature term $-K/a^2$ in the Friedmann 
equation as an effective fluid, one must also attribute to it a 
pressure $P_K$. This is done by imposing that the covariant conservation 
equation~(\ref{conservation}) be satisfied for this effective component 
of the cosmic fluid, which yields
\be
P_K=-\frac{\rho_K}{3} = \frac{K}{8\pi a^2} 
\ee
(this remark appears to have been made by Hughston \& Shepley 
\cite{HughstonShepley1970}, McIntosh \cite{McIntosh1972}, and McIntosh \& 
Foyster \cite{McIntoshFoyster1972} 
long ago and then apparently forgotten). Similarly, it is customary to use 
the effective energy density $\rho_{\Lambda}$ and pressure $P_{\Lambda}$, 
with  $P_{\Lambda}=-\rho_{\Lambda}= 
-\Lambda/(8\pi )$, for the cosmological constant.\footnote{Oddly, 
Harrison (who found many of the solutions reported in this review) 
distinguished between the cosmological constant and a perfect 
fluid with $w=-1$, reporting solutions where only these two components 
source  the Einstein-Friedmann equations simultaneously 
\cite{Harrison1967}.}   
The acceleration 
equation is automatically satisfied since the combination $\rho_K+3P_K$ 
vanishes (as it should be, since the acceleration 
equation~(\ref{acceleration}) does not contain the curvature index $K$). 
By defining the total pressure
\be
P_\text{tot}=P+P_{\Lambda}+P_K \,,
\ee
the Einstein-Friedmann equations can be rewritten as 
\begin{eqnarray}
&& H^2= \frac{8\pi}{3} \, \rho_\text{tot}\,,\label{Friedmann1}\\
&&\nonumber\\
&& \frac{\ddot{a}}{a} = -\frac{4\pi}{3} \left( \rho_\text{tot} +3P_\text{tot} 
\right) \,,\label{acceleration1}\\
&&\nonumber\\
&& \dot{\rho}_\text{tot}+3H\left( P_\text{tot}+\rho_\text{tot} \right) =0 \,. 
\label{conservation1}
\end{eqnarray}
Formally, we have reduced spatially curved ($K\neq 0$) universes, 
potentially with a cosmological constant $\Lambda$, to a fictitious 
spatially flat universe with vanishing $\Lambda$. The topology of the 
three-dimensional spatial sections cannot change, of course, and this is 
only a formal trick used in the cosmology literature and textbooks to 
introduce the total density parameter $\Omega_\text{tot} \equiv 
\rho_\text{tot}/\rho_c$. This formal reduction to a single effective fluid 
is normally not completed by including the pressure $P_K$ in modern 
literature and textbooks. Here, this extended definition of effective 
curvature fluid is useful to study ``simple'' analytical two-fluid 
solutions 
of FLRW universes with any spatial curvature, with or without $\Lambda$.

\subsection{Scalar field}

Scalar fields are the simplest fundamental physical fields: they abound 
in particle physics and are widely used in cosmology. Indeed, much of 
modern cosmology is based on scalar fields as the form of matter 
propelling the expansion of the universe. Inflation in the early universe 
\cite{KolbTurner, LiddleLyth, Slava, 
Martin:2013tda} is usually described as the effect of the dynamics of a single 
scalar field, dubbed inflaton. Starobinsky inflation 
\cite{Starobinsky:1980te, StarobinskyI}, which historically was the first 
inflationary scenario (although Guth's scenario \cite{Guth:1980zm} was 
much better known initially), is currently favoured by cosmological 
observations \cite{Ade:2015lrj}. It relies on quadratic corrections to the 
Einstein-Hilbert Lagrangian, but those can be reduced to an effective 
scalar field, as in all $f(R)$ theories of gravity \cite{Sotiriou:2008rp, 
DeFelice:2010aj, Nojiri:2010wj}. In both standard and 
Starobinsky inflation, a single scalar degree of freedom acts 
as the source of the Einstein-Friedmann equations. What is more, the 
current acceleration of the cosmic expansion is commonly 
explained with a quintessence scalar field acting as dark energy, or with 
modifications of gravity, among which $f(R)$ gravity is very popular. In 
both cases, we have again a scalar field dominating the dynamics of the 
universe. In the late universe, this scalar field acts together with the 
dark matter fluid modelled as a zero pressure dust. 
Analytical solutions of the Einstein-Friedmann equations describing a 
scalar 
field and a fluid will be discussed in Sec.~\ref{subsec:sf+fluid}: here we 
focus on a single scalar field.

The cosmological literature offers also scalar field models of dark matter 
and unified dark energy-dark matter models 
\cite{Matos:1998vk, Matos:1999et, 
Alcubierre:2001ea, Matos:2001ps, Matos:2001ps, Padmanabhan:2002sh,  
Guzman:2003kt, Fuchs:2004xe,  
Arbey:2006it, Bernal:2006ci, Bertacca:2007ux, Bertacca:2007fc, Lee:2008jp, 
Matos:2008ag, 
Bertacca:2008uf, Lee:2008ab, 
Suarez:2011yf, Robles:2012uy, Rindler-Daller:2013zxa,  
Suarez:2013iw, Li:2013nal, Bertolami:2016ywc, Cosme:2018nly, 
Robles:2018fur}.  If 
scalar fields are truly responsible for early inflation 
and late-time acceleration, then their physical nature amounts to two big 
mysteries of cosmology and fundamental physics. Certainly, scalar fields 
abound in high energy theories and gravitational scalars arise naturally 
in modifications of general relativity. The prototypical alternatives to 
Einstein theory are Jordan-Brans-Dicke gravity \cite{Jordan38, 
Jordan:1959eg, Brans:1961sx}  
and its scalar-tensor 
generalizations \cite{Bergmann:1968ve, Wagoner:1970vr, Nordtvedt:1970uv}, 
in which the gravitational coupling strength becomes a 
dynamical scalar field. While, with the exception of 
Sec.~\ref{subsec:Einsteinframe}, we do not 
discuss these alternative theories 
of gravity here, there is little doubt that the scalar fields appearing in 
these theories, as well as in high energy physics, have contributed to the 
increasing use of scalars in cosmology.

The energy-momentum tensor of a scalar field minimally coupled to the 
curvature is
\be
T_{ab}^{(\phi)}= \nabla_a \phi \nabla_b \phi -\frac{1}{2} \, 
g_{ab} \nabla^c \phi\nabla_c \phi -V(\phi) g_{ab} \,,\label{scalarTab}
\ee
where $V(\phi)$ is the potential energy density or self-interaction 
potential. The stress-energy tensor~(\ref{scalarTab}) is equivalent to 
that of a perfect 
fluid \cite{Madsen:1988ph, Madsen2,Pimentel,Faraoni:2012hn, 
Semiz:2012zz,Ballesteros:2016kdx,Faraoni:2018qdr} with four-velocity
\be
u^a = \frac{ \nabla^a \phi}{ \sqrt{ -\nabla^c \phi \nabla_c \phi}} \,,
\ee
provided that the gradient $\nabla^c \phi$ is timelike ($ u^c $ is 
correctly normalized, $u^c u_c=-1$). This condition is always satisfied 
in  an unperturbed FLRW universe, where $\phi=\phi(t)$ in comoving 
coordinates 
in order to 
respect spatial homogeneity. In general, if the scalar field has a 
potential $V(\phi)$, the effective equation of 
state of the equivalent fluid is dynamical.  Choosing a potential 
$V(\phi)$ determines a dynamical equation of state   
but there isn't a one-to-one correspondence \cite{Madsen2, Faraoni:2000vg,   
Bayin:1994nz}.  

The covariant conservation equation $\nabla^b T_{ab}^{(\phi)}=0$ gives  
the Klein-Gordon equation obeyed by $\phi$
\be
\Box \phi-\frac{dV}{d\phi}=0 \,,
\ee
which reduces to
 \be
\ddot{\phi}+3H\dot{\phi}+\frac{dV}{d\phi}=0 \label{KleinGordon}
\ee
for a scalar $\phi(t)$ in FLRW space, or to 
\be
\phi_{,\eta\eta} +2{\cal H} \phi_{,\eta} +a^2 \, \frac{dV}{d\phi}=0
\ee
in terms of conformal time, where $ {\cal H} \equiv a_{,\eta}/a$.
In general, this equation is non-linear and relatively few analytical  
solutions are known.
 
The effective energy density and pressure of a cosmological scalar field 
are
\begin{eqnarray}
\rho_{\phi} &=& T_{ab}^{(\phi)} u^a u^b = \frac{ \dot{\phi}^2}{2} +V(\phi) 
\,, \label{rhophi}\\
&&\nonumber\\
P_{\phi} &=& h^{ab} T_{ab}^{(\phi)} = \frac{T_{ii} }{g_{ii}} = \frac{ 
\dot{\phi}^2}{2} -V(\phi) \,, \label{Pphi}
\end{eqnarray}
where ${h^a}_b$ is the projection operator on the 3-dimensional spatial 
sections with Riemannian metric $h_{ab} = g_{ab}+u_a u_b$. 
In general, the equation of state of the  effective fluid equivalent to 
$\phi$ is 
dynamical and time-dependent, however one can in principle impose that it 
remains constant and determine the potential $V(\phi)$ that achieves this 
peculiar situation (which may or may not not be physically motivated), as 
done in \cite{Barrow:1993ah, StarkovichCooperstock92}. One obtains 
\be
w_{\phi} \equiv \frac{P_{\phi}}{\rho_{\phi}} = \mbox{const.}
\ee
if the kinetic and potential energy densities are proportional to each 
other, $\dot{\phi}^2/2=\alpha V(\phi)$ with 
\be
\alpha = \frac{-\left( w_{\phi}+1 \right)}{w_{\phi}-1} 
\ee
for $w_{\phi}\neq 1$. The case $w_{\phi}=1$ corresponds to $V=0$ and to a 
stiff fluid, for which introducing the parameter $\alpha$ does not make 
sense.

Interest in analytical solutions of the Einstein-Friedmann 
equations~(\ref{Friedmann})-(\ref{conservation}) describing FLRW universes 
sourced by a single scalar field ranged from mathematical curiosity to the 
desire of investigating dynamics beyond the standard slow-roll 
approximation to inflation \cite{Martin:2013tda}, or to showing that 
almost any behaviour of the scale factor $a(t)$ can be obtained with a 
scalar field with a suitable potential \cite{EllisMadsen1991}. It was 
hoped that, since any 
inflationary scenario motivated by a particle physics theory eventually 
amounts to specifying the scalar field potential $V(\phi)$, by spanning a 
range of possibilities for $V(\phi)$ and the resulting cosmic history 
$a(t)$ and reconstructing the latter through cosmography, 
one could use the universe as a hot laboratory to investigate physics at 
energies unreachable by particle accelerators. Unfortunately, the 
reconstruction of the potential $V(\phi)$ by means of the spectral indices 
of scalar (and maybe, in the future, of tensor) perturbations provides 
only the values of $V$ and of its first two derivatives at the value of 
$\phi$ when these perturbations cross outside the horizon 
\cite{Lidsey:1995np}. This information is insufficient to reconstruct the 
functional form of the inflationary potential $V(\phi)$.

\subsection{Multi-fluid cosmology}

Suppose that there are $n$ non-interacting fluids with densities $\rho_i$ 
and pressures $P_i=w_i \rho_i$, with $w_i=$~const. ($i=1,2, ..., n$). Then 
the total density is 
\be
\rho_\text{tot}=\sum_{i=1}^n \rho_i
\ee
and, according to Dalton's law, the pressure is the sum of the partial 
pressures
\be
P_\text{tot}=\sum_{i=1}^n w_i \rho_i \,.
\ee
The total effective equation of state parameter is defined as 
\begin{eqnarray}
w_\text{tot}(a) &\equiv & \frac{P_\text{tot}}{\rho_\text{tot}} 
= \frac{ \sum_{i=1}^n P_i}{\sum_{j=1}^n \rho_j} \nonumber\\
&&\nonumber\\ 
&=& \frac{ \sum_{i=1}^n w_i \rho_i^{(0)} a^{-3(w_i+1)} }{\sum_{j=1}^n 
\rho_j^{(0)} a^{-3(w_j+1)}} \,.  \label{weffective}
\end{eqnarray} 
It follows that, in the presence of two or more perfect fluids, the 
effective equation of state of the mixture is time-dependent even 
if the equation of state of each component is constant.  The time 
variation of the total effective equation of state parameter is 
given by \cite{McIntosh1972}
\be
\dot{w}_\text{tot} =  -\frac{3H}{\rho_\text{tot}^2} \, \sum_{i<j} 
\left(w_i-w_j\right)^2 
\rho_i \rho_j  \label{wdot}
\ee
(this equation is derived in Appendix~\ref{AppendixA}).
 
A consequence of Eq.~(\ref{weffective}) is that, if dark energy is made of 
two distinct components with constant equations of state 
\begin{eqnarray} 
P_1 &=& w_1 \rho_1 (a) = w_1 \, \frac{ \rho_1^{(0)} }{a^{3(w_1+1)} } \,,\\
&&\nonumber\\
P_2 &=& w_2 \rho_2 (a) = w_2 \, \frac{ \rho_2^{(0)} }{a^{3(w_2+1)} } \,,
\end{eqnarray}
the effective equation of state parameter 
\be
w_\text{tot}= \frac{ w_1 \rho_1^{(0)} + w_2 \rho_2^{(0)} a^{3(w_1-w_2)} }{
\rho_1^{(0)} + \rho_2^{(0)} a^{3(w_1-w_2)} }
\ee
is not constant (even when all the $w_i$ are) and is non-linear in the 
scale factor or the redshift $z 
\equiv \frac{a_0}{a} -1 $ (where $a_0$ is the present value of the scale 
factor). Therefore, it cannot be reproduced by the linear parametrizations 
$w(z)=w_0+w_1 z $ or $w(a)=w_0 +w_1(1-a)$ (with $w_{1,2}$ constants) 
common in observational cosmology.

\subsection{Reduction to quadratures}

Rewriting the Friedmann equation as~(\ref{Friedmann1}), one 
obtains 
\be
\dot{a} = \pm \sqrt{\frac{8\pi}{3}} \, a \sqrt{ \rho_\text{tot}} \,,
\ee
where $\rho_\text{tot}=\rho_\text{tot}(a)$. 
Formal integration yields
\be
\int \frac{da}{ a \sqrt{ \rho_\text{tot}(a)} } = \pm \sqrt{ \frac{8\pi}{3} } 
\left( t-t_0 \right) \,, \label{direi1.23}
\ee
where $t_0$ is an integration constant. Usually, one cannot compute the 
integral in the left hand side explicitly in terms of elementary 
functions. Even when this is possible, one will have an 
analytical solution $t=t(a) $ which, in general, cannot be inverted to 
obtain $a=a(t)$ explicitly. 

While physically there may be a big difference between a source composed 
of two fluids (for example, a cosmic radiation background that is still 
hot, inducing significant microphysical effects in a universe already 
dominated by dust) and a universe filled by a  single fluid with exotic 
non-linear equation of state, the mathematics ruling the Einstein 
equations does not know the difference. From the purely formal point 
of view, one can define the total equation of state parameter for a 
universe sourced by two distinct fluids as in Eq.~(\ref{weffective}).

\subsection{Symmetries of the Einstein-Friedmann equations}

In any physical theory, symmetries are important and the symmetries of the 
Einstein-Friedmann equations have been the subject of many works  
(\cite{Chimento:2002gb,Aguirregabiria:2003uh,Chimento:2003qy, 
Dabrowski:2003jm, Aguirregabiria:2004te, Calcagni:2004wu, 
Szydlowski:2005nu, Chimento:2005au, Chimento:2006gk, Chimento:2005xa, 
Dabrowski:2006dd, Cai:2006dm, Chimento:2007fx, Cataldo:2005gb, 
Capozziello:2009te, Wang:2009ag, CapozzielloFaraoni,Cai:2009zp, 
Pucheu:2009jw, 
Chimento:2010un, Faraoni:2011ut} and referenes therein). A 
connection has been made between Einstein-Friedmann equations and  methods 
of supersymmetric quantum mechanics \cite{Rosu:2010if, Rosu:2005rz, 
Nowakowski:2002kh}. As in other areas of physics, also in FLRW  cosmology 
Lie,  Noether, and other symmetries are often studied with the ultimate 
goal  of generating first integrals of motion and exact solutions    
\cite{Capozziello:2009te, 
CapozzielloFaraoni, Chimento:2002gb,Pailas:2020xhh,
Christodoulakis:2018swq, Tsamparlis:2018nyo, Gionti:2017ffe, 
Paliathanasis:2015arj, 
Paliathanasis:2016heb, 
Zampeli:2015ojr, 
Paliathanasis:2014yfa, 
Christodoulakis:2013xha, 
Dimakis:2013oza, 
Basilakos:2011rx, 
Paliathanasis:2011jq,
Borowiec:2014wva, Capozziello:2014bna,  
Capozziello:2016eaz, Piedipalumbo:2019snr}, sometimes in relation with 
Penrose's cyclic time cosmologies or conformal time 
\cite{Vazquez:2012ag, BenAchour:2019ufa, BenAchour:2020xif, 
BenAchour:2020njq}. The existence of such a symmetry does not 
automatically make the corresponding solution a physical one.

A symmetry studied in \cite{Chimento:2002gb, Pailas:2020xhh} (see also 
\cite{Aguirregabiria:2003uh, Chimento:2003qy, Aguirregabiria:2004te, 
Chimento:2006gk, Chimento:2005xa, Chimento:2007fx, Cataldo:2005gb, 
Chimento:2010un}) and generalized in \cite{Dussault:2020uvj} applies to 
$K=0$ universes sourced by perfect fluids. It consists of a rescaling of 
the comoving time $t$, the Hubble function $H$ and all first time 
derivatives, and of the fluid energy density and pressure which leaves the 
Einstein-Friedmann equations invariant in form.  The pressure is rescaled 
differently than the energy density, which changes the equation of state. 

The symmetry transformations are given by \cite{Dussault:2020uvj} 
\begin{eqnarray}
dt  \rightarrow   d\bar{t} &=& f(\rho)dt \,, \label{eq:7}\\
&&\nonumber\\ 
\rho  \rightarrow \bar{\rho} &=& \frac{\left(1-f^2\right)}{8\pi G f^2} 
\Lambda + \frac{\rho}{f^2}  
= \frac{(1-f^2)}{f^2} \, \rho_{\Lambda} +\frac{\rho}{f^2} \,, 
\nonumber\\
&&\label{eq:8}\\
&& \nonumber\\
P  \rightarrow  \bar{P } &=& -\bar{\rho}+ 
\frac{\left[4\pi G f-\left(8\pi G \rho+\Lambda\right) f'\right]}{ 
4\pi G f^3} \left(P+\rho\right)  \nonumber\\
&&\nonumber\\
 &=& -\bar{\rho} + \frac{ \left[ f-2\left( \rho+\rho_{\Lambda} \right) 
f'\right]}{f^3} \left(P+\rho\right) \,, \label{eq:9}
\end{eqnarray}
with $f(\rho) >0 $  a dimensionless regular  function (the differential 
$d\bar{t}=f(\rho(t)) dt$ is exact because $\rho=\rho(t)$).  
The inverse  is 
\begin{eqnarray}
dt&=& \frac{d\bar{t}}{f} \,, \label{1inverse}\\
&&\nonumber\\
\rho &=& f^2 \bar{\rho} + \left(f^2-1\right) \rho_{\Lambda} 
\,, \label{2inverse} \\
&&\nonumber\\
P+\rho &=& \frac{f^3}{ f- 2\left( \rho + \rho_{\Lambda} \right) 
f'} \left(\bar{P}+\bar{\rho} \right) \,. \label{3inverse}
\end{eqnarray}
Using 
\begin{eqnarray}
\dot{a} &\equiv& \frac{da}{dt}= \frac{da}{d\bar{t}}\, 
\frac{d\bar{t}}{dt} =f \, \frac{da}{d\bar{t}} \,, \label{eq:adot}\\
&&\nonumber\\
H &=& f \bar{H} \equiv \frac{f}{a} \, \frac{da}{d\bar{t}} \,,
\end{eqnarray} 
and Eq.~(\ref{2inverse}), the Friedmann equation~(\ref{Friedmann}) for the 
spatially flat universe changes into 
\be
\bar{H}^2 = \frac{8\pi}{3} \, \bar{\rho} +\frac{\Lambda}{3} \,,
\label{Friedmanneqbarred}
\ee
while the covariant conservation  equation~(\ref{conservation}) becomes 
\cite{Dussault:2020uvj}
\begin{eqnarray}
&& \frac{d\bar{\rho}}{d\bar{t}} +3\bar{H} \left( \bar{P}+\bar{\rho} 
\right) 
\nonumber\\
&&\nonumber\\
&& = \left[ \frac{ f- 2\left( \rho + \rho_{\Lambda} \right) f'}{f^4} 
\right] 
\left[ \dot{\rho}+3H \left(P+\rho\right) \right] =0 \,. \nonumber\\
&& \label{conservationbarred}
\end{eqnarray}
When $\Lambda=0$, one has 
\be
f(\rho)=\sqrt{ \frac{\rho}{\bar{\rho}} } \,,
\ee
and the symmetry (\ref{eq:7})-(\ref{eq:9}) reduces to the one of 
Ref.~\cite{Chimento:2002gb} (see also \cite{Pailas:2020xhh}) given by 
\begin{eqnarray}
 \rho &\rightarrow &\bar{\rho}(\rho) \,,\label{chimento1}\\
&&\nonumber\\
H &\rightarrow& \bar{H}= \sqrt{ \frac{ \bar{\rho}}{\rho}} \, H \,, 
\label{chimento2}\\
&&\nonumber\\
P+\rho &\rightarrow &\bar{P}+\bar{\rho}= \left(P+\rho \right) \sqrt{ 
\frac{ \rho}{\bar{\rho}} } \, \frac{d\bar{\rho}}{d\rho} 
\,,\label{chimento3}\\
&&\nonumber\\
f(\rho) &=& \sqrt{ \frac{ \rho}{ \bar{\rho}} }\,.
\end{eqnarray}
de Sitter universes with $\Lambda>0$ are fixed points   of the 
transformation because, if $P=\rho=0$, then  
$\bar{P}=-\bar{\rho}=$~const., 
\be
\bar{\rho}=\frac{(1-f^2)}{f^2} \, \rho_{\Lambda} 
\ee
with $f=$~const. and, by changing units of time one can set $f$ to unity. 
These symmetries map  perfect fluids FLRW universes  with equation of 
state~(\ref{barotropic}) into universes with cosmological constant 
and fluids with non-linear equations of state  
\cite{Dussault:2020uvj}
\be
\bar{w}+1= \left( w+1\right) \frac{\rho}{\rho + (1-f^2) \rho_{\Lambda}} \, 
\frac{ f-2\left( \rho +\rho_{\Lambda} \right) f'}{f} \,.
\ee
One can impose that the new equation of state be constant, {\em i.e.}, 
$\bar{w} \equiv \bar{P}/\bar{\rho} =$~const. Then, denoting 
\be
s \equiv \frac{\bar{w}+1}{w+1} = \mbox{const.} \,,
\ee
the non-linear ordinary differential equation 
\be
f' + \frac{(s-1) f}{2\left( \rho+\rho_{\Lambda} \right)} + s f\left( 1-f^2 
\right) \frac{ 
\rho_{\Lambda}}{2\rho \left( \rho +\rho_{\Lambda} \right)} =0  
\label{ODE}
\ee
must be satisfied by the transformation function $f$. One possible 
solution is \cite{Dussault:2020uvj}
\be
f(\rho) = \sqrt{ \frac{\rho+\rho_{\Lambda}}{ \gamma \rho^s  
+\rho_{\Lambda} } } \,,\label{putative}
\ee
where $\gamma$ is a constant with  dimensions 
$ \left[ \gamma \right] = \left[ \rho^{1-s} \right]$.

The energy density and pressure transform according to  
\begin{eqnarray}
\bar{\rho}&=& \gamma\, \rho^s \,,\\
&&\nonumber\\
\bar{P}&=& \gamma\left[ \alpha(w+1)-1\right]  \rho^s \,,
\end{eqnarray}
and the equation of state parameter becomes
\be
\bar{w}= s \left( w+1\right)-1 \,.
\ee
For example, a de Sitter universe with  $ w = -1$ is mapped into another 
one for any  $\alpha$; if $s=1$, a  perfect fluid does not change 
the equation of state parameter ($\bar{w}=w$). If $s=4/3$, dust with  
$w=0$ is mapped into radiation.  If $s=3/2$ radiation becomes a stiff 
fluid. This symmetry is useful to generate fluid solutions  with 
non-linear equation of state from seed solutions with 
linear one, 
but there is always the question of how physical the result of a 
formal technique is. Generating new scalar field 
solutions for new potentials starting from a given one, which would be 
more interesting, proves difficult or impossible \cite{Dussault:2020uvj}. 

Another symmetry exists for spatially flat universes 
\cite{Faraoni:2020tpe}: the transformation 
\begin{eqnarray}
a & \rightarrow & \tilde{a} = a^\sigma, \,\label{change1}\\
&& \nonumber\\
dt & \rightarrow & d\tilde{t} = \sigma \, a^{ 
\frac{3(w+1)(\sigma-1)}{2} } dt =  \sigma \, 
\tilde{a}^{ \frac{3(w+1)(\sigma-1)}{2\sigma} }dt,   
\, \label{change2}\\ 
&&\\
\rho & \rightarrow & \tilde{\rho} = a^{-3(w+1)(\sigma-1)} \rho, 
\,\label{change3}
\end{eqnarray}
leaves the form of the Einstein-Friedmann equations unchanged and the 
perfect fluid retains its equation of state~(\ref{barotropic})  with the 
same equation of state parameter $w$, in contrast with the previous 
symmetries. The symmetry with $\sigma=1$, which leaves the comoving time 
unchanged, is the best known ({\em e.g.}, \cite{Chimento:2002gb, 
Aguirregabiria:2003uh, Dabrowski:2003jm, Faraoni:2011ut}).

\section{Two fluids with constant equations 
of state---comoving time}
\setcounter{equation}{0}
\label{sec:3}

Let us consider the situation in which there are two non-interacting 
fluids with constant equations of state $P_1=w_1 \rho_1$, $ P_2=w_2 
\rho_{2}$ with $w_{1,2}=$~const. We use the subscripts m,~r, and~s to 
denote dust (non-relativistic matter), radiation, and a stiff fluid, 
respectively. It is assumed that the two fluids are collinear and not 
tilted with respect to each other, {\em i.e.}, that they have the same 
four-velocity $u^c$ in their stress-energy tensors, in agreement 
with spatial homogeneity and isotropy.
Since the spatial curvature and $\Lambda$ are now treated as effective 
perfect fluids with constant equations of state, we include spatially 
curved FLRW universes, with or without $\Lambda$, in the discussion.

In this section we restrict to the use of comoving time $t$ and we look 
for solutions of the Einstein-Friedmann 
equations~(\ref{Friedmann1})-(\ref{conservation1}) of the form $a=a(t)$ or 
$t=t(a)$. The total energy density of the mixture is 
$\rho_\text{tot}=\rho_1+\rho_2$,  where the energy densities of the 
components scale as
\be
\rho_1 (a) =\frac{ \rho_1^{(0)} }{a^{3(w_1+1)}} \,, \;\;\;\;\;\;\;
\rho_2 (a) =\frac{ \rho_2^{(0)} }{ a^{3(w_2+1)}} \,,
\ee
and $\rho_i^{(0)}$ are constants. Then
\be
\dot{a}= \pm \sqrt{ \frac{8\pi}{3}} \sqrt{ 
\frac{ \rho^{(0)}_1}{ a^{3w_1+1}} + 
\frac{ \rho^{(0)}_2}{ a^{3w_2+1}}  } 
\ee
and
\begin{eqnarray}
I & \equiv &\int da \, a^{\frac{3w_1+1}{2}}  \left[  \rho^{(0)}_1 + 
\rho^{(0)}_2  a^{3(w_1 - w_2)} \right]^{-1/2}  \nonumber\\
&&\nonumber\\
&=& \pm \sqrt{ \frac{8\pi}{3}} \left(t-t_0 \right)\,, \label{eq:2.3}
\end{eqnarray}
where $t_0$ is an integration constant. This expression reduces to the 
single fluid equation~(\ref{direi1.23}) in the limit of a single fluid 
$\rho_2^{(0)} \rightarrow0$.

For general values of the parameters $w_{1,2}$, this integral can be 
expressed in terms of the Gauss hypergeometric function ${}_2F_1$ as 
\cite{AssadLima88}
\be 
I=\frac{\alpha^q}{p+1} \, a^{p+1} \, {}_2F_1 \left( -q; \frac{p+1}{r}; 
\frac{p+1}{r} +1; -\frac{\beta}{\alpha} \, a^r \right) 
\ee
where $\alpha$ and $\beta$ are constants and 
\begin{eqnarray}
p &=& (3w_1 +1)/2 \,,\\
&&\nonumber\\
q &=& -1/2 \,,\\
&&\nonumber\\ 
r &=& 3(w_1-w_2) \,.
\end{eqnarray} 
In general, this representation is not useful for practical purposes in 
cosmology. However, as several authors (Jacobs when one of the fluids is a 
stiff 
fluid \cite{Jacobs1968}; McIntosh \cite{McIntosh1972} and MacIntosh \& 
Foyster \cite{McIntoshFoyster1972}; and, more recently, Chen, Gibbons, Li 
\& 
Yang \cite{Chen:2014fqa})  have noticed, for special values of the parameters 
$w_{1,2}$ the integral can be expressed in terms of elementary functions 
thanks to the truncation of the hypergeometric series or to the Chebysev 
theorem of integration \cite{Chebysev,MarchisottoZakeri}. This theorem 
states that\\\\ {\em The integral
\be
J \equiv \int dx \, x^p \left( \alpha + \beta x^r\right)^q 
\;\;\;\;\;\;\;\;\;
r\neq 0, \, p,q,r \in \mathbb{Q} \label{integral}
\ee
admits a representation in terms of elementary functions if and only if at 
least one of $ \frac{p+1}{r}, q, \frac{p+1}{r}+q$ is an integer.} 

In our situation, we have 
\begin{eqnarray}
\frac{p+1}{r}&=& \frac{w_1+1}{2(w_1-w_2)} \,, \\
&&\nonumber\\
q&=&-\frac{1}{2} \,,\\
&&\nonumber\\
\frac{p+1}{r}+q &=& \frac{w_2+1}{2(w_1-w_2)} \,. \label{second}
\end{eqnarray}

The authors of \cite{Jacobs1968, McIntosh1972, McIntoshFoyster1972, 
Chen:2014fqa} have studied systematically the simple integrability cases.  To 
proceed we need $w_1$ and $w_1$ 
to be rational numbers, which always 
happens in cosmology, or else one can use the rational approximation to 
the 
equation of state parameter $w$. In practice, we only need to assume that 
one of the parameters $w_i \in \mathbb{Q}$; since we impose one of the 
conditions~(\ref{acci1}), (\ref{acci2}) below in our search for 
integrability, the second $w$-parameter is automatically rational if the 
first one is.

\subsection{First condition: $ \frac{p+1}{r}=n \in \mathbb{Z} $}

Imposing $\frac{p+1}{r}=n \in \mathbb{Z}$ in order to find integrability 
cases in terms of elementary functions \cite{McIntosh1972, Chen:2014fqa}, one 
obtains 
\be
\left(2n-1\right) w_1 -2nw_2 =1 \,. \label{acci1}
\ee

We can fix the first fluid ({\em i.e.}, fix $w_1\in \mathbb{Q}$) and look 
for physically interesting values of 
the equation of state parameter $w_2$ of the second fluid, as $n\in 
\mathbb{Z}$ runs.

\subsubsection{Dust plus a second (real or effective) fluid}

Suppose that the first fluid is a dust, $w_1=0$, and we look 
for a second fluid that gives ``simple'' integrability. Then, 
\be
-\frac{1}{2} \leq  w_2=-\frac{1}{2n} \leq \frac{1}{2} 
\ee
and, as $n=-\infty, \, ... \, , -3, -2, -1, 1, 2, 3, \, ... \,, +\infty$, 
we 
obtain the  pairs
\be
\begin{array}{rcll}
\left( w_1, w_2 \right) &=& \left( 0, 0 \right) & \mbox{(single dust 
fluid)},\\
 & ... &   & \\
\left( w_1, w_2 \right) &=& \left( 0,  1/6 \right) ,& \\
\left( w_1, w_2 \right) &=& \left( 0, 1/4  \right) ,&\\
\left( w_1, w_2 \right) &=& \left( 0, 1/2 \right) ,&\\
\left( w_1, w_2 \right) &=& \left( 0, -1/2 \right) ,&\\
\left( w_1, w_2 \right) &=& \left( 0,  -1/4 \right) ,&\\
\left( w_1, w_2 \right) &=& \left( 0, -1/6 \right) ,&\\
&  ... & &\\
\left( w_1, w_2 \right) &=& \left( 0 , 0\right) & \mbox{(again, a 
single dust fluid)}.
\end{array}\label{list1}
\ee
Both limits $n\rightarrow \pm \infty$ produce a second dust, {\em i.e.}, 
there is a  single dust fluid in the FLRW universe. No value of $w_2$ 
particularly interesting 
from the physical point of view is obtained; although quintessence 
(but not phantom) fluids are obtained this way, their equation of state 
parameter $w_2$ is not close to $-1$, as required by current observations 
\cite{Planck}.

\subsubsection{Radiation plus a second (real or effective) fluid}

Now suppose that the first fluid is radiation, $w_1=1/3$, and 
we look  for a second fluid giving integrability in terms of elementary 
functions. Then, 
\be
-\frac{1}{3} \leq   w_2= \frac{n-2}{3n} \leq 1 
\ee
and, letting $n=-\infty, \, ... \,, -3, -2, -1, 1, 2,3, \, ... \,,  
+\infty$,  we obtain the pairs 
\be
\begin{array}{rcll}
\left( w_1, w_2\right) &=&  \left(1/3  , 1/3 \right) &\mbox{(a 
single radiation fluid)}, \\
&  ... & &\\  
\left( w_1, w_2 \right) &=&  \left( 1/3  , 5/9 \right) ,&\\ 
\left( w_1, w_2 \right) &=&  \left( 1/3 , 2/3 \right) ,&\\
\left( w_1, w_2 \right) &=&  \left( 1/3  ,  1 \right) &\mbox{(radiation plus 
stiff fluid)}, \\
\left( w_1, w_2 \right) &=&  \left( 1/3  , -1/3 \right) & 
\mbox{(radiation plus spatial curvature)},\\ 
\left( w_1, w_2 \right) &=& \left( 1/3  ,  0 \right) & \mbox{(radiation 
plus dust)},\\
\left( w_1, w_2 \right) &=&  \left( 1/3  ,  1/9 \right) ,& \\
 &  ...& &\\
\left( w_1, w_2 \right) &=&   \left( 1/3  , 1/3 \right) 
&\mbox{(again, a single  radiation fluid)} .
\end{array}\label{list2}
\ee

\subsubsection{Cosmological constant plus a second (real or effective) 
fluid}

In this case $w_1=-1$, $n$ disappears from Eq.~(\ref{acci1}), which gives 
$w_2=-1$, producing only a cosmological constant with no other fluids.

\subsubsection{Stiff matter plus a second (real or effective) fluid}

In this case $w_1=1$, 
\be
0\leq w_2= \frac{(n-1)}{n} \leq 2 \,,
\ee
and we have the pairs 
\be
\begin{array}{rcll}
\left( w_1, w_2 \right) &=& \left( 1, 1 \right) & 
\;\;\;\;\mbox{(single stiff fluid)},\\
&  ... &   &\\
\left( w_1, w_2 \right) &=& \left( 1, 4/3  \right) ,& \\
\left( w_1, w_2 \right) &=& \left( 1, 3/2  \right) ,&\\
\left( w_1, w_2 \right) &=& \left( 1,  2\right) ,&\\
\left( w_1, w_2 \right) &=& \left( 1, 0 \right) &  \mbox{(stiff 
matter plus dust)},\\
\left( w_1, w_2 \right) &=& \left( 1, 1/2 \right) ,&\\
\left( w_1, w_2 \right) &=& \left(  1, 2/3 \right) ,&\\
&  ... & \\
\left( w_1, w_2 \right) &=& \left( 1 , 1 \right) &\mbox{(again, a 
single stiff fluid)}.
\end{array} \label{list3}
\ee

\subsection{Second condition: $ \frac{p+1}{r} +q =n \in 
\mathbb{Z} $}

Let us explore now the second possibility (\ref{second}) of expressing the 
integral 
$I$ in 
terms of elementary functions. Imposing $\frac{p+1}{r} +q =n \in 
\mathbb{Z}$ \cite{Chen:2014fqa}, one obtains
\be
w_2=\frac{2n w_1 -1}{2n+1} \,. \label{acci2}
\ee

Solving the Friedmann equation with an hypergeometric function, McIntosh 
\cite{McIntosh1972} and later McIntosh and Foyster 
\cite{McIntoshFoyster1972} 
noted that the latter truncates and simplifies to elementary functions 
when one of its 
arguments is an integer (which is the situation when the Chebysev theorem 
arises, and the route followed by the authors of \cite{Chen:2014fqa}, who 
apparently were unaware of Refs.~\cite{McIntosh1972,McIntoshFoyster1972}). 
This happens when \cite{McIntosh1972,McIntoshFoyster1972} (cf. Eqs.~(22) 
and~(24) of Ref.~\cite{McIntosh1972})
\be
\frac{\gamma_1}{\gamma_2} = 1-\frac{1}{m}\,,
\ee
where $\gamma_{1,2} \equiv w_{1,2}+1$ and $m\in \mathbb{Z}$. This relation 
is 
equivalent to 
\be
m w_1 +(1-m) w_2 = -1 \,.
\ee
Since $m\in \mathbb{Z}$, setting $m=2n \in \mathbb{Z}$ gives 
back Eq.~(\ref{acci2}).  

Let us fix the first fluid and look for a second fluid that guarantees 
``simple integrability''.

\subsubsection{Dust plus a second (real or effective) fluid}

If the first fluid is a dust, $w_1=0$, we have 
\be
-1 \leq  w_2=-\frac{1}{2n+1} \leq 1 
\ee 
and the pairs
\be
\begin{array}{rcll}
\left(w_1, w_2\right) &=&   \left(0 ,0 \right) & \mbox{(a 
single dust fluid)},\\
& ... & & \\  
\left(w_1, w_2\right) &=&   \left( 0 , 1/5 \right) ,\\ 
\left(w_1, w_2\right) &=&    \left( 0 , 1/3 \right) & \mbox{(dust 
plus radiation)}, \\
\left(w_1, w_2\right) &=&   \left( 0 , 1 \right) & \mbox{(dust plus 
stiff fluid)},\\
\left( w_1, w_2\right) &=&   \left( 0 , -1 \right) & \mbox{(dust plus 
$\Lambda$)},\\
\left(w_1, w_2\right) &=&   \left( 0 , -1/3 \right) &  
\mbox{(dust plus spatial curvature)},\\ 
\left(w_1, w_2\right) &=&   \left( 0 ,  - 1/5 \right) ,&\\
\left(w_1, w_2\right) &=&   \left( 0 ,  - 1/7 \right) ,&\\
& ... & &\\
\left(w_1, w_2\right) &=&   \left( 0 , 0 \right)  & \mbox{(again, 
a single dust fluid)}. 
\end{array} \label{list4}
\ee

\subsubsection{Radiation plus a second (real or effective) fluid}

If we start from radiation, $w_1=1/3$, we obtain 
\be
-1 \leq w_2=\frac{2n-3}{3(2n+1)} \leq \frac{5}{3} 
\ee
and the pairs
\be
\begin{array}{rcll}
\left(w_1, w_2\right) &=&  \left( 1/3  ,1/3 \right) 
& \mbox{(single radiation fluid )},\\
& ... & &\\
\left(w_1, w_2\right) &=&   \left( 1/3 , 3/5 \right) ,&\\ 
\left(w_1, w_2\right) &=&  \left( 1/3  , 7/9 \right) ,&\\
\left(w_1, w_2\right) &=&   \left( 1/3 , 5/3 \right) ,&\\
\left(w_1, w_2\right) &=&   \left( 1/3  , 1 \right) &  
\mbox{(radiation plus stiff fluid)},\\ 
\left(w_1, w_2\right) &=&   \left( 1/3  , -1/9 \right) ,& \\
\left(w_1, w_2\right) &=&   \left( 1/3  , 1/15 \right) ,&\\
\left(w_1, w_2\right) &=&   \left( 1/3  , 1/7 \right) ,&\\
& ... & &\\
\left(w_1, w_2\right) &=&    \left( 1/3  , 1/3 \right)  
&\mbox{(again, a single radiation fluid)}. 
\end{array} \label{list5}
\ee

\subsubsection{$ \Lambda$ plus a second (real or effective) 
fluid}

Again, setting $w_1=-1$ makes $n$ disappear from Eq.~(\ref{acci2}) and  
produces only $w_2=-1$: there is only a cosmological constant in a 
spatially flat FLRW universe.

\subsubsection{Stiff matter plus a second (real or effective) fluid}

Setting $w_1=1$ (the equation of state parameter of a stiff fluid) yields 
\be
-1 \leq  w_2= \frac{2n-1}{2n+1} \leq 3
\ee
and the pairs
\be
\begin{array}{rcll}
\left( w_1, w_2 \right) &=& \left( 1, 1 \right)& \mbox{(single stiff 
fluid)},\\
&  ... &  &\\
\left( w_1, w_2 \right) &=& \left( 1, 7/5  \right) ,& \\
\left( w_1, w_2 \right) &=& \left( 1,  5/3 \right) ,&\\
\left( w_1, w_2 \right) &=& \left( 1, 3 \right) ,&\\
\left( w_1, w_2 \right) &=& \left( 1, -1 \right) & \mbox{(stiff fluid plus  
$\Lambda$)},  \\
\left( w_1, w_2 \right) &=& \left( 1, 1/3 \right) &
\mbox{(stiff fluid plus radiation)},\\
\left( w_1, w_2 \right) &=& \left( 1, 3/5 \right) ,&\\
\left( n, w_2 \right) &=& \left(  1, 5/7 \right) ,&\\
&  ... & &\\
\left( w_1, w_2 \right) &=& \left( 1 , 1 \right) & \mbox{(again, a 
single stiff fluid)}.
\end{array}\label{list6}
\ee

\subsection{Summary}

In all cases, when $n\rightarrow \pm \infty $, the solutions degenerate  
into a single fluid solution. If both conditions (\ref{acci1}) and 
(\ref{acci2}) hold simultaneously, one necessarily has $w_1=w_2=-1$, which 
is the case of a single cosmological constant  with $K=0$ and no real 
fluids.

Based on the tables of values $\left(w_1, w_2\right)$ obtained, in which 
we always find $w_2\geq -1$, one concludes that, when one the fluids is 
phantom and the other is dust, radiation, or stiff matter, there are no 
analytical solutions in the form $t=t(a)$ or $a=a(t)$ expressed in terms 
of 
elementary functions (there can still be simple analytical solutions in 
parametric form, or solutions given by a hypergeometric function or 
elliptic integral, or solutions with a time-dependent equation of state). 
The same conclusion applies to quintessence fluids with realistic values 
of the equation of state parameter, other than $\Lambda$. In the tables 
reported, ``real'' quintessence fluids (other than $\Lambda$)  have 
equation of state parameter $w_2 $ far from $-1$, while current 
observations give $w \simeq -1 $ \cite{Planck}.

Looking at the previous tables, the situations involving physically 
interesting fluids in which the two-fluid solution can 
be expressed in terms of elementary functions are:

\begin{itemize}

\item dust plus spatial curvature, $\Lambda=0$;

\item radiation plus spatial curvature, $\Lambda=0$;

\item $K=0, \Lambda=0$, dust plus radiation;

\item $K=0$, dust plus $\Lambda$;

\item $K=0, \Lambda=0$, dust plus stiff matter;

\item $K=0, \Lambda=0$, radiation plus stiff matter.

\end{itemize}

Both single fluid solutions for $K\neq 0, \Lambda=0$ (studied in 
\cite{Chen:2014fqa} in $D$ spatial dimensions) and $K=0$ FLRW universes with a 
single fluid and $\Lambda\neq 0$ can be obtained by regarding them as 
two-fluid solutions in a fictitious spatially flat universe, provided that 
they fall within the list of cases above.

\subsection{Exotic fluids}

In addition to the classic dust, radiation, non-diluting $\Lambda$ (with 
$\dot{\rho}_{\Lambda}=0$), and hypothetical phantom dark energy which 
concentrates with the cosmic expansion ({\em i.e.}, 
$\dot{\rho}=-3(w+1)\rho >0 $ for $w<-1$), other exotic equations of state 
are of physical interest.

A stiff fluid with $w=1$  is believed to be appropriate to 
describe matter at high (nuclear) densities and it is used in astrophysics 
to model dense nuclear matter in the core of neutron stars. It is 
reasonable that, as the universe cooled from high temperature and energy 
scales, it underwent a period at lower temperatures and nuclear densities 
described by the stiff equation of state.

A stiff fluid also corresponds to a free scalar field $\phi$ minimally 
coupled to the Ricci curvature. By setting 
$V(\phi)=0$ in Eqs.~(\ref{rhophi}) and (\ref{Pphi})  it is 
$P_{\phi}=\rho_{\phi}$ and the scalar field behaves as a stiff perfect 
fluid. This regime, called ``kination'' in early universe 
literature, is achieved exactly when the potential $V$ is absent and 
approximately when $\dot{\phi}^2 \ll V(\phi)$. Then the Klein-Gordon 
equation~(\ref{KleinGordon})  admits the 
first integral
\be
\dot{\phi}= \frac{C_0}{a^3} \,,\label{KGfirstintegral}
\ee
or
\be
\phi_{, \eta}= \frac{C_0}{a^2} \,,\label{KGfirstintegral-eta}
\ee
where $C_0$ is an integration constant. Authors from the 1960s and 1970s, 
who were not particularly interested in scalar fields,\footnote{Interest 
in scalar field cosmology arose with inflationary scenarios in the 1980s 
\cite{KolbTurner, LiddleLyth, Slava, Lidsey:1995np, Martin:2013tda} and, 
again, with the discovery of the present acceleration of the universe in 
1998 \cite{Perlmutter:1997zf, Riess:1998cb, Perlmutter:1998np}.} did not 
regard a stiff fluid as the realization of a scalar field and did not 
integrate Eq.~(\ref{KGfirstintegral}) or Eq.~(\ref{KGfirstintegral-eta}). 
We will report the explicit form of $\phi$ in this kination regime,  when 
it is 
given explicitly in terms of elementary functions. A stiff fluid can also 
mimic anisotropy in the universe \cite{Misner:1967uu,HughstonShepley1970, 
Jacobs1968, Shikin1968, McIntosh1972}.

Finally, certain string-inspired effective gravitational field theories 
of the primordial universe \cite{Basilakos:2019acj, Basilakos:2020qmu, 
Basilakos:2019mpe, Mavromatos:2020crd, Mavromatos:2020kzj} contain a 
massless Kalb-Ramond axion of gravitational nature which is 
equivalent to an effective fluid with a stiff equation of state, and other 
stringy axions are possible. These fields are linked to matter-antimatter 
asymmetry in the universe after inflation and to axionic dark matter. It 
is speculated that a stiff matter era precedes inflation in these 
theories \cite{Mavromatos:2020kzj}.

The value $-1/3 $ of the equation of state parameter is also motivated in 
cosmology: a frustrated cosmic string network produces an effective fluid 
with this equation of state \cite{Vilenkin:1984rt, Spergel:1996ai}.

A fluid with equation of state parameter $w=-2/3$ describes domain walls 
and is interesting in cosmology \cite{Bucher:1998mh, Battye:1999eq, 
Conversi:2004pi} but, unfortunately, it does not appear in the lists of 
integrable cases found above.  In general, a network of non-intercommuting 
$n$-dimensional topological defects produces a fluid with effective 
equation of state parameter $ w=-n /3 $ \cite{Vilenkin:1984rt, 
Spergel:1996ai}.

The other values of the equation of state parameter $w$ for which one has 
integrability in terms of elementary functions are not particularly 
interesting from the physical point of view, except perhaps as toy models. 
Among these, solutions involving a $w=2/3$ fluid were given by Vajk 
\cite{Vajk1969}.

\section{Two fluids with constant equations 
of state---parametric solutions with conformal time}
\setcounter{equation}{0}
\label{sec:4}

Several exact solutions, for both single- and two-fluid solutions, can be 
obtained in parametric form using the conformal time $\eta$ defined by $dt 
\equiv a d\eta$ as a parameter.  Two-fluid solutions in this form are 
perhaps not 
as widely known as single-fluid solutions (which are instead reported in 
the pedagogical literature \cite{LandauLifschitz, Wald, Carroll, 
EllisMaartensMacCallum, Faraoni:1999qu, Lima:2001fi}).

In terms of conformal time, the Friedmann equation for a FLRW universe 
with a single matter fluid (satisfying $P=w\rho$, $w=$~const.),  
cosmological constant $\Lambda$, and spatial curvature reads 
\be
\left(a_{,\eta}\right)^2= \frac{8\pi}{3} \, \frac{ \rho^{(0)}}{a^{3w-1} }
+\frac{\Lambda a^4}{3} -Ka^2 \,,
\ee
which gives, along the lines of the discussion already made using comoving 
time,
\be
\int \frac{da}{ \sqrt{ \frac{8\pi}{3} \, \frac{ \rho^{(0)}}{a^{3w-1} }
+\frac{\Lambda a^4}{3} -Ka^2 } }=\pm \left( \eta-\eta_0 \right) 
\,,\label{eq:3.2}
\ee
where $\eta_0$ is an integration constant. From now on, this section 
follows Refs.~\cite{McIntoshFoyster1972,Chen:2014fqa} and we assume that the 
equation of state parameter $w$ is rational.

\subsection{$ K=0, \Lambda\neq 0 $, any $ w\in \mathbb{Q}$}

For this combination of parameters, the Einstein-Friedmann 
equations~(\ref{Friedmann1})-(\ref{conservation1}) can always be integrated in 
terms of simple functions \cite{Harrison1967,Vajk1969,McIntosh1972, 
Chen:2014fqa}.  
Let us assume that $w\neq -1, -1/3$, otherwise we fall into one of 
the cases previously discussed. 
The integral on the left hand side of Eq.~(\ref{eq:3.2}) has the form 
\be
\sqrt{3} \int da \, a^{-2} \left[ 8\pi \rho^{(0)} a^{-3(w+1)} 
+\Lambda \right]^{-1/2} \,,
\ee
{\em i.e.}, the form~(\ref{integral}) with 
\be
p=-2 \,, \;\;\;\;\;\; r=-3(w+1) \neq 0 \,, \;\;\;\;\;\; q=-\frac{1}{2}\,,
\ee
which are all rational if $w$ is. The Chebysev theorem applies if one of 
\be
\frac{p+1}{r}= \frac{1}{3(w+1)} \,, \;\;\;\;\;\;\;\;\;\;
\frac{p+1}{r}+q = -\frac{3w+1}{6(w+1)} 
\ee
is an integer. These conditions correspond, respectively, to 
\begin{eqnarray}
w &=& -1+\frac{1}{3n} \,,\label{condition1prime}\\
&&\nonumber\\
w &=& -1 +\frac{2}{3(2m+1)}  \,, \label{condition2prime}
\end{eqnarray}
with $n,m \in \mathbb{Z}$. These two conditions are mutually exclusive and 
they both allow to get arbitrarily close to the phantom divide $w=-1$ if 
$|n| $ or $|m|$ are  sufficiently large, spanning both quintessence 
($-1\leq w< 
-1/3$)  and  phantom ($w<-1$) equations of state near the $w=-1$ limit. 

Imposing the first condition~(\ref{condition1prime}) for   
integrability in terms of elementary functions, as $n$ spans the 
values $n=-\infty , \, ... \,, -3, 
-2, -1, 1, 2, 3, \, ... \,, +\infty$, one obtains the corresponding 
equation of state parameters 
\be
w=-1, \, ... \,,  -\frac{10}{9}, -\frac{7}{6}, - \frac{ 4}{3}, 
-\frac{2}{3}, 
- \frac{5}{6}, -\frac{8}{9}, \, ... \,, -1 \,.
\ee

Imposing instead the second condition~(\ref{condition2prime}), for  
$m= -\infty , \, ... \,, -3, -2, -1, 0, 1, 2, 3, \, ... \,, +\infty$ one  
obtains 
\be
w=-1, \, ... \,, -\frac{17}{15}, -\frac{11}{9}, - \frac{5}{3}, 
-\frac{1}{3}, -\frac{7}{9}, -\frac{13}{15}, -\frac{19}{21}, \, ... \,, -1 
\,.
\ee
The degenerate case $w=-1$ reproduces the empty, spatially flat, de Sitter 
universe, while $w=-1/3$ gives a spatially curved universe filled 
with a string gas and with $\Lambda$ already discussed. None of the other 
values of $w$ correspond to dust, radiation, or stiff matter, but they 
cover quintessence and phantom equations of state. There is no general 
formula expressing the solution that is valid for all values of $w$ 
corresponding to integrability in terms of elementary functions.

\subsection{Any $K, \Lambda = 0$, any $ w\in \mathbb{Q}$}

This case gives rise to more physically interesting situations. When the 
fluid is radiation or dust, this is a standard solution found in all 
cosmology 
textbooks \cite{LandauLifschitz, Weinberg, Wald, Carroll, 
Liddle, EllisMaartensMacCallum}. For $K=0$, the solution is elementary and 
given by Eq.~(\ref{usual}). When $K=\pm 1$ and there is only radiation, 
one can eliminate the parameter $\eta$ to obtain $a(t)$.

For $K=\pm 1$ and dust, one can express both $a$ and $t$ as simple 
functions of $\eta$, but eliminating this parameter  can only be done to 
obtain a relation $t(a)$ that cannot be inverted explicitly 
to obtain $a(t)$.

For $K=\pm 1$ and $w\neq 0, 1/3$, in general, one can compute $a (\eta)$ 
explicitly in terms of elementary functions, while $t(\eta)$ is reduced to 
a quadrature and written as an integral that may not have an explicit 
expression (see Sec.~\ref{sec:5}).

Explicitly, for any $K$, $\Lambda=0$, and $w\in \mathbb{Q}$, it is 
possible to express $a(\eta)$ in terms of elementary functions. Various 
solutions have been proposed by many authors over the years (see the next 
sections). The general solution is 
given by \cite{Chen:2014fqa}
\be
\pm \left( \frac{3w+1}{2} \right) \left(\eta-\eta_0 \right) = 
\left\{ 
\begin{array}{lc}
-\tan^{-1} \left[ f(a) \right] & \mbox{if }\, K=+1 \,,\\
& \\
\frac{1}{f(a)} & \mbox{if }\, K=0 \,,\\
& \\
\tanh^{-1} \left| \frac{f(a)+1}{f(a)-1} \right| & \\
& \mbox{if }\, K=-1 \,,\\
=\frac{1}{2} \ln \left| \frac{f(a)+1}{f(a)-1} \right| & 
\end{array} \right.  \label{general}
\ee
where 
\be
f(a) = \sqrt{ \frac{8\pi}{3} \, \rho^{(0)} a^{-(3w+1)} -K} \,.
\ee

\subsubsection{$K=0, \Lambda=0, w$-fluid}

For $K=0, \Lambda=0$, Eq.~(\ref{general}) yields 
\cite{Harrison1967,Vajk1969, Chen:2014fqa}
\be \left\{
\begin{array}{lll}
a(\eta) &=& \left( \frac{2\pi}{3} \, \rho^{(0)} \right)^{ 
\frac{1}{3w+1} } 
\left[ \left(3w+1\right) \left( \eta-\eta_0 \right) \right]^{ 
\frac{2}{3w+1} 
}  \,,\\
&&\\
t(\eta) &=& 
\frac{  \left( 3w+1 \right)^{ \frac{3(w+1)}{3w+1} } }{3 (w+1) }
\left( \frac{2\pi}{3} \, \rho^{(0)} \right)^{ \frac{1}{3w+1} } 
\left( \eta-\eta_0 \right)^{ \frac{ 3(w+1) }{ 3w+1} } + t_0 \,. 
\end{array} \right. 
\ee
The initial condition at $t=t_0$ (corresponding to $\eta=\eta_0$) is 
$a(0)=0$; this universe expands forever and the fluid redshifts away if 
$w>-1$.

\subsubsection{$K=+1, \Lambda=0, w$-fluid}

In this case Eq.~(\ref{general}) yields
\be
f(a) = \pm \tan\left[  \left(\frac{3w+1}{2} \right) \left( 
\eta-\eta_0\right) \right] \,.
\ee
Using the expression of $f(a)$ and squaring yields 
\be
\frac{8\pi}{3} \, \rho^{(0)} a^{-(3w+1)} = \cos^{-2}  
\left[  \left(\frac{3w+1}{2} \right) \left( \eta-\eta_0\right) \right] 
\ee
and finally \cite{Harrison1967,Vajk1969, Chen:2014fqa}
\be\left\{
\begin{array}{lll}
a(\eta) &=& \left( \frac{8\pi}{3} \, \rho^{(0)} \right)^{ 
\frac{1}{3w+1}} \left\{ \cos 
\left[ \frac{3w+1}{2} \left( \eta-\eta_0 \right) \right]\right\}^{ 
\frac{2}{3w+1}} 
\,,\nonumber\\
&&\\
t(\eta) &=& \int a(\eta) d\eta \,.
\end{array} \right.
\ee
If the initial condition $a(\eta=0)=0$ is imposed, it must be $\eta_0=\pi 
/(3w+1)$. The special cases of this solution for dust and radiation 
were found by Friedmann in 1922 \cite{Friedmann1922} and Tolman in 1931 
\cite{Tolman1931}, respectively.

\subsubsection{$K=-1, \Lambda=0, w$-fluid}

Two-(effective) fluid solutions for $K\Lambda\neq 0$ for dust and for 
radiation (separately) were reported in terms of elliptic functions by 
Lema\^itre \cite{Lemaitre1933}, Edwards \cite{Edwards1972} and Kharbedya 
\cite{Kharbedya1976}.  The solution for $K=-1, \Lambda=0$ and any 
$w$-fluid was reported by Tauber \cite{Tauber1967}. In this case, the 
exponentiation of Eq.~(\ref{general}) gives
\be
\left| \frac{f(a)+1}{f(a)-1} \right| = 
\mbox{e}^{ \pm (3w+1) (\eta-\eta_0)} 
\ee
and it is straightforward to obtain
\begin{eqnarray}
f(a) &=& - \frac{1+ \mbox{e}^{ \pm (3w+1) \left( \eta-\eta_0 \right)}  }{
1- \mbox{e}^{ \pm (3w+1) (\eta-\eta_0)}  } \nonumber\\
&&\nonumber\\
&=& -\coth
\left[  \pm \frac{(3w+1)}{2} \left( \eta-\eta_0 \right) \right] \,.
\end{eqnarray}
Using the expression of $f(a)$ and squaring yields 
\cite{Vajk1969, Chen:2014fqa}
\be \left\{
\begin{array}{lll}
a(\eta) &=& \left( \frac{8\pi}{3} \, \rho^{(0)} \right)^{ 
\frac{1}{3w+1}} \left\{ \sinh 
\left[ \frac{3w+1}{2} \left( \eta-\eta_0 \right) \right]\right\}^{ 
\frac{2}{3w+1}} \,,\\
&&\\
t(\eta) &=& \int a(\eta) d\eta \,.
\end{array} \right.
\ee
If $w>-1/3$, the initial condition is $a( \eta_0)=0$. If $w=-1/3$, we have 
an empty universe with hyperbolic 3-sections and zero cosmological 
constant: this is the Milne universe, which is nothing but Minkowski space 
sliced with a hyperbolic foliation ({\em e.g.}, 
\cite{EllisMaartensMacCallum, Slava, Wald}).

The most well known two-fluid solution in parametric form is the one for 
dust plus radiation and $K=0$ (see Eq.~(\ref{K=0dust+radiation})).

\section{Explicit solutions}
\setcounter{equation}{0}
\label{sec:5}

In the following, we present the known explicit analytical solutions 
of the Einstein-Friedmann 
equations~(\ref{Friedmann1})-(\ref{conservation1})    expressible in terms 
of elementary functions, 
for two real or effective fluids with constant, linear, barotropic 
equation of state $P=w\rho$, $w=$~const.  
Not all the situations, listed in the previous two sections, in which 
integration in finite terms by means of elementary functions is 
possible are physically motivated. Therefore, we limit ourselves to the 
most physically significant 
values of the equation of state parameters for the real fluids 
(dust, radiation, or stiff fluid).  
We first provide the solutions using comoving time $t(a)$ or,
whenever possible, their inverse $a(t)$. Then, we provide the solution in 
the parametric form 
\be \left\{
\begin{array}{lll}
a&=& a(\eta) \,,\\
&&\\
t&=&t(\eta) \,,
\end{array} 
\right.
\ee 
using the conformal time $\eta$ as the parameter, which may be useful in 
certain applications where conformal time is preferred for computational 
convenience (for example, to solve the equations for scalar or tensor 
perturbations in slow-roll inflation \cite{KolbTurner, LiddleLyth, Slava, 
Lidsey:1995np, Martin:2013tda}).  We 
also report initial 
conditions, asymptotics, and single-fluid limits for these solutions.

\subsection{Dust, $K=0, \Lambda=0$}

This well known single fluid solution is 
\begin{eqnarray}
a(t) &=& \left( 6\pi \rho_\text{m}^{(0)} \right)^{1/3} \left( 
t-t_0\right)^{2/3}\\
&&\nonumber\\
\rho(a) &=& \frac{\rho_\text{m}^{(0)} }{a^3}= \frac{ 1}{ 6\pi (t-t_0)^2} 
\,,
\end{eqnarray}
or, in the less commonly encountered parametric form,
\be \left\{
\begin{array}{lll}
a(\eta) & = & \frac{2\pi}{3} \, \rho_\text{m}^{(0)} \left( 
\eta-\eta_0\right)^2 
\,,\\
&&\\
t(\eta) & = & \frac{2\pi}{9} \, \rho_\text{m}^{(0)} \left( \eta-\eta_0 
\right)^3 
+t_0 \,.
\end{array} \right.
\ee
The initial condition is a Big Bang $a=0$ at $t=t_0$, corresponding to 
$\eta=\eta_0$.

\subsection{Dust plus spatial curvature, $\Lambda=0$}

In its parametric form using conformal time, the solution is well known 
(Tauber \cite{Tauber1967} and Gilman \cite{Gilman:1970zv} gave the solution 
for $K=0, \pm 1$) and is found in standard cosmology textbooks, {\em 
e.g.}, \cite{Wald,Liddle,Carroll}. It is rarer to find it expressed in 
terms of comoving time, in which case it reads 
\cite{Vajk1969,Aldrovandi:2005ya}
\be
t(a)=- \sqrt{ \frac{8\pi}{3} \, \rho_\text{m}^{(0)} a-a^2} 
+\frac{8\pi}{3} \, \rho_\text{m}^{(0)} \sin^{-1} \left( \sqrt{ 
\frac{3a}{8\pi \rho_\text{m}^{(0)} } } \, \right) +t_0  
\label{DustCurvPos}
\ee 
for $K=+1$ and 
\be
t(a)= \sqrt{ \frac{8\pi}{3} \, \rho_\text{m}^{(0)} a+ a^2} 
-\frac{8\pi}{3} \, \rho_\text{m}^{(0)} \sinh^{-1} \left( \sqrt{ 
\frac{3a}{8\pi \rho_\text{m}^{(0)} } } \, \right) +t_0 \label{DustCurvNeg}
\ee 
for $K=-1$ \cite{Vajk1969,Aldrovandi:2005ya}. Both solutions satisfy the 
Big Bang initial condition $a(t_0)=0$. 

The parametric forms of these solutions are \cite{Weinberg,Wald,Carroll, 
Liddle}
\be \left\{
\begin{array}{lll}
a(\eta) &=& \frac{4\pi}{3} \, \rho_\text{m}^{(0)} \left(1-\cos\eta\right) 
\,,\\
&&\\
t(\eta) &=& \frac{4\pi}{3} \, \rho_\text{m}^{(0)} \left(\eta 
-\sin\eta\right) 
+t_0  \,,
\end{array} \right.
\ee
for $K=+1$ and
\be \left\{
\begin{array}{lll}
a(\eta) &=& \frac{4\pi}{3} \, \rho_\text{m}^{(0)} \left(\cosh\eta -1 
\right) 
\,,\\
&&\\
t(\eta) &=& \frac{4\pi}{3} \, \rho_\text{m}^{(0)} \left( \sinh\eta - \eta 
\right) 
+t_0  \,,
\end{array} \right.
\ee
for $K=-1$ where, in both cases, $t=t_0$ corresponds to $\eta=0$ and 
$a=0$ and 
\be
\rho_\text{m} (a) = \frac{\rho_\text{m}^{(0)} }{a^3} \,, \quad \quad 
\rho_K (a) = \frac{-3K}{8\pi a^2} \,.
\ee
The parametric forms make it clear that the solution for $K=+1$ 
reaches a maximum size $a_\text{max}= \frac{8\pi}{3}\, 
\rho_\text{m}^{(0)}$ at 
$\eta=\pi$ (or $t=\frac{4\pi^2}{3} \, \rho_\text{m}^{(0)}$), while the 
$K=-1$ 
universe expands forever becoming curvature-dominated.

\subsection{Radiation, $K=0, \Lambda=0$}

This is another classic textbook solution
\begin{eqnarray}
a(t) &=& \left( \frac{32\pi}{3} \, \rho_\text{r}^{(0)} \right)^{1/4} 
\sqrt{ 
t-t_0} \,,\\
&&\nonumber\\
\rho_\text{r}(a) &=& \frac{\rho_\text{r}^{(0)} }{a^4} \,,
\end{eqnarray}
or, in parametric form,
\be \left\{
\begin{array}{lll}
a(\eta) &=& \sqrt{ \frac{8\pi}{3} \, \rho_\text{r}^{(0)} } \left( 
\eta-\eta_0 
\right) \,,\\
&&\\
t(\eta) &=& \sqrt{ \frac{2\pi}{3} \, \rho_\text{r}^{(0)} } \left( 
\eta-\eta_0\right)^2 +t_0 \,,
\end{array} \right.
\ee
where $t_0$ is an integration constant. The Big Bang initial condition 
$a=0$ at $t=t_0$ (or $\eta=\eta_0$) has been imposed and the universe 
expands forever.

\subsection{Radiation plus spatial curvature, $\Lambda=0$}

This solution is well-known and found in standard cosmology textbooks, 
{\em e.g.}, \cite{Wald,Liddle,Carroll} and as a special case of more 
complicated solutions \cite{Aldrovandi:2005ya}. For $K=+1$, the scale factor 
is
\be 
a(t)= 
\sqrt{ \frac{8\pi}{3} \, \rho_\text{r}^{(0)} -\left( t-t_0\right)^2} 
\label{RadCurvPos}
\ee 
or, in parametric form, 
\be 
\left\{ \begin{array}{lll} 
a(\eta) &=& \sqrt{ \frac{8\pi}{3} \, \rho_\text{r}^{(0)} } \, \sin \eta 
\,,\\ 
&&\\ 
t(\eta) &=& t_{0}- \sqrt{ \frac{8\pi}{3} \, \rho_\text{r}^{(0)} } \, \cos 
\eta \,. 
\end{array}  \right. 
\ee 
This solution has a Big Bang at $t=t_0 -\sqrt{8\pi 
\rho_\text{r}^{(0)}/3}$ (or $\eta=0$) and a Big Crunch at 
$t= t_{0}+ \sqrt{8\pi \rho_\text{r}^{(0)}/3}$ (or $\eta=\pi$).

For $K=-1$, the scale factor is
\be 
a(t)= \sqrt{ \left( t-t_0\right)^2 - \frac{8\pi}{3} \, \rho_\text{r}^{(0)} 
} 
\label{RadCurvNeg}
\ee
or, in parametric form,
\be \left\{
\begin{array}{lll}
a(\eta) &=& \sqrt{ \frac{8\pi}{3} \, \rho_\text{r}^{(0)} } \, \sinh \eta 
\,,\\
&&\\
t(\eta) &=& \sqrt{ \frac{8\pi}{3} \, \rho_\text{r}^{(0)} } \, \cosh \eta 
+t_0 
\,.
\end{array} \right.
\ee
There is a Big Bang at $\eta=0$, corresponding to $t= t_0+\sqrt{ 8\pi 
\rho_\text{r}^{(0)}/3} $.  At late times $t\rightarrow + \infty$, the 
solution is approximately linear, 
$a(t) \sim t$.

\subsection{$K=0, \Lambda=0$, dust plus radiation} 

Solutions describing dust plus radiation are, no doubt, some of the 
most well-motivated because of the need to describe the simultaneous 
presence of non-relativistic matter and cosmic background 
radiation or, more in general, neutrinos or ultra-relativistic particle 
species. 
FLRW universes containing dust and radiation have been studied  
in the years immediately following the discovery of the cosmic microwave 
background \cite{Penzias:1965wn}, beginning with Alpher \& Herman 
\cite{Alpher:1949sef} and 
continuing with the works of Chernin \cite{Chernin1966}, Jacobs 
\cite{JacobsNature1967}, Cohen \cite{CohenNature1967}, Roeder 
\cite{Roeder1967}, McIntosh \cite{McIntosh1968}, Vajk \cite{Vajk1969}, 
Sapar \cite{Sapar1970}, Sistero \cite{Sistero1972}, May \cite{May1975}, 
Coquereaux \& Grossman \cite{CoquereauxGrossman1982}, and Dabrowski \& 
Stelmach \cite{DabrowskiStelmach1986}. However this exact solution for two 
non-interacting fluids cannot reproduce the complicated microphysics of 
ultrarelativistic particle species decaying, electrons combining with 
protons, {\em etc.} When an accurate description is needed, for example 
to study imprints left in the cosmic microwave background to probe the 
thermal history of the universe, the 
physics must inserted in numerical calculations in detailed scenarios 
\cite{Peebles:1968ja, Zeldovich:1969ff, Sunyaev:1970er, Hu:1992dc, 
Chluba:2010ca, AliHaimoud:2010dx, Chluba:2015gta}.

The solution with non-interacting dust and radiation and $\Lambda=0$ was 
found by Tolman for $K=+1$ \cite{Tolman} and by Chernin \cite{Chernin1966} 
and McIntosh \cite{McIntosh1968b} for any $K$.

The spatially flat, $\Lambda=0$, radiation plus dust solution  due to 
Jacobs \cite{JacobsNature1967} 
(see also \cite{CohenNature1967,Vajk1969}) is probably the most well 
known solution 
with two real (as opposed to effective) fluids \cite{Chen:2014fqa}.  In 
spite of 
its importance for the transition from the radiation to the dust era, it 
rarely appears \cite{EllisMaartensMacCallum} in modern textbooks. Setting 
$w_1=0, w_2=1/3$, the relevant integral~(\ref{eq:2.3}) becomes
\begin{eqnarray} 
\int \frac{da \, a}{\sqrt{ 
\rho_\text{m}^{(0)} a +\rho_\text{r}^{(0)} }} &=& \frac{2}{3 \left( 
\rho_\text{m}^{(0)} 
\right)^2 } \left( \rho_\text{m}^{(0)} a-2\rho_\text{r}^{(0)} \right) 
\sqrt{ 
\rho_\text{m}^{(0)} a+ \rho_\text{r}^{(0)} } \nonumber\\
&&\nonumber\\
&=& \sqrt{ \frac{8\pi}{3}} \left( t-t_0 \right) \,.\label{100salcaz}
\end{eqnarray}
This $K=0$ solution was given by Jacobs in 
1967 \cite{JacobsNature1967} (see also \cite{Vajk1969}) as
\be
t(a)= A \left( \rho_\text{m}^{(0)} a-2\rho_\text{r}^{(0)} 
\right) \sqrt{ \rho_\text{m}^{(0)}  a+ \rho_\text{r}^{(0)} } +t_0 \,,  
\label{K=0dust+radiation}
\ee
which coincides with~(\ref{100salcaz}) for $ A= \left[\sqrt{6\pi} \, 
(\rho_\text{m}^{(0)})^2\right]^{-1}$.

In parametric form, the solution is\footnote{The authors of 
Ref.~\cite{Mukhanov:1990me,Aldrovandi:2005ya} were apparently unaware of 
Vajk's paper \cite{Vajk1969, Mukhanov:1990me, Aldrovandi:2005ya}.} 
\cite{Vajk1969, Mukhanov:1990me, Aldrovandi:2005ya}  
\be
\left\{
\begin{array}{lll}
a (\eta) &=& \frac{2\pi}{3} \, \rho_\text{m}^{(0)}  
\eta^2  -\frac{\rho_\text{r}^{(0)}}{\rho_\text{m}^{(0)}} \,   \,,\\
&& \label{flatdust+radiationparametric}\\
t (\eta) &=& \frac{2\pi}{9} \, \rho_\text{m}^{(0)} \,  \eta^3  - \frac{\rho_\text{r}^{(0)}}{\rho_\text{m}^{(0)}}    \, \eta  \,,
\end{array} \right. 
\ee
with the initial condition $a(t=0)=0$.   
This universe expands forever. The radiation fluid dominates at early 
times, when  $a(t) \simeq \sqrt{t-t_0}$, while the dust dominates at 
late times, with $a(t) \simeq \left(t-t_0\right)^{2/3}$.

Let us consider the single-fluid limits. The limit to dust is obtained 
trivially by setting $\rho_\text{r}^{(0)}=0$ in Eq.~(\ref{100salcaz}), 
obtaining the well known power-law $a(t)=a_0 \left(t-t_0\right)^{2/3} $, 
{\em i.e.}, Eq.~(\ref{usual}) with $w=0$.

The radiation-only limit is not as obvious: letting $ \rho_\text{m}^{(0)} 
\rightarrow 0$ in Eq.~(\ref{100salcaz}) (or in the parametric form of 
this solution) does not produce a meaningful result. Instead, one has to 
take the limit $\rho_\text{m}^{(0)}\rightarrow 0$ in the 
integral~(\ref{eq:2.3}), which reduces to 
\be
\int da \, \frac{a}{ \sqrt{\rho_\text{r}^{(0)} } }=\sqrt{\frac{8\pi}{3}}  
\left(t-t_0\right) \,,
\ee
giving $a(t)=a_0 \sqrt{t-t_0}$ (or Eq.~(\ref{usual}) with $w=1/3$). In 
general, the single-fluid limit of a two- or three- (real or effective) 
fluid solution in which radiation survives is more problematic than the 
analogous limit in which a non-radiative fluid remains, as will be seen in 
the following.

The parametric form of this spatially flat universe was rediscovered by 
Barrow \& Saich \cite{Barrow:1993ah} as 
a solution with radiation plus a scalar field in the 
potential\footnote{Barrow \& Saich do not seem to have made the 
connection between their solution and one of the exact integrability cases 
(in the  list~(\ref{list2}) of the 
Einstein-Friedmann equations---the results of 
\cite{Jacobs1968, McIntosh1972, McIntoshFoyster1972} discussed in 
our Sec.~\ref{sec:3} are not mentioned in \cite{Barrow:1993ah}.}
\be
V(\phi)=\frac{V_0 \, \mbox{e}^{\frac{2\phi}{\phi_0}} }{\left( 
\mbox{e}^{ \frac{2\phi}{\phi_0} } + V_1\right)^2} \label{BarrowSaichV}
\ee
with $V_{0,1}, \phi_0 $ constants, but imposing that the equation of state 
of the effective fluid equivalent of the scalar field be constant, 
\be
\frac{ \dot{\phi}^2}{2} =\alpha V(\phi) \,,\label{maiocheneso}
\ee
which gives the effective equation of 
state parameter $w_{\phi} = \frac{\alpha-1}{\alpha+1}$ for the scalar 
field fluid. As expected from the list~(\ref{list2}), 
the case $w_{\phi}=0$ equivalent to $\alpha=1$ is integrable and gives the  
exact 
solution~(\ref{flatdust+radiationparametric})  for the 
potential~(\ref{BarrowSaichV}) \cite{Barrow:1993ah}. The scalar field (not 
reported in \cite{Barrow:1993ah}) can be found by noting that the 
Klein-Gordon equation admits the first integral  \cite{Barrow:1993ah}
\be
\dot{\phi}= \frac{ \phi_0}{ a^{\frac{3\alpha}{\alpha+1}} } 
\ee
which, in our case, yields $\rho_{\phi}= \phi_0^2 a^{-3}$, $d\phi/d\eta 
=\phi_0/\sqrt{a}$ and 
\be
\phi(\eta) = 
\phi_0 \sqrt{\frac{3}{2\pi \rho_\text{m}^{(0)} }} \, \text{arccosh} \left( 
\sqrt{ \frac{2\pi }{3\rho_\text{r}^{(0)} } } \,\, \rho_\text{m}^{(0)} \, 
\eta \right) +\phi_1 \,,
\ee
where $\phi_{0,1}$ are integration constants.

The $K=0$ dust-plus-radiation universe was generalized by Jacobs 
\cite{Jacobs1968} to anisotropic Bianchi~I models, but it is given in 
terms of elliptic integrals. The dust-plus-radiation solution describing  
$K=\pm 1$ FLRW universes is given by Eqs.~(\ref{+1dust+radiation}) and 
(\ref{-1dust+radiation}) below.

\subsection{$K=0$, dust plus $\Lambda$}

If $\Lambda>0$, this is the standard $\Lambda$CDM model (with $\Lambda$ as 
the simplest form of dark energy) and the analytical solution of the 
Einstein-Friedmann equations~(\ref{Friedmann1})-(\ref{conservation1})  is 
well-known ({\em e.g.}, \cite{Aldrovandi:2005ya, 
Chen:2014fqa,Chavanis2015}):

\begin{eqnarray}
a(t)&=& \frac{a_0}{2^{2/3}} \left[ \left(1+\sqrt{ 1+\frac{3M}{a_0^3 
\Lambda}} \, \right) \mbox{e}^{ \frac{\sqrt{3\Lambda}}{2}\, t} 
\right.\nonumber\\
&&\nonumber\\
&\, & \left. + \left(1-\sqrt{ 1+\frac{3M}{a_0^3 \Lambda}} \, \right) 
\mbox{e}^{ -\, \frac{\sqrt{3\Lambda}}{2}\, t} 
\right]^{2/3} \,, \label{LCDM}
\end{eqnarray}
where $a_0=a(t_0)$ at the present time $t_0$,  
\begin{eqnarray}
M &=& \Omega_{m0} H_0^2 a_0^3 =\frac{8\pi}{3} \, 
\rho_\text{m}^{(0)} \,,
\end{eqnarray}
and 
\be
\Omega_\text{m} (t) \equiv \frac{\rho_\text{m}(t) }{\rho_c (t)}=
\frac{8\pi \rho_\text{m}(t)}{3H^2(t)} 
\ee 

\noindent is the dimensionless dust density parameter.  At late times the 
cosmological constant with 
constant energy density $\Lambda/3$ 
dominates over the dust that redshifts away as $a^{-3}$ and the 
solution~(\ref{LCDM}) converges to the de Sitter universe
\be
a(t\rightarrow +\infty) \simeq 
\frac{a_0}{2^{2/3}} \left(1+\sqrt{ 1+\frac{3M}{a_0^3 
\Lambda}} \, \right)^{2/3} \mbox{e}^{H_0 t}
\ee
(where $H_0=\sqrt{\Lambda/3\,}$), which is a phase space attractor. 
Indeed, this is the solution obtained by letting $M\rightarrow 0$ 
in Eq.~(\ref{LCDM}).

The limit to a spatially flat, dust-filled universe without cosmological 
constant is obtained straightforwardly by a formal expansion of 
Eq.~(\ref{LCDM}) as $\Lambda \rightarrow 0$, which produces $a(t) \sim 
t^{2/3}$. 

The solution for  dust and $\Lambda <0$ is obtained as  the special case 
$w_2=0$ of Eq.~(\ref{negLambdaanyw}):
\be
a(t) = \left( \frac{\rho_\text{m} }{|\rho_{\Lambda}|}\right)^{1/3} 
\sin^{2/3} \left[ \frac{ \sqrt{3|\Lambda|} }{2} \left(t-t_0\right) \right] 
\,.
\ee
There are a Big Bang at $t=0$ and a Big Crunch at $t=t_0 +\frac{2\pi}{ 
\sqrt{3|\Lambda|}} $.

\subsection{$K=0$, radiation plus $\Lambda$}

 The solution for the spatially flat FLRW universe containing  
radiation and with $\Lambda > 0 $ is \cite{Harrison1967}
\be
a(t)= \left( \frac{8\pi \rho_\text{r}^{(0)} }{\Lambda} \right)^{1/4} 
\sqrt{ \sinh \left( 2\sqrt{ \frac{\Lambda}{3}} \, t \right)  } 
\,. 
\ee
It was rediscovered in the more complicated form \cite{Aldrovandi:2005ya}
\begin{eqnarray}
a(t) &=& \frac{1}{\sqrt{2}} \Bigg[
\left( a_0^2 +\sqrt{\frac{3\Gamma}{\Lambda} +a_0^4 }\, \right)\mbox{e}^{ 
2\sqrt{\frac{\Lambda}{3} } \, t }  \nonumber\\
&&\nonumber\\
&\, & 
+ \left( a_0^2 -\sqrt{\frac{3\Gamma}{\Lambda} +a_0^4} \, \right) 
\mbox{e}^{- 2\sqrt{ \frac{\Lambda}{3} } \, t } \Bigg]^{1/2} \\
&&\nonumber\\
&=&  \sqrt{a_0^2 
\cosh \left( 2\sqrt{\frac{\Lambda}{3}} \, t \right) + 
\sqrt{ \frac{\rho_\text{r}^{(0)}}{\rho_{\Lambda}} + a_0^4 } \, \sinh  
\left( 2\sqrt{\frac{\Lambda}{3} } \, t \right) } \nonumber\\
&&  \label{eq:RadLambdaNoK}
\end{eqnarray}
where 
\begin{eqnarray}
a_0 &=& a(0) \,,\\
&&\nonumber\\
\Gamma &=& \Omega_{r0} H_0^2 a_0^4 =\frac{8\pi}{3} \, 
\rho_\text{r}^{(0)} \,,
\end{eqnarray}
and a zero subscript denotes quantities evaluated at the present time.
 Again, this universe expands forever and the cosmological 
constant dominates over the radiation at late times. 

As for the single-fluid limits, a Taylor expansion 
of Eq.~(\ref{eq:RadLambdaNoK}) as $\Lambda\rightarrow 0$ reproduces the 
spatially flat radiative universe with zero cosmological constant and with 
scale factor $a(t) \sim \sqrt{t}$. Likewise, setting 
$\rho_\text{r}^{(0)}=0$ 
reduces the solution to the de Sitter space with pure $\Lambda$ and scale 
factor $a(t)=a_0 \, \mbox{e}^{\sqrt{\frac{\Lambda}{3}} \, t}$. 

The solution for radiation and cosmological constant can be presented also 
in the simple form \cite{Vazquez:2012ag}
\be
a(t)= a_\text{eq} \sqrt{ \sinh \left(2\sqrt{\frac{\Lambda}{3}} \, t 
\right)} 
\ee
where 
\be
a_\text{eq} =\left(  \frac{8\pi \rho_\text{r}^{(0)} }{\Lambda} 
\right)^{1/4}
\ee
is the value of the scale factor at the time of equivalence between the 
energy densities of radiation and of $\Lambda$ (this is a special case 
of the more general solution~(\ref{anyfluidLambdapos})).  The expression 
of the solution in terms of  conformal time requires 
elliptic integrals.

\subsection{$\Lambda=0$, stiff fluid, any $K$} 

These solutions are again given by Vajk \cite{Vajk1969}. For $K=0$, we 
have the usual, forever expanding, solution~(\ref{usual}) 
\begin{eqnarray}
a(t) &=& \left( 24\pi \rho_\text{s}^{(0)} \right)^{1/6} \left( t-t_0 
\right)^{1/3} \,,\\
&&\nonumber\\
\rho(a) &=& \frac{\rho_\text{s}^{(0)} }{a^6} \,,
\end{eqnarray}
or, in parametric form,
\be \left\{
\begin{array}{lll}
a(\eta) &=& \left( \frac{32\pi}{3} \, \rho_\text{s}^{(0)} \right)^{1/4}  
\sqrt{\eta}   \,,\\
&&\\
t(\eta) &=& \frac{4}{3} \left(  \frac{2\pi}{3} \, \rho_\text{s}^{(0)} 
\right)^{1/4}  \, \eta^{3/2} +t_0  \,.
\end{array} \right.
\ee

The scalar field equivalent to the stiff fluid is 
\be
\phi(t) = \phi_0 \ln \left(t-t_0 \right) +\phi_1 \,,
\ee
or
\be
\phi(\eta) = \frac{3}{2} \, \phi_0 \ln \eta +  \phi_1 +  \phi_0 \ln 
\left[  \frac{4}{3} \left( \frac{2\pi}{3} \rho_\text{s}^{(0)} 
\right)^{1/4} 
\right]  \,.
\ee
Again, for $K=\pm 1$ one cannot integrate $t(\eta)$ in finite form. For 
$K=+1$ we have \cite{Vajk1969}
\be \left\{
\begin{array}{lll}
a(\eta) &=& \left( \frac{8\pi}{3} \, \rho_\text{s}^{(0)} \right)^{1/4}  
\sqrt{ \cos( 2\eta)}    \,,\\
&&\\
t(\eta) &=& \int d\eta \, a(\eta)   \,,
\end{array} \right. 
\ee
a universe with finite size that begins in a Big Bang and ends in a Big 
Crunch.   For $K=-1$, the solution is instead 
\be \left\{
\begin{array}{lll}
a(\eta) &=& \left( \frac{8\pi}{3} \, \rho_\text{s}^{(0)} \right)^{1/4}  
\sqrt{ \sinh( 2\eta)}   \,,\\
&&\\
t(\eta) &=& \int d\eta \, a(\eta)   \,,
\end{array} \right.
\ee
describing a universe beginning in a Big Bang and eternally expanding.

\subsection{$K=0$, stiff fluid plus $\Lambda$}

One can alternatively regard this solution as being sourced by a  stiff 
fluid plus $\Lambda$, or by a free scalar field plus $\Lambda$, or by a 
scalar field with constant potential  $V(\phi)=\Lambda/(8\pi) $. 
The solution is \cite{Faraoni:2000vg,Chavanis2015}
\be
a(t)= a_0 \sinh^{1/3} \left( \sqrt{3\Lambda} \, t \right)  
\label{myAmJPsol1}
\ee
where 
\be
a_0= \left( \frac{8\pi \rho_\text{s}^{(0)} }{\Lambda} \right)^{1/6} =  
\left( 
\frac{4\pi C_0^2}{\Lambda} \right)^{1/6}  \label{myAmJPsol2}
\ee
and $C_0$ is the constant appearing in the first 
integral~(\ref{KGfirstintegral}) of the Klein-Gordon equation. If a scalar 
field is the stiff fluid source, then \cite{Faraoni:2000vg}
\be 
\phi(t) = \phi_0 \ln \left[ \tanh \left( \frac{\sqrt{3\Lambda}\, 
t}{2}\right)\right] + \phi_1 \label{myAmJPsol3}
\ee
with $\phi_0= \pm \left( 12\pi \right)^{-1/2}$ and $\phi_1$ is an 
integration constant. This solution has a Big 
Bang singularity at $t=0$, where
\begin{eqnarray}
a(t) &\simeq & t^{1/3} \,,\\
&&\nonumber\\
\phi(t) &\simeq & \phi_0 \ln \left( \frac{\sqrt{3\Lambda}\, t}{2} 
\right)\,.
\end{eqnarray}
At late times $t\rightarrow +\infty$, the solution asymptotes to the 
phase space de Sitter attractor with constant scalar field
\begin{eqnarray}
a(t) &\simeq & a_0 \, \mbox{e}^{\sqrt{ \frac{\Lambda}{3}} \, t} \,,\\
&&\nonumber\\
\phi(t) &\simeq & \phi_1 \,.
\end{eqnarray}
The total energy density and pressure are
\begin{eqnarray}
\rho_\text{tot} &=& \frac{\dot{\phi}^2}{2} - \frac{\Lambda}{8\pi} \,,\\
&&\nonumber\\
P_\text{tot} &=& \frac{\dot{\phi}^2}{2} + \frac{\Lambda}{8\pi} \,.
\end{eqnarray}
The effective equation of state of the mixture is time-dependent,
\be
w(t)\equiv \frac{P_\text{tot}}{\rho_\text{tot}}= 1-2 \tanh^2 \left( 
\sqrt{3\Lambda}\, t \right) \,,
\ee
interpolating between the stiff equation of state $P\simeq \rho$ at early 
times (when the cosmological constant has not yet had the time to 
influence 
the dynamics) and $P_\text{tot}\simeq -\rho_\text{tot}$ at late times when 
$\Lambda$ dominates \cite{Faraoni:2000vg}.

The limit $\Lambda\rightarrow 0$ in Eqs.~(\ref{myAmJPsol1}) and 
(\ref{myAmJPsol1}) reproduces the stiff fluid (or free scalar field) 
solution
\be
a(t)= a_0 t^{1/3}
\ee
({\em i.e.}, the scale factor~(\ref{usual}) with $w=1$). The scalar 
field~(\ref{myAmJPsol3}), however, diverges in the limit 
$\Lambda\rightarrow 0$; the correct free $\phi(t)$ can be obtained from 
the first integral~(\ref{KGfirstintegral}) which yields $\dot{\phi}=C_0/t$ 
and, finally, $\phi(t) =C_0 \ln \left(t/t_0\right)$. Similarly, setting 
$\rho_\text{s}^{(0)}=0$ does not automatically recover the de Sitter space 
corresponding to the surviving $\Lambda$: one must notice that removing 
the stiff fluid is equivalent to setting $\phi=$~const. while retaining  
its potential $V(\phi)=\Lambda/(8\pi)$, which generates $a(t) \propto 
\mbox{e}^{\sqrt{\frac{\Lambda}{3}}\, t}$.

\subsection{$K=0$, any real fluid plus $\Lambda$}

Indeed, when $K=0$ and $\Lambda \neq 0$, not only dust but {\em any} 
single fluid with constant equation of state parameter $w_2$ can be 
integrated explicitly using comoving time\footnote{A similar 
solution with a specific value of the exponent of the hyperbolic sine for 
dust plus dark energy with a time-dependent equation of state is obtained 
in Ref.~\cite{Pradhan13} after an {\em ad hoc} assumption on the 
functional form of 
the deceleration parameter $q\equiv -\ddot{a} a/\dot{a}^2$ (which 
essentially amounts to a choice of the scale factor).} 
\cite{Harrison1967,McIntosh1972, Chen:2014fqa,Chavanis2015}. For 
$\Lambda>0$, the solution is  
\be
a(t) = \left( \frac{\rho_2^{(0)}}{\rho_{\Lambda}} \right)^{ 
\frac{1}{3(w_2+1) }} 
\left\{ \sinh \left[ \frac{(w_2+1)}{2} \, \sqrt{3\Lambda}\,  
\left(t-t_0\right) \right] \right\}^{\frac{2}{3(w_2+1)} } 
\,, \label{anyfluidLambdapos}
\ee
where $\rho_{\Lambda}=\Lambda/(8\pi)$ is the effective energy density of 
the cosmological constant.  If $w_2>-1$, there is a Big Bang at $t=t_0$. 
At late times the solution converges to the de Sitter phase space 
attractor
\be
a(t) \simeq \left( \frac{\rho_2^{(0)}}{\rho_{\Lambda}} \right)^{ 
\frac{1}{3(w_2+1) }} \, \mbox{e}^{\sqrt{\Lambda/3} \left( t-t_0\right)} 
\,.
\ee
 The limit $\Lambda\rightarrow 0$ reproduces, through a 
straightforward Taylor expansion of the hyperbolic sine, the single-fluid, 
no-$\Lambda$ solution $a(t)=a_0 \left(t-t_0\right)^{ \frac{2}{3(w_2+1)} } 
$. However, regarding the other single-fluid limit, one cannot simply take 
$\rho_2^{(0)} \rightarrow 0 $ in Eq.~(\ref{anyfluidLambdapos}) to obtain 
the empty de Sitter space generated by $\Lambda$.

For $\Lambda<0$ we have \cite{Harrison1967} 
\begin{eqnarray}
a(t)&=&\left( \frac{\rho_2^{(0)}}{| \rho_{\Lambda}| } \right)^{ 
\frac{1}{3(w_2+1) }} \nonumber\\
&&\nonumber\\
&\, & \times \left\{ \sin \left[ \frac{(w_2+1)}{2} \, \sqrt{3  
|\Lambda| }\,  
\left(t-t_0\right) \right] \right\}^{\frac{2}{3(w_2+1)} } 
\label{negLambdaanyw}
\end{eqnarray}
and there are a Big Bang at $t_0$ and  a Big Crunch at 
\be
t=t_0  +\frac{2\pi}{\sqrt{3  |\Lambda| } \left( w_2+1\right)} 
\ee
if $w_2>-1$. 

Again, the limit $\Lambda\rightarrow 0$ reproduces the single-fluid 
universe with scale factor $a(t)=a_0 \left(t-t_0\right)^{ 
\frac{2}{3(w_2+1)} } $, but the simple limit $\rho_2^{(0)} \rightarrow 0 $ 
in Eq.~(\ref{anyfluidLambdapos}) fails to reproduce anti-de Sitter space.

For $\Lambda=0$ we have the usual, forever-expanding, 
solution~(\ref{usual})
\be
a(t)= \left[ (w_2+1)  \sqrt{ 6\pi \, \rho_2^{(0)} }
\left(t-t_0\right)\right]^{\frac{2}{3(w_2+1)}} \,,
\ee
with $a=0$ at $t=t_0$ if $w_2>-1$.

\subsection{$K=0, \Lambda=0$, dust plus stiff matter} 

This possibility appears in both lists~(\ref{list3}) and 
(\ref{list4}). As a function of comoving time, the scale factor   
assumes the simple form found by Vajk \cite{Vajk1969}
\begin{eqnarray}
a(t) &=&  \left[ 6\pi \rho_\text{m}^{(0)} \left(t-t_0\right)^2 
-\frac{\rho_\text{s}^{(0)}}{\rho_\text{m}^{(0)} } \right]^{1/3} \,, 
\label{urca}\\
&&\nonumber\\
\rho_\text{m} (a) &=& \frac{ \rho_\text{m}^{(0)} }{a^3} \,, \quad \quad 
\rho_\text{s} (a) = \frac{ \rho_\text{s}^{(0)} }{a^6} \,.
\end{eqnarray}
This universe expands forever, with the stiff fluid redshifting away 
considerably faster than the dust. This solution is a special 
case of the one found by Chavanis 
\cite{Chavanis2015} and Dariescu {\em et al.} \cite{Dariescu2017} for dust 
plus a stiff fluid plus $\Lambda$, which is given by  
Eq.~(\ref{Dariescu}) with the constants $\alpha$ and $\beta$ as in 
Eqs.~(\ref{alphaDariescu}) and 
(\ref{betaDariescu}). By  taking 
the limit $\Lambda \rightarrow 0$ and using the second order expansions  
$\sinh x = x+\, ...$, $ \cosh x = 1 +x^2/2 + \, ...$ as $x\rightarrow 0$, 
Eq.~(\ref{Dariescu}) yields  
\begin{eqnarray}
a(t) &=& \left[ \frac{9\beta t^2}{4} +3\sqrt{\alpha} \, t \right]^{1/3} \\
&&\nonumber\\
&=& a_0 \left[ 6\pi \rho_2^{(0)} t^2 +\sqrt{24\pi \rho_1^{(0)} } \, t +\, 
... \right]^{1/3} \,,  
\end{eqnarray}
which reproduces the solution~(\ref{urca}) for
\be
t_0= \sqrt{ \frac{\rho_\text{s}^{(0)} }{6\pi}} \, 
\frac{1}{\rho_\text{m}^{(0)}} 
\,.
\ee  
Although the authors of both Refs.~\cite{Vajk1969,Dariescu2017} do not 
regard this as a  free scalar field solution, it is straightforward to 
integrate 
Eq.~(\ref{KGfirstintegral}), which yields
\be
\phi(t) = \frac{\phi_0}{3\sqrt{\alpha}} \, \ln \left( 
\frac{9\beta}{4} 
+\frac{3\sqrt{\alpha}}{t} \right) + \phi_1 \,,
\ee
where $\phi_{0,1}$ are integration constants.

The single-fluid limit $\rho_\text{s}^{(0)} \rightarrow 0$ leaving only a 
dust, 
applied to the scale factor~(\ref{urca}) reproduces the correct $a(t)=a_0 
\left(t-t_0\right)^{2/3}$ and $\phi=$~const. disappears since 
$\rho_\text{s}=\rho_{\phi}=P_{\phi}= \dot{\phi}^2/2 $ vanishes. However, 
one 
cannot take the limit $\rho_\text{m}^{(0)} \rightarrow 0$  in 
Eq.~(\ref{urca}) to 
obtain the pure stiff fluid solution.

\subsection{$K=0, \Lambda=0$, radiation plus stiff matter}

This solution for $K=0, \Lambda=0$, and radiation plus a stiff fluid   
was given by Vajk\footnote{This solution has 
recently been 
rediscovered in its parametric form, using conformal time as the 
parameter, in Ref.~\cite{Alvarenga:2016yxh}.} \cite{Vajk1969}. We have 
$\left(w_1, w_2 \right)=\left( 1/3, 1 \right)$, the Friedmann equation 
yields
\be
I= \int \frac{da \, a^2}{\sqrt{ \rho_\text{r}^{(0)} a^2 
+\rho_\text{s}^{(0)} } } = \pm 
\sqrt{ \frac{8\pi }{3} } \left(t-t_0\right)  \label{Hellointegral}
\ee
and, integrating,
\begin{eqnarray}
t(a) &=& t_0 \pm \sqrt{\frac{3}{32\pi \rho_\text{r}^{(0)} } } 
\left\{ 
a\sqrt{  a^2+ \frac{ \rho_\text{s}^{(0)} }{ \rho_\text{r}^{(0)} } }   
\right.\nonumber\\
&&\nonumber\\
&\, & \left. 
-   \frac{\rho_\text{s}^{(0)} }{ \rho_\text{r}^{(0)} }  \ln \left[ 
\sqrt{  
 a^2 +  \frac{  \rho_\text{s}^{(0)} }{ \rho_\text{r}^{(0)} } } +a   
 \right]  \right\} \label{moh}
\end{eqnarray}
(in this $K=0$ case, the scale factor $a$ is dimensionless). At late times 
this universe is dominated by radiation and $a(t) \sim \sqrt{t}$.  
The parametric form of this solution is \cite{Vajk1969}
\be \left\{
\begin{array}{lll}
a(\eta) &=& \sqrt{  \frac{8\pi}{3} \, \rho_\text{r}^{(0)} \eta^2  
-\frac{ \rho_\text{s}^{(0)}}{\rho_\text{r}^{(0)} }   }   \,,\\
&&\\
t(\eta) &=& \frac{\eta}{2} \, \sqrt{  \frac{8\pi}{3} \, 
\rho_\text{r}^{(0)} 
\eta^2 -\frac{ \rho_\text{s}^{(0)}}{\rho_\text{r}^{(0)} }   }\\
&&\\
&\, &  -\frac{ \rho_\text{s}^{(0)}}{\rho_\text{r}^{(0)} } \sqrt{ 
\frac{3}{32\pi\rho_{r}^{(0)}} \,  }    
 \ln \left[ \sqrt{ \frac{8\pi}{3} \, \rho_\text{r}^{(0)} } \, \eta 
\right.\\
&&\\
&\, & \left. + \sqrt{ \frac{8\pi}{3} \, \rho_\text{r}^{(0)} \eta^2 
-\frac{\rho_\text{s}^{(0)}}{\rho_\text{r}^{(0)}} } \right] +t_0   \,,
\end{array} \right.
\ee
from which one sees that this universe begins in a  Big Bang. 

Equation~(\ref{KGfirstintegral}) for the scalar field equivalent of the 
stiff fluid can be integrated giving $\phi(a)$. Using 
$ d\phi/dt=\dot{a} \, d\phi/da $, one obtains
\be
\frac{d\phi}{da} =  \frac{C_0}{ { a} 
\sqrt{ \rho_\text{r}^{(0)}a^2 +\rho_\text{s}^{(0)}  } } \,,
\ee
which integrates to
 
\be
\phi(a) = \phi_0 \tanh^{-1} \left( \sqrt{ 1+\frac{\rho^{(0)}_\text{r} }{ 
\rho_\text{s}^{(0)} } \, ~a^2  } \, \right) +\phi_1 \,,
\ee  
 where $\phi_0 = -C_0/\sqrt{ \rho_\text{s}^{(0)} }  $ and $\phi_1$ 
is an integration constant. 

The limit $\rho_\text{s}^{(0)} \rightarrow 0$ in which the stiff 
fluid 
disappears, leaving only radiation, transforms the relation~(\ref{moh})  
into the correct scale factor $ a(t)=a_0 \sqrt{t-t_0}$. The other limit 
$\rho_\text{r}^{(0)}\rightarrow 0$ in Eq.~(\ref{moh}) does not produce the 
corresponding stiff fluid solution, which has to be recovered directly 
from the limit $\rho_\text{r}^{(0)}\rightarrow 0$ of the 
integral~(\ref{Hellointegral}).

\subsection{$w=2/3, \Lambda=0$}

This solution was given by Vajk \cite{Vajk1969} without apparent physical 
motivation and was reported by other authors \cite{McIntosh1972}. For 
$K=0$, it is the single-fluid universe~(\ref{usual}),
\begin{eqnarray}
a(t) &=& \left( \frac{50\pi}{3} \, \rho^{(0)} \right)^{1/5} \left( t-t_0 
\right)^{2/5} \,,\\
&&\nonumber\\
 \rho(a) &=&  \frac{\rho^{(0)} }{a^5} \,, 
\end{eqnarray}
or, in parametric form,
\be \left\{
\begin{array}{lll}
a(\eta) &=& \left( 6\pi \, \rho^{(0)} \right)^{1/3}  \,  \eta^{2/3} 
\,,\\
&&\\
t(\eta) &=&  \frac{3}{5} \left( 6\pi \, \rho^{(0)} \right)^{1/3}  
\,  \eta^{5/3} +t_0 \,.
\end{array} \right.
\ee
There is a Big Bang singularity $a=0$ at $t=t_0$ (or $\eta=0$) and the 
universe expands forever.

For $K=\pm 1$, the comoving time cannot be expressed explicitly in terms 
of simple functions. When $K=+1$ we have \cite{Vajk1969} 
\be \left\{
\begin{array}{lll}
a(\eta) &=& \left( \frac{8\pi}{3} \, \rho^{(0)} \right)^{1/3}  \,  
\cos^{2/3} \left( \frac{3\eta}{2} \right)  \,,\\
&&\\
t(\eta) &=&  \int d\eta \,  a(\eta) \,,
\end{array} \right. 
\ee
which has a Big Bang at $\eta=-\pi/3$  and a Big Crunch at $\eta=\pi/3$.

For  $K=- 1$ the solution is \cite{Vajk1969} 
\be \left\{
\begin{array}{lll}
a(\eta) &=& \left( \frac{8\pi}{3} \, \rho^{(0)} \right)^{1/3}  \,  
\sinh^{2/3} \left( \frac{3\eta}{2} \right) \,,\\
&&\\
t(\eta) &=&  \int d\eta \,  a(\eta) \,.
\end{array} \right.
\ee
In  this case there is a Big Bang at $\eta=0$ and the universe expands  
forever.

\subsection{$K=0, \Lambda=0, w=2/3$ fluid plus radiation}

This case, with $w_1=1/3, w_2= 2/3$, is the third entry in the 
list~(\ref{list2}).  The comoving time form of this solution is again 
given by Vajk as\footnote{An error in the square root $\sqrt{ a 
+\rho_\text{s}^{(0)}/\rho_\text{r}^{(0)}} $ in the first term on the right 
hand side of Eq.~(\ref{VajikCorrectedAgain}), present in the solution of 
Ref.~\cite{Vajk1969}, was corrected in \cite{McIntosh1972}.} 
\cite{Vajk1969}
 
\begin{eqnarray}
t(a) &=& \sqrt{ \frac{3}{32\pi \rho_\text{r}^{(0)} } }   \left[ \left( 
a-\frac{3 \rho_2^{(0)} }{2\rho_\text{r}^{(0)} } \right) \sqrt{ a^2 + 
\frac{\rho_2^{(0)} }{\rho_\text{r}^{(0)} } \, a } \right. \nonumber\\
&&\nonumber\\
&\, & \left. + \frac{3}{ {4}} \left(
\frac{ \rho_2^{(0)} }{\rho_\text{r}^{(0)} }\right)^2 \ln \left( \sqrt{a^2 
+ \frac{\rho_2^{(0)} }{\rho_\text{r}^{(0)} }\, a} +a +
\frac{\rho_2^{(0)} }{2\rho_\text{r}^{(0)} } \right)  \right] \nonumber\\
&\,& +t_0 \,, \label{VajikCorrectedAgain}\\
&&\nonumber\\
\rho_\text{r} (a) &=& \frac{\rho_\text{r}^{(0)} }{a^4} \,, \quad\quad 
\rho_2 (a) = \frac{\rho_2^{(0)} }{a^5} \,.
\end{eqnarray} 
The Big Bang occurs at 
\be
t= t_0 + \frac{3\sqrt{3}}{4\sqrt{\pi}} \left( 
\frac{ \rho_2^{(0)} }{ \rho_\text{r}^{(0)} } \right)^2 \ln \left( 
\frac{\rho_2^{(0)} }{ 2\rho_\text{r}^{(0)} } \right) 
\ee
and the universe expands forever, becoming radiation-dominated at late 
times. The parametric form of this solution in terms of conformal time is 
\cite{Vajk1969} 
\be \left\{
\begin{array}{lll}
&&  \sqrt{ a^2 + \frac{\rho_2^{(0)} }{ \rho_\text{r}^{(0)} } \, a }
-\frac{ \rho_2^{(0)} }{ 2\rho_\text{r}^{(0)} }  \cosh^{-1} 
\left( \frac{ 2 \rho_\text{r}^{(0)} }{ \rho_2^{(0)} }  +1 
\right)\\
&&\\
&& = \sqrt{ \frac{8\pi}{3} \, \rho_\text{r}^{(0)} } \,  \eta \,,\\
&&\\
&& t(\eta) =  \int d\eta \,  a(\eta) \,.
\end{array} \right. \label{VajikCorrectedAgainParametric}
\ee
The limit $\rho_2^{(0)} \rightarrow 0$ leaves only radiation; taking this 
limit in Eq.~(\ref{VajikCorrectedAgain}) reproduces the correct scale 
factor $a(t)=a_0 \sqrt{t-t_0}$ for the single-fluid radiative universe. 
The limit $\rho_\text{r}^{(0)} \rightarrow 0$, instead, fails to reproduce 
the corresponding scale factor, which must be recovered directly from the 
usual integral.

\subsection{$K=0, \Lambda =0, w_1=2/3$ fluid plus stiff fluid}

This possibility appears as the second last entry of the 
list~(\ref{list3}). This  
solution was again given by Vajk \cite{Vajk1969} using comoving 
time\footnote{Vajk's solution has a sign error in the square root, which 
was corrected by McIntosh \cite{McIntosh1972}.}
\begin{eqnarray}
t(a)&=& \frac{1}{15} \sqrt{ \frac{3}{2\pi \rho_1^{(0)} } }
\sqrt{ a+\frac{ \rho_\text{s}^{(0)} }{ \rho_1^{(0)} } }\nonumber\\
&&\nonumber\\
&\, & \times 
\left[ 3a^2 -\frac{4\rho_\text{s}^{(0)} }{ \rho_1^{(0)} } \, a + 8 \left( 
\frac{ \rho_\text{s}^{(0)} }{ \rho_1^{(0)} }\right)^2  \right] 
+t_0\nonumber\\
&&\label{eq:sailc100}\\
 \rho_1(a) &=&  \frac{ \rho_1^{(0)} }{a^5} \,, \quad \quad 
\rho_\text{s} (a) =\frac{ \rho_\text{s}^{(0)} }{ a^6} \,,
\end{eqnarray}
or, in parametric form \cite{Vajk1969},
\be \left\{
\begin{array}{lll}
&& \sqrt{ a + \frac{\rho_\text{s}^{(0}}{\rho_1^{(0)}} }
\left( a- \frac{2\rho_\text{s}^{(0}}{\rho_1^{(0)}} \right)
=\sqrt{ 6\pi \rho_1^{(0)} } \, \eta \,,\\
&&\\
&&t(\eta) =  \int d\eta \,  a(\eta) \,.
\end{array} \right.
\ee
The Big Bang occurs at
\be
t =t_0 +\frac{4}{5} \, \sqrt{ \frac{2}{3\pi \rho_1^{(0)} }} 
\left( \frac{ \rho_\text{s}^{(0)} }{\rho_2^{(0)} } \right)^{5/2}  \,,
\ee
or at $t=0$ if the integration constant is chosen as 
\be
t_0= -\frac{4}{5} \sqrt{ \frac{2}{3\pi \rho_1^{(0)} } }  
\left( \frac{ \rho_\text{s}^{(0)} }{
\rho_1^{(0)} } \right)^{5/2} \,.
\ee
 This universe expands eternally and is dominated by the $w_1=2/3$ 
fluid, which decays slower than the stiff fluid, in its late history. 

We can find the scalar field $\phi$ associated with the stiff fluid 
as a function of the scale factor $a$. By differentiating 
Eq.~(\ref{eq:sailc100}), one obtains 
\be
\frac{dt}{da}=(\dot{a})^{-1} = \sqrt{ \frac{3}{8\pi \rho_1^{(0)} }} \, 
\frac{a^2}{ \sqrt{ a+
\frac{ \rho_\text{s}^{(0)} }{ \rho_1^{(0)} }}} 
\ee
which, substituted into $ d\phi/dt = \dot{a} d\phi/da $, gives 
\be
\frac{d\phi}{da} = \sqrt{\frac{3}{8\pi \rho_1^{(0)} }} \,    
\frac{d\phi}{dt} \, \frac{a^2}{\sqrt{a+
\frac{ \rho_\text{s}^{(0)} }{ \rho_1^{(0)}} } 
} \,.
\ee
Equation~(\ref{KGfirstintegral}) now yields  
\be
\frac{d\phi}{da} = \sqrt{\frac{3}{8\pi \rho_1^{(0)} }} \, \frac{ 
C_0}{a\sqrt{a+  \frac{ \rho_\text{s}^{(0)} }{ \rho_1^{(0)} }} } \,,
\ee
which integrates to
\be
\phi(a)=\phi_0 \tanh^{-1} \left( \sqrt{ 1+\frac{\rho_1^{(0)} 
}{\rho_\text{s}^{(0)} } 
\, a} \,  \right)+\phi_1 \,, \label{sailc101}
\ee
where $\phi_{0,1}$ are integration constants.  

 The single-fluid limit $\rho_\text{s}^{(0)}\rightarrow 0$ in 
Eq.~(\ref{eq:sailc100})  reproduces the correct scale factor $a(t)=a_0 
\left(t-t_0 \right)^{2/5} $, but fails on the expression~(\ref{sailc101}) 
of the scalar field: this happens because the limit corresponds to the 
value $C_0=0$ of the integration constant in the first 
integral~(\ref{KGfirstintegral}) leading to Eq.~(\ref{sailc101}).

\section{Three (real or effective) fluids solutions}
\setcounter{equation}{0}
\label{sec:6}

Known three-fluid solutions  include:

\begin{itemize}

\item $\left(w_1,w_2, w_3 \right)= \left( 1/3, -1 , -1/3\right)$, 
radiation plus $\Lambda$ plus spatial curvature; 

\item $\left(w_1,w_2, w_3 \right)= \left( -1/3, 0, 1/3\right)$, spatial 
curvature $K\neq 0$ plus dust plus 
radiation;

\item $\left(w_1,w_2, w_3 \right)= \left( -1, 0, 1 \right)$, or $\Lambda$ 
plus a dust plus a stiff fluid; 

\item $\left(w_1,w_2, w_3 \right)= \left( 1/3, 2/3, 1 \right)$, or 
radiation plus $P_2=2\rho_2/3$ fluid plus a stiff fluid.

\end{itemize}

\subsection{Radiation plus $\Lambda$ plus spatial curvature $K\neq 0$}

These solutions were found by Harrison \cite{Harrison1967} and 
subsequently rediscovered \cite{Aldrovandi:2005ya}. For $K=+1$ and 
$\Lambda>0$, we have \cite{Harrison1967,StephaniExact}
\begin{eqnarray}
a(t) &=& \sqrt{\frac{3}{2\Lambda}} \left\{1-\cosh \left[2 \sqrt{ 
\frac{\Lambda}{3} } \, \left( t-t_0\right) \right] \right. \nonumber\\
&&\nonumber\\
&\, & \left. + 
2\sqrt{\frac{\Lambda}{\Lambda_c}} \, \sinh\left[ 2  
\sqrt{\frac{\Lambda}{3}} \left(t-t_0\right) \right] \right\}^{1/2} \,,
\end{eqnarray}
where 
\be
\Lambda_c = \frac{9}{32\pi \rho_\text{r}^{(0)} } 
\ee
is the critical value of the cosmological constant required to have a 
static Einstein universe with the same amount of radiation.

For $K=-1$ and $\Lambda>0$,
\begin{eqnarray}
a(t) &=& \sqrt{\frac{3}{2\Lambda}} \left\{\cosh \left[2 \sqrt{ 
\frac{\Lambda}{3} } \, \left( t-t_0\right) \right] -1\right. \nonumber\\
&&\nonumber\\
&\, & \left. + 
2 \sqrt{\frac{\Lambda}{\Lambda_c}} \, \sinh\left[ 2  
\sqrt{\frac{\Lambda}{3}} \left(t-t_0\right) \right] \right\}^{1/2} \,,
\end{eqnarray}
Both solutions begin with a  Big Bang at $t=t_{0}$ and asymptote to the de 
Sitter  space with scale factor $ a(t)=a_{0}\exp\left( 
\sqrt{\frac{\Lambda}{3}} \, t \right)$ at late times.

\noindent For $K=+1$ and $\Lambda<0$, the solution is \cite{Harrison1967}
\begin{eqnarray}
a(t) &=& \sqrt{\frac{3}{2|\Lambda|}} \left\{1-\cosh \left[2 \sqrt{ 
\frac{|\Lambda|}{3} } \, \left( t-t_0\right) \right] \right. \nonumber\\
&&\nonumber\\
&\, & \left. + 
2\sqrt{\frac{|\Lambda|}{\Lambda_c}} \, \sinh\left[ 2  
\sqrt{\frac{|\Lambda|}{3}} \left(t-t_0\right) \right] \right\}^{1/2} \,.
\end{eqnarray}

\subsection{$K=0, \Lambda$ plus a stiff fluid plus dust}

The FLRW universe with stiff matter plus dust plus $\Lambda$ was found by 
Chavanis \cite{Chavanis2015} and, like several others, was rediscovered 
recently \cite{Dariescu2017}. Using the slightly different notation 
\begin{eqnarray}
\rho_1 &=& \rho_1^{(0)} \left( \frac{a_0}{a} \right)^6 \,,\\
&&\nonumber\\
\rho_2 &=& \rho_2^{(0)} \left( \frac{a_0}{a} \right)^3 \,,\\
&&\nonumber\\
\alpha &=& \frac{8\pi}{3} \, \rho_1^{(0)} a_0^6 
\,,\label{alphaDariescu}\\
&&\nonumber\\
\beta &=& \frac{8\pi}{3} \, \rho_2^{(0)} a_0^3 \,,\label{betaDariescu}
\end{eqnarray}
the solution  for $\Lambda >0$ is
 
\be 
a(t) = \left\{ \frac{3\beta}{2\Lambda} \left[ \cosh\left( 
\sqrt{3\Lambda} \, t \right) -1 \right]  +\sqrt{ 
\frac{3\alpha}{\Lambda} } \sinh\left( \sqrt{3\Lambda} \, t 
\right)  \right\}^{1/3} \,.\label{Dariescu}
\ee
This universe, which begins with a Big Bang at $t=0$, asymptotes 
to the de Sitter phase space attractor $a(t)=a_0 
\, \mbox{e}^{\sqrt{ \frac{\Lambda}{3}} \, t} $ at late times.

\subsection{$K\neq 0$, dust plus radiation} 

The solution for dust plus radiation in a spatially flat universe is given 
by Eq.~(\ref{K=0dust+radiation}) using comoving time or by 
Eqs.~(\ref{flatdust+radiationparametric}) in parametric form using 
conformal time. The corresponding solutions for spatially curved FLRW 
universes were found in \cite{Chernin1966, CohenNature1967, Vajk1969, 
JacobsNature1967, McIntosh1968b,McIntoshFoyster1972} (see also 
\cite{EllisMaartensMacCallum,Slava}). The solution for $K=+1$ is given by 
({\em e.g.}, \cite{CohenNature1967})
\begin{eqnarray}
t(a) &=& t_0 - \sqrt{ \frac{8\pi}{3}\,  \rho_\text{r}^{(0)} 
+\frac{8\pi}{3}\, 
\rho_\text{m}^{(0)} a -a^2 } \nonumber\\
&&\nonumber\\
& \, & 
+\frac{4\pi}{3} \, \rho_\text{m}^{(0)} \sin^{-1} \left[ 
\frac{ -\frac{4\pi}{3} \, \rho_\text{m}^{(0)} +a }{ \sqrt{
\left( \frac{4\pi}{3} \, \rho_\text{m}^{(0)} \right)^2 +\frac{8\pi}{3} \, 
\rho_\text{r}^{(0)} } } \right] \,. \nonumber\\
&& \label{DustRadPositive}
\end{eqnarray}
 This universe begins at a Big Bang, reaches a maximum size, and 
ends in a Big Crunch singularity. The parametric form of this 
solution is \cite{Vajk1969, Mukhanov:1990me, Aldrovandi:2005ya}
\be
\left\{ 
\begin{array}{lll}
a (\eta) & = &  \frac{4\pi}{3}\, \rho_\text{m}^{(0)}  \left( 
1-\cos\eta\right) 
+ \sqrt{ \frac{8\pi}{3} \, \rho_\text{r}^{(0)} } \sin \eta  \,,  \\
& & \label{+1dust+radiation}\\
t (\eta) & = & \frac{4\pi}{3} \, \rho_\text{m}^{(0)}  \left( \eta-\sin 
\eta\right) 
+\frac{8\pi}{3} \, \rho_\text{r}^{(0)}  \left( \cos  \eta -1 \right) \,, 
\end{array}  \right.
\ee

For $K=-1$ the solution is \cite{CohenNature1967, Vajk1969, 
McIntoshFoyster1972}
\begin{eqnarray}
&&t(a) = t_0 + \sqrt{
\frac{8\pi}{3} \, \rho_\text{r}^{(0)} +\frac{8\pi}{3} \, 
\rho_\text{m}^{(0)} a 
+a^2 } 
\nonumber\\
&&\nonumber\\
& & -\frac{4\pi}{3} \, \rho_\text{m}^{(0)} \ln\left\{ C \left[ \sqrt{
\frac{8\pi}{3} \, \rho_\text{r}^{(0)} +\frac{8\pi}{3} \, 
\rho_\text{m}^{(0)} a
+a^2 } +a \right.\right.\nonumber\\
&&\nonumber\\
&\, & \left. \left.  + \frac{4\pi}{3}\rho_\text{m}^{(0)}  \right]\right\} 
\,,   \label{CohenDustRadCurvLog}
\end{eqnarray}
where $C$ is a constant (missing in \cite{CohenNature1967}) 
needed to make the argument of the logarithm dimensionless.  This 
universe begins at a Big Bang and expands forever, becoming 
dust-dominated. The  
expression~(\ref{CohenDustRadCurvLog}) can be cast in the alternative form 
presented by Vajik \cite{Vajk1969} 
\begin{eqnarray}
t(a) &=& \sqrt{ a^2 +\frac{8\pi}{3} \, 
\rho_\text{m}^{(0)} a +\frac{8\pi}{3} \, 
\rho_\text{r}^{(0)} } \nonumber\\
&& \nonumber\\
&\, &  -\frac{4\pi}{3} \, \rho_\text{m}^{(0)} \sinh^{-1} \left[
\frac{ a+4\pi\rho_\text{m}^{(0)} /3}{ 
\sqrt{ \frac{4\pi}{3} \, \rho_\text{m}^{(0)} \left[ 
\frac{-4\pi}{3} \, \rho_\text{m}^{(0)} 
+\frac{2 \rho_\text{r}^{(0)} } 
{\rho_\text{m}^{(0)}}
\right] } } \right]   +t_0 \label{-1dust+radiation}
\end{eqnarray}
(see Appendix \ref{AppendixB} for a proof of the equivalence),  
or in the parametric form\footnote{Ref.~\cite{Aldrovandi:2005ya} contains 
an error in this parametric solution.}  \cite{Vajk1969, Mukhanov:1990me}\\

\be
\left\{
\begin{array}{lll}
a (\eta) &=& \sqrt{ \frac{8\pi}{3}\,  \rho_\text{r}^{(0)} } \sinh\eta  + 
\frac{4\pi}{3}\, \rho_\text{m}^{(0)}  \left( \cosh \eta-1 \right) \,,\\
&&\\
t (\eta) &=& \sqrt{ \frac{8\pi}{3}\, \rho_\text{r}^{(0)} } \left( 
\cosh\eta -1\right)  
+\frac{4\pi}{3}\, \rho_\text{m}^{(0)} 
\left( \sinh \eta - \eta \right) \,,
\end{array} \right.
\ee
with the initial condition $a(t=0)=0$.

When $\rho_\text{m}^{(0)} $ or $\rho_\text{r}^{(0)}$ are set to zero, the 
solutions 
reduce, of course, to single-fluid spatially curved universes filled only 
with radiation or matter, respectively. In the case of matter-only, 
however, it is not trivial to reproduce the single-fluid expressions 
(\ref{DustCurvPos}) for $K=+1$ and (\ref{DustCurvNeg}) for $K=-1$ as 
$\rho_\text{r}^{(0)} \rightarrow 0$.  Appendix~\ref{AppendixC} discusses 
these limits. 

The solutions of the Einstein-Friedmann equations for dust plus radiation 
with spatial curvature and $\Lambda\neq 0$ were classified qualitatively 
by Payne \cite{Payne70}.

\subsection{$ K\neq 0, \Lambda=0$, radiation plus stiff fluid}  

The spatially curved FLRW universes filled with  radiation 
plus a stiff fluid were given by Vajk in parametric form \cite{Vajk1969}. 
For $K=+1$, 
\be \left\{
\begin{array}{lll}
a(\eta) &=& \sqrt{ \frac{4\pi}{3} \, \rho_\text{r}^{(0)} +\sqrt{
\frac{4\pi}{3} \, \rho_\text{r}^{(0)}  \left( \frac{4\pi}{3} \, 
\rho_\text{r}^{(0)} + 
\frac{2 \rho_\text{s}^{(0)} }{  \rho_\text{r}^{(0)} } \right) } \, 
\sin(2\eta)} 
\,,\\
&&\\
t(\eta) &=&  \int d\eta \,  a(\eta) \,.
\end{array} \right.
\ee
This cosmos begins at a Big Bang, expands to a maximum size, and 
collapses in a Big Crunch. 
The scalar field equivalent of the stiff fluid can only be expressed in 
terms of $\eta$, or of $a$, by means of elliptic integrals.

For $K=-1$, instead, we have the forever-expanding universe
\be \left\{
\begin{array}{lll}
a(\eta) &=& \left( \sqrt{\frac{4\pi}{3} \, \rho_\text{r}^{(0)}  \left( 
\frac{4\pi}{3} \, \rho_\text{r}^{(0)} +
\frac{2 \rho_\text{s}^{(0)}}{  \rho_\text{r}^{(0)} } \right) } \, 
\sinh(2\eta) \right. \\
&\, & \left. -\frac{4\pi}{3} \, \rho_\text{r}^{(0)} \right)^{1/2} \,,\\
&&\\
t(\eta) &=&  \int d\eta \,  a(\eta) 
\end{array} \right.
\ee
which, during its history, is first dominated by the stiff fluid,  
then becomes radiation-dominated, and ends in a  curvature-dominated era.

\subsection{$K=0, \Lambda=0$, radiation plus 
$P_2=2\rho_2/3$ fluid plus stiff fluid}  

The solution for radiation plus a $ P_2=2\rho_2/3$ fluid plus a stiff 
fluid was again given by Vajk \cite{Vajk1969}. It reads
\begin{eqnarray}
t(a) &=& t_0 +\sqrt{ \frac{3}{32\pi \rho_\text{r}^{(0)} } } \left\{
\left( a-\frac{3\rho_2^{(0)} }{2 \rho_\text{r}^{(0)} } \right) \sqrt{ 
a^2 + \frac{\rho_2^{(0)} }{\rho_\text{r}^{(0)}} \, a  + 
\frac{\rho_\text{s}^{(0)} 
}{ 
\rho_\text{r}^{(0)} } } \right. \nonumber\\  
&&\nonumber\\
&\, & \left.   + \left[ \frac{3}{4} \left( \frac{\rho_2^{(0)} 
}{\rho_\text{r}^{(0)}} 
\right)^2  - \frac{\rho_\text{s}^{(0)} }{\rho_\text{r}^{(0)}} \right] 
 \ln \left[ \sqrt{a^2 + \frac{\rho_2^{(0)} }{\rho_\text{r}^{(0)}}\, a
+ \frac{\rho_\text{s}^{(0)} }{\rho_\text{r}^{(0)}}  }   \right.\right. 
\nonumber\\
&&\nonumber\\
&\, & \left. \left. +a + 
\frac{\rho_2^{(0)} }{2\rho_\text{r}^{(0)} }   \right]  \right\}  
\label{eq:5.19}
\end{eqnarray}
 and describes a Big Bang universe which expands forever, with the 
stiff fluid, then the $w_2=2/3$ fluid, and then radiation becoming 
dominant.  
In the limit $\rho_\text{s}^{(0)} \rightarrow 0$ in which the 
matter 
content reduces to the $w_2=2/3$ fluid plus radiation, 
Eq.~(\ref{VajikCorrectedAgain}) is reproduced. 

The parametric form corresponding to the universe~(\ref{eq:5.19}) using 
conformal time as the parameter is 
\cite{Vajk1969}
\be \left\{
\begin{array}{lll}
&&  \sqrt{ a^2 +    
\frac{\rho_2^{(0)}}{\rho_\text{r}^{(0)} } \, a +
\frac{\rho_\text{s}^{(0)}}{\rho_\text{r}^{(0)} } }
-\frac{\rho_2^{(0)}}{2\rho_\text{r}^{(0)} }\\
&&\\
&& \times \cosh^{-1} \left[
\frac{ 2a +\rho_2^{(0)}/\rho_\text{r}^{(0)} }{
\sqrt{ \left( \frac{\rho_2^{(0)}}{\rho_\text{r}^{(0)} } \right)^2 -
\frac{4\rho_\text{s}^{(0)}}{\rho_\text{r}^{(0)} } } } \right] 
= \sqrt{ \frac{8\pi}{3} \, \rho_\text{r}^{(0)} } \,  \eta \,,\\
&&\\
&& t(\eta) =  \int d\eta \,  a(\eta) \,.
\end{array} \right.
\ee
In the limit $\rho_\text{s}^{(0)} \rightarrow 0$ it reproduces 
Eq.~(\ref{VajikCorrectedAgainParametric}) for 
a $w_2=2/3$ fluid plus radiation.


\section{Scalar field solutions}
\setcounter{equation}{0}
\label{sec:7}

\subsection{Single scalar field}

We have already seen that a minimally coupled scalar field is equivalent 
to a perfect fluid which, in general, has a dynamical effective equation 
of state and that a free scalar field is equivalent to a stiff fluid. 
Therefore, all stiff fluid solutions are automatically free scalar 
solutions, and we have provided the corresponding expression of $\phi(t)$ 
when it is given explicitly in terms of elementary functions.

Originally, the study of analytical solutions of scalar field cosmology 
was motivated by early universe inflation.  When the scalar field 
potential $V(\phi)$ is part of a high energy theory, exact solutions are 
usually not available because of the non-linearity of the Klein-Gordon 
equation~(\ref{KleinGordon}) and phase space analyses are most informative 
about the dynamics ({\em e.g.}, \cite{StabellRefsdal, WainwrightEllisBook, 
Coleybook}). Several formal solutions exist in the literature, although 
some of them are mostly of mathematical interest. Only solutions that are 
attractors in phase space, or that display particular physical properties 
(such as evading slow-roll constraints part of the time, providing exact 
spectral indices or desired expansion histories $a(t)$, {\em etc.}) are 
usually important from the physical point of view. Many of the analytical  
solutions found are only valid in restricted regions of initial conditions 
and parameters, usually due to the fact that variables changes are needed 
which are restricted to regions in which the scalar field $\phi(t)$ is 
monotonic and invertible. Therefore, most of these solutions should be 
regarded as toy models (which may still be quite useful), with the notable 
exception of solutions that behave as attractors in phase space.

The following solutions\footnote{These solutions have been used as 
seeds to generate corresponding universes in scalar-tensor gravity by 
means of the conformal transformation from the Einstein to the Jordan 
frame ({\em e.g.}, \cite{Abreu}).} have been found in the context of 
inflationary scenarios of the early universe \cite{KolbTurner, LiddleLyth, 
Martin:2013tda} (we do not report approximate solutions here).

\subsubsection{Exponential potentials}

The power-law inflationary scenario is described by the exact solution  
\cite{Lucchin:1984yf, Abbott:1984fp, Barrow:1987ia, 
Burd:1988ss,Liddle:1988tb, Muslimov:1990be}
\begin{eqnarray}
a(t)&=& a_0 t^p \,,\\
&&\nonumber\\
\phi(t) &=& \phi_0+\alpha \ln \left(  \, t \right)
\,,\\
&&\nonumber\\
V(\phi)&=& V_0 \, \mbox{e}^{\pm \sqrt{ \frac{ 4\pi }{p}} \, 
\frac{\phi}{ m_\text{pl}} } \,,
\end{eqnarray}
where $m_\text{pl}$ is the Planck mass, $V_0$, $\alpha$, and $p$ are 
constants, with $p>1$ in order to have cosmic acceleration $\ddot{a}>0$. 
The de Sitter solution is obtained if $\phi=$~const., which corresponds to 
a cosmological constant \cite{Barrow:1987ia, Burd:1988ss, Muslimov:1990be}. 
The Klein-Gordon equation with exponential potential has a long history in 
mathematics, beginning with early studies of the non-linear wave equation 
by Liouville in 1853 \cite{Liouville1853}. Power-law inflation has the 
advantage that the spectral indices of scalar and tensor perturbations are 
also determined exactly.

Spatially curved ($K\neq 0$) versions of power-law inflation are possible, 
but these solutions cannot be put in explicit form even though the field 
equations can be integrated explicitly \cite{EllisMadsen1991}.

The exponential potential is a trademark of higher-dimensional 
compactified theories and of the subsequent transformation to the 
Einstein conformal frame and potentials consisting of a sum of 
exponential terms appear in supergravity and superstring theories. A sum 
of exponentials is typical of perturbation expansions in superstring 
theories \cite{Ozer:1992wh}. Two-exponential potentials
\be
V(\phi)= V_0 \, \mbox{e}^{2\lambda \phi} + V_1 \, \mbox{e}^{-2\lambda 
\phi} -2V_0 V_1
\ee
(with $V_{0,1}$ positive constants) were studied in relation with Noether 
and Hojiman symmetries and other conserved quantities 
\cite{deRitis:1990ba, deRitis2, deRitis3, Chimento:1995da, 
Chimento:2011dw, Capozziello:2013bma, Dimakis:2016mip}.

Kruger and Norbury \cite{Kruger:2000nra} found the spatially flat 
FLRW solution
\begin{eqnarray}
V(\phi) &=& V_0 \left[ 1+\cosh (\lambda \phi) \right] \,,\\
&&\nonumber\\
\phi(t) &=& \frac{1}{\lambda} \, \ln \left( \frac{ 
\mbox{e}^{ \lambda \sqrt{V_0}\, t} +1}{
\mbox{e}^{ \lambda \sqrt{V_0}\, t} -1} \right) \,,\\
&&\nonumber\\
a(t) &=& \left[ \exp \left( 2\lambda \sqrt{V_0}\, t\right) -1 
\right]^{1/3} \,,\\
&&\nonumber\\
\rho_{\phi}(a) &=& 2V_0 \left( 1+\frac{2}{ a^3} +\frac{1}{a^6} 
\right) \,.
\end{eqnarray}

Ellis and Madsen \cite{EllisMadsen1991} found a solution, for any 
curvature index $K$, for the potential
\be
V(\phi) = \frac{3H_0^2}{8\pi} +\phi_1^2 \sinh^2 \left[ \frac{2H_0}{\phi_1} 
\left( \phi-\phi_0\right)\right] 
\ee
(determined {\em a posteriori}), which produces
\begin{eqnarray}
a(t)&=& a_0 \sinh\left( H_0t\right) \,,\\
&&\nonumber\\
\phi(t)&=& \phi_0 \pm \frac{\phi_1}{H_0} \ln \left( 
\frac{ \mbox{e}^{H_0t} -1}{\mbox{e}^{H_0t} +1} \right)\,,
\end{eqnarray}
where 
\be
\phi_1^2 =\frac{1}{4\pi} \left( H_0^2 +\frac{K}{a_0^2}\right) \,,
\ee
$H_0$ and $a_0$ are positive constants, and $\phi_0$ is an integration 
constant. The dimensionless density parameter 
\be
\Omega(t) = \left( 1+ \frac{4\pi \phi_1^2}{H_0 \sinh^2 (H_0t)} \right) 
\tanh^2 (H_0t) 
\ee 
approaches unity at late times when the solution enters a slow-roll 
regime. 

Yet another exact solution by Ellis \& Madsen \cite{EllisMadsen1991} for 
any curvature index $K$ exhibits the unusual linear expansion
\begin{eqnarray}
a(t)&=& a_0t \,,\\
&&\nonumber\\
\phi(t) &=& \phi_0 \pm \phi_1 \ln t \,,\\
&&\nonumber\\
V(\phi) &=& \phi_1^2 \, \mbox{e}^{\frac{2(\phi-\phi_0)}{\phi_1}} \,,
\end{eqnarray}
with $a_0$ a positive constant,
\be
\phi_1^2 = \frac{1}{4\pi} \left( 1+\frac{K}{a_0^2} \right) 
\ee
and constant dimensionless density parameter $\Omega$. This 
(non-slow-roll) solution can be regarded as a modern version of the Milne 
universe which is filled by a scalar field and allows for any spatial 
curvature \cite{EllisMadsen1991}. By contrast, the original Milne universe 
is a hyperbolic ($K=-1$) foliation of empty Minkowski spacetime 
\cite{Slava}.

Easther \cite{Easther:1993qg} studied a potential of the form
\be 
V(\phi)= \sum_{j=1}^N V_j \, \mbox{e}^{ -\lambda_j \gamma\phi}
\ee
and gave solutions for $K=\pm 1$ for $a,t, \phi$ in parametric form, 
recovering  a previous solution of Ref.~\cite{Ozer:1992wh} and for $K=0$.

\subsubsection{Other exact solutions}

The intermediate inflationary scenario contains another exact solution  
\cite{Barrow:1990vx, Muslimov:1990be, Barrow:1993ah}:

\begin{eqnarray} 
V(\phi) & = &\frac{m^2}{ \phi^{\beta} } \left( 
1-\frac{ \beta^2 }{6\phi^2}\right) \,,\\
&&\nonumber\\
\rho(a) &=& \frac{m^2}{ \left( 2\beta\right)^{\beta/2} }\left( \ln a 
\right)^{-\beta/2} \,,\\
&&\nonumber\\
\phi &=& \left[ \phi_0^{\frac{\beta+4}{2}} \pm A 
(t-t_0)\right]^{\frac{2}{\beta+4} } \,,\\
&&\nonumber\\
a(t)&=& a_0 \, \exp \left[ \alpha (t-t_0)^{\gamma} \right] \,,
\end{eqnarray}
where $m$ is a mass scale, $\beta$ is a constant, and  
\begin{eqnarray}
A &=& \frac{\beta\left( \beta+4\right) m}{2\sqrt{3}} \,,\\
&&\nonumber\\
\gamma &=&  1- \frac{\beta+2}{\beta+4} \,,\\
&&\nonumber\\
\alpha &=& \frac{A^{\frac{2}{\beta+4}} }{2\beta} \,.
\end{eqnarray}

Various solutions were provided by Ellis \& Madsen 
\cite{EllisMadsen1991} with the purpose of obtaining any cosmic history 
$a(t)$ that could be reconstructed from cosmological observations, or to 
evade slow-roll. Their method consists of imposing the form of $a(t)$ and 
then solving for $\phi(t)$ and $V(t)$, inverting to obtain 
$t(\phi)$, and deriving  $V(\phi)=V(t(\phi))$ {\em a posteriori}. Their 
first solution is de Sitter expansion for $  K\ge0 $, given by
\begin{eqnarray}
a(t) &=& a_0 \, \mbox{e}^{H_0 t} \,,\label{EllisMadsen1a}\\
&&\nonumber\\
\phi(t) &=& \phi_0 \pm \frac{\phi_1}{H_0} \, \mbox{e}^{-H_0t} \,, 
\label{EllisMadsen1phi}\\
&&\nonumber\\
V(\phi) &=& \frac{ 3H_0^2}{8\pi} +H_0^2 \left( \phi-\phi_0\right)^2 
\,,\label{EllisMadsen1V}
\end{eqnarray}
where
\be
\phi_1= \sqrt{ \frac{K}{4\pi a_0^2}} 
\ee
and $a_0, H_0 $ are positive constants, while $\phi_0$ is an integration 
constant. The density parameter is \cite{EllisMadsen1991}
\be
\Omega(t) = 1+\frac{4\pi \phi_1^2}{H_0^2} \, \mbox{e}^{-2H_0t} \,.
\ee
This solution enters a slow-roll regime only at late times, as 
$\Omega\rightarrow 1$ and reduces to the usual de Sitter solution with 
cosmological constant (contained in the potential~(\ref{EllisMadsen1V})) 
and constant scalar field $\phi$ if $K=0$.

Another solution for $K>0$ is \cite{EllisMadsen1991} 
\begin{eqnarray}
a(t) &=& a_0 \, \cosh (H_0 t) \,,\\
&&\nonumber\\
\phi(t) &=& \phi_0 \pm \frac{2\phi_1}{H_0} \, \arctan \left( 
\mbox{e}^{H_0t} \right) \,,\\
&&\nonumber\\
V(\phi) &=& \frac{3H_0^2}{8\pi} + \phi_1^2 \sin^2 \left[ 
\frac{2H_0}{\phi_1} \left( \phi-\phi_0 \right) \right]
\,,
\end{eqnarray}
where $a_0, H_0$ are positive constants, 
\be
\phi_1^2= \frac{1}{4K} \left( \frac{K}{ a_0^2 }-H_0^2\right) 
\ee
and the density parameter
\be
\Omega(t) = \left[ 1+ \frac{4\pi \phi_1^2}{ H_0^2 \cosh^2 (H_0t)} \right] 
\text{coth}^2 (H_0t) 
\ee
goes from unity to infinity and back to unity.

Maartens, Taylor \& Roussos used the number of $e$-foldings (or $\ln 
(a/a_i)$, where $a_i$ is the scale factor at the beginning of inflation), 
or ultimately the scale factor itself, as the independent variable in 
order to find new analytical solution of scalar field cosmology 
\cite{Maartens:1995uz}. Their solutions interpolate between exponential or 
power-law inflationary expansion and the radiation era, providing an 
effective description of the exit from inflation. Although reheating and 
entropy production at the end of inflation cannot be taken into account by 
such a simple model, these analytical solutions offer a quick way to model 
a 
complicated transition.

Methods to solve the coupled Einstein-Friedmann-Klein-Gordon equations 
produced other exact inflationary solutions \cite{Salopek:1990jq,
Lidsey:1991dz, Lidsey:1991zp, Lidsey:1992wk, StarkovichCooperstock92, 
Barrow:1993hn, Barrow:1993zq, Carr:1993aq,
Lidsey:1994qa, Barrow:1994nt, Schunck:1994yd,  Ellis:1988jw,  
Zhuravlev:1998ff,  Liddle:1998xm, Kruger:2000nra, 
Yurov:2008sy, Charters:2009ku,  Andrianov:2011fg, Harko:2013gha, 
Paliathanasis:2014zxa, Chervon:2017kgn, Fomin18, Joseph:2019icj}. In 
general, all these solution-generating methods have some restrictions and 
limitations, and they necessarily focus on the mathematics, leading to new 
solutions which may not be very relevant for the physical aspects of the 
accelerated cosmic expansion during inflation or late quintessence 
domination.  We refer the reader to the recent review by Martin, Ringeval 
\& Vennin \cite{Martin:2013tda} for a comprehensive survey of inflationary 
scenarios with exact and approximate solutions of the relevant field 
equations.

\subsection{Scalar field plus fluid}
\label{subsec:sf+fluid}

Since a stiff fluid is equivalent to a free scalar field, solutions 
describing universes sourced by a single perfect fluid decoupled from a 
free scalar field have already been given as two-fluid solutions one of 
which obeys the stiff equation of state $P=\rho$. Whenever the scalar 
field can be integrated explicitly, its expression has been provided. Let 
us turn now to the physical motivation for such solutions.

Early universe inflation, which is believed to be driven by a scalar field 
in the early universe \cite{LiddleLyth, KolbTurner, 
Lidsey:1995np, Martin:2013tda}, 
provided the first motivation for searching solutions that describe FLRW 
universes sourced by a single fluid plus a scalar field. However, until 
1998 the interest was more mathematical than physical. Following 
the 1998 discovery of the present acceleration of the universe made with 
type Ia supernovae \cite{Perlmutter:1997zf, Riess:1998cb, 
Perlmutter:1998np} and the introduction of dark energy in theoretical 
cosmology, many models of scalar field (modelling dark energy) and 
a 
single fluid (a dust modelling dark matter) were introduced. The most 
popular 
ones (at least before the idea of modifying gravity at large scales to get 
rid of the {\em ad hoc} dark energy altogether \cite{Sotiriou:2008rp, 
DeFelice:2010aj, Nojiri:2010wj}) were, and remain, models 
consisting of a scalar field (the ``quintessence'') interacting with the 
dark matter fluid (a dust) only gravitationally. Indeed, the study of 
these models with inverse power-law potentials 
$V(\phi)=V_0/\phi^{\alpha}$, $\alpha>0$, begun with the work of Peebles \& 
Ratra  \cite{Peebles:1987ek} predating the discovery of the cosmic 
acceleration. We refer the reader to the 
excellent book by Amendola \& Tsujikawa \cite{Amendola:2015ksp} for a 
proper account of these quintessence models.

Pre-1998 analytical solutions include those of Barrow \& Saich 
\cite{Barrow:1993ah} and Chimento \& Jakubi \cite{Chimento:1995da}. 
Barrow \& Saich solved the Einstein-Friedmann equations for a spatially 
flat 
FLRW universe sourced by a scalar field in a potential $V(\phi)$ and a 
perfect fluid. They imposed that the scalar field effective equation of 
state be constant and determined the potential {\em a posteriori}. 

Attempts to solve the Einstein-Friedmann-Klein-Gordon equations with a 
potential usually amount to imposing a certain scale factor (usually, but 
not always, power-law or exponential), substituting it into the field 
equations, and solving them while determining a scalar field potential 
$V(\phi)$. The last step requires expressing $V$ as a function of time $t$ 
and inverting the functional relation $\phi=\phi(t)$ to find $t(\phi)$ and 
then $V(\phi)=V(t(\phi))$. In practice, this is not always possible to do 
analytically. In this approach \`a la Synge \cite{Synge}, the potential 
$V(\phi)$ is not determined by physical considerations but is {\em ad 
hoc}.

For example, with loose motivation from inflation, M\'endez 
\cite{Mendez:1996ug} imposed $a(t)=a_0 \, 
\mbox{e}^{Ht}$ while allowing for spatial curvature and a perfect fluid 
with constant equation of state parameter in the range $0\leq w\leq 1$. He 
found no solutions for $K\leq 0$ since \cite{Mendez:1996ug} 
\be 
\phi(t)=\phi_0 \pm \int_0^t dt' \sqrt{ \frac{K\, \mbox{e}^{-2Ht'} }{4\pi 
a_0^2} -\left(w+1\right) \rho_0 \, \mbox{e}^{-3(w+1) Ht'} } 
\,,\label{Mendez} 
\ee 
and found one solution for $K>0$, but the relation 
$\phi(t)$ cannot be inverted to give $V(\phi)$ \cite{Mendez:1996ug}. There 
are actually more analytical solutions $\phi(t)$ for $K>0$ that can be 
expressed in terms of elementary functions, as can be seen by applying the 
Chebysev theorem to Eq.~(\ref{Mendez}). Moreover, one can in principle 
find solutions for $K\leq 0$ by allowing a phantom equation of state 
($w<-1$) but invertibility of $\phi(t)$ is usually not possible. M\'endez 
turned instead to a viscous imperfect fluid with bulk viscosity 
proportional to a power of its energy density, finding analytical 
solutions 
with scalar field that is an inverse exponential of time and quadratic 
potential $V=V_0 +V_1 \left(\phi-\phi_0 \right)^2$. (This reproduces the 
solution (\ref{EllisMadsen1a})-(\ref{EllisMadsen1V}) of Ellis \& Madsen 
\cite{EllisMadsen1991}.)  However, the chain of assumptions jeopardizes 
the physical significance of these solutions.

Hawkins \& Lidsey discovered that the Einstein-Friedmann 
equations~(\ref{Friedmann1})-(\ref{conservation1})  describing the 
dynamics of a spatially flat universe sourced by a perfect fluid plus a 
scalar field can be reformulated in terms of the non-linear Ermakov-Pinney 
equation \cite{Hawkins:2001zx}. Ermakov systems consisting of two coupled, 
second order, non-linear ordinary differential equations arise in many 
fields of physics. In one dimension (the case of FLRW cosmology, in which 
the scale factor $a(t)$ depends on only one variable), the Ermakov system 
reduces to a single (``Ermakov-Pinney'' or ``Milne-Pinney'') equation 
\cite{Ermakov1880, Pinney50} of the form
\be
\frac{d^2b}{d\tau^2} +Q( \tau) b =\frac{\lambda}{b^3} 
\,,\label{ErmakovPinney}
\ee
where $\lambda$ is a constant and $Q(t)$ is an arbitrary function. In 
fact, considering a spatially flat FLRW universe sourced by a perfect 
fluid with a constant equation of state 
$P=w\rho=w\rho^{(0)}/a^{3(w+1)} $ and a minimally coupled scalar field 
$\phi$ not interacting directly with 
it and  with potential 
$V(\phi)$,  the Einstein-Friedmann 
equations~(\ref{Friedmann1})-(\ref{conservation1}) can be combined to give
\be
\frac{\ddot{a} }{a} -\frac{\dot{a}^2}{ a^2} =-4\pi \left[ \dot{\phi}^2 + 
(w+1)\frac{\rho^{(0)} }{a^{3(w+1)}} \right] \,. \label{combo}
\ee
Setting $n\equiv 3(w+1)$ and $ a\equiv b^{2/n}$ (with $b>0$) and changing 
the time coordinate to $\tau$ defined by $d\tau=b dt$, Eq.~(\ref{combo}) 
becomes \cite{Hawkins:2001zx}
\be
\frac{d^2b}{d\tau^2} +2\pi n \left( \frac{d\phi}{d\tau} \right)^2 b = 
-\frac{2\pi n^2 \rho^{(0)}}{3b^3} \,,
\ee
which does not contain $V(\phi)$ and is of the form~(\ref{ErmakovPinney}) 
with 
\begin{eqnarray}
Q(\tau) &=& 2\pi n \left( \frac{d\phi}{d\tau} \right)^2 \,,\\
&&\nonumber\\
\lambda &=& -\frac{2\pi n^2 \rho^{(0)}}{3} \,.
\end{eqnarray}
Given two linearly independent solutions $b_{1,2}(\tau)$ of the associated 
linear homogeneous equation
\be
\frac{d^2 b}{d\tau^2}  + Q( \tau) \, b = 0 \,,\label{EPhomogeneous}
\ee
the general solution of the non-linear Ermakov-Pinney 
equation~(\ref{ErmakovPinney}) is \cite{Pinney50}
\be
b_P(\tau) = \left[ A b_1^2(\tau) + B b_2^2(\tau)+2C b_1(\tau)b_2(\tau) 
\right]^{1/2} \,,
\ee
where $A,B$, and $C$ are constants satisfying
\be
AB-C^2 =\frac{\lambda}{W^2}
\ee
and the Wronskian 
\be
W= b_1 \, \frac{db_2}{d\tau} - b_2 \, \frac{db_1}{d\tau} 
\ee
is constant \cite{Pinney50}. 

The Ermakov-Pinney equation is closely related to the one-dimensional 
Schr\"odinger equation. Setting now 
$\psi \equiv a^{-n/2}$, changing the variable $t$ to $\sigma$ defined by 
$d\sigma=\psi(t) dt$, Eq.~(\ref{combo}) assumes the Schr\"odinger form
\be
\frac{d^2\psi}{d\sigma^2} +\left[ E-P(\sigma)\right] \psi (\sigma)=0 \,,
\ee
where $ E\equiv -2\pi n^2 \rho^{(0)} /3$ and 
\be
P(\sigma)= 2\pi n \left( \frac{d\phi}{d\sigma}\right)^2
\ee
is the potential. 

In practice, the Ermakov-Pinney equation is solved  by choosing 
the scale factor $a(t)$, solving for $\phi(t)$, inverting this relation 
to obtain $t(\phi)$, and then determining the potential 
$V(\phi)=V(t(\phi))$ \cite{Hawkins:2001zx}. As a simple check, for  
$\phi=$~const.  (which implies $V(\phi)=$~const., equivalent to a 
cosmological constant $\Lambda$), one recovers Harrison's 
solution~(\ref{anyfluidLambdapos}) for a 
single fluid with $\Lambda$ \cite{Hawkins:2001zx}. In general, the 
potential $V(\phi)$ may turn out to be physically unmotivated and care 
must be taken to restrict to situations of physical interest. 

Several authors have used the Hawkins-Lidsey reduction to an 
Ermakov-Pinney equation  \cite{Gumjudpai:2007qq, Gumjudpai:2007bx, 
Phetnora:2008mf, Gumjudpai:2008mg, 
Gumjudpai:2009ws}. D'Ambroise studied further the equivalence between the 
Einstein-Friedmann-Klein-Gordon system with a perfect fluid and non-linear 
Schr\"odinger equations \cite{DAmbroise:2010dgl}.  Seven exact solutions 
for a scalar field and a single barotropic fluid $P=w\rho$ with 
$w=$~const. were found \cite{DAmbroise:2010dgl}.

The case of a scalar field and two perfect fluids with constant equation 
of state $P_i=w_i \rho_i$ was studied in \cite{Kritpetch:2019eof} using a 
non-linear Schr\"odinger equation representation, and the D'Ambroise 
single-fluid solutions were generalized to include the second fluid. 
Again, the potentials $V(\phi)$ 
found do not appear to be physical. 

The reductions to the Ermakov-Pinney equation and to linear and non-linear 
Schr\"odinger equations have been applied also to spatially curved FLRW 
universes \cite{Williamsetal06, Gumjudpai:2007bx}, anisotropic Bianchi 
models \cite{DAmbroise:2007zhp}, braneworld models \cite{Hawkins:2001zx}, 
and quantum cosmology \cite{Rosu:1999ud, 
Christodoulakis:2003pg,DAmbroiseWilliams}. There is also a connection with 
analogue gravity: in the $K=+1$ case, the Ermakov-Pinney equation 
establishes an analogy with two-dimensional Bose-Einstein condensates 
\cite{Lidsey:2003ze}.

\subsection{Multiple non-interacting scalar fields}

Inflation with multiple non-interacting scalar fields was studied before 
the 1998 discovery of the present acceleration of the universe 
\cite{Perlmutter:1997zf, Riess:1998cb, Perlmutter:1998np}, when it 
seemed that the total dimensionless density parameter $\Omega_\text{tot}$ 
was approximately $0.3$, {\em i.e.}, before the inclusion in the standard 
$\Lambda$CDM model of dark energy bringing $\Omega_\text{tot}$ up to 
unity. In fact, $\Omega_\text{tot}=1$ is a robust prediction of single 
scalar field inflation and a natural way to obtain inflation with 
$\Omega_\text{tot} <1 $ consists of using multiple inflaton fields.  
These models of multiple field inflation with $\Omega_\text{tot} <1 $ were 
practically abandoned soon after 1998. Modern interest in multiple field 
inflation 
originates from the fact that these two-field models \cite{Linde:1996gt, 
Bartolo:2001cw, Bernardeau:2002jy, Bernardeau:2002jf, Lyth:2002my, 
Enqvist:2004bk, Bartolo:2003jx, Linde:2005yw, Alabidi:2006wa, 
Barnaby:2006cq, Alabidi:2006hg, Sasaki:2006kq, Choi:2007su, 
Barnaby:2006km, Byrnes:2008wi, Sasaki:2008uc, Vincent:2008ds, Wang:2010si, 
Meyers:2010rg, Naruko:2008sq, Byrnes:2008zy, Peterson:2010mv} and multiple 
field models \cite{Rigopoulos:2005ae, Rigopoulos:2005us, Battefeld:2006sz, 
Kim:2006te, Yokoyama:2007dw, Yokoyama:2007uu, Battefeld:2007en, 
Huang:2009xa, Kim:2010ud}
 predict non-Gaussianity in the spectrum of density 
perturbations. The search for non-Gaussianity by measuring $n$-point 
correlation functions (with $n\geq 3$) is an active line of research in 
observational cosmology \cite{Linde:1996gt, Bartolo:2001cw, 
Bernardeau:2002jy, Bernardeau:2002jf, Lyth:2002my, Enqvist:2004bk, 
Bartolo:2003jx, Linde:2005yw, Alabidi:2006wa, Barnaby:2006cq, 
Alabidi:2006hg, Sasaki:2006kq, Choi:2007su, Barnaby:2006km, Byrnes:2008wi, 
Sasaki:2008uc, Vincent:2008ds, Wang:2010si, Meyers:2010rg, Naruko:2008sq, 
Byrnes:2008zy, Peterson:2010mv, Rigopoulos:2005ae, Rigopoulos:2005us, Battefeld:2006sz, Kim:2006te,
Yokoyama:2007dw, Yokoyama:2007uu, Battefeld:2007en, 
Huang:2009xa, Kim:2010ud}.

Since, in principle, one can mimic a perfect fluid with constant equation 
of state with a scalar field $\phi$ by imposing that its effective 
equation of state be constant and determining its potential $V(\phi)$ 
accordingly, as done by Barrow \& Saich \cite{Barrow:1993ah}, all 
two-fluids exact solutions presented in Secs.~\ref{sec:5} and~\ref{sec:6} 
can in principle be reproduced in this way, although the two potentials 
determined in this way for the two scalar fields will, in general, not be 
physically motivated. 
A potential is necessary because a free scalar field with $V(\phi)\equiv 
0$ plus a stiff fluid is equivalent to a single stiff fluid and does not 
give rise to new solutions beyond what already seen in 
Sec.~\ref{sec:5}.

\section{Two interacting fluids}
\setcounter{equation}{0}
\label{sec:8}

The modern interest in two interacting fluids in cosmology arises in 
relation with the dark energy problem. Early interest in the 1960s and 
1970s was motivated by the need to model the conversion of radiation into 
non-relativistic matter in the early universe. We proceed with some 
general considerations applied to interacting dark energy scenarios before 
moving to the early literature.

\subsection{Interacting dark energy and dark matter}
\label{subsec:DEDM}

The standard $\Lambda$CDM model of the universe 
contains, in addition to a small fraction of baryonic matter, dark energy 
and dark matter, the two most abundant constituents the nature of which is 
completely unknown \cite{Planck:2013pxb}. Dark energy and dark matter, the 
two biggest mysteries of $\Lambda$CDM cosmology, are normally treated as 
two separate physical sectors. However, 
it has been speculated \cite{Billyard:2000bh} that these two sectors 
could couple explicitly in the so-called ``interacting dark energy'' 
scenarios. The phenomenological implications of such an interaction would 
be significant: the coincidence problem of why the current densities 
$\Omega_\text{DE}$ and $\Omega_\text{DM}$ of dark energy and dark matter 
are of the same order of magnitude today, although they evolve 
differently, would be alleviated. In interacting dark energy models, the 
interaction is modelled by modifying arbitrarily the equations of motion 
for dark energy and dark matter, described as interacting perfect fluids. 
In this picture, these fluid stress-energy tensors are not covariantly 
conserved 
separately but the total $T_{ab}^\text{(tot)} = T_{ab}^\text{(DE)} + 
T_{ab}^\text{(DM)} $ is:
\begin{eqnarray}
\nabla^b T_{ab}^\text{(DM)} &=& Q_a \,, \nonumber\\
&& \label{conservsepar}\\
\nabla^b T_{ab}^\text{(DE)} &=& -Q_a \,,\nonumber
\end{eqnarray}
so that
\be
\nabla^b T_{ab}^{(tot)} =0 
\,.\label{conservtotal}
\ee
The four-vector $Q^a$ describes  phenomenologically the 
interaction. Its detailed form is picked by hand in the literature, 
reflecting our ignorance about the dark sectors, while a proper 
description would express the unknown physics of the interaction 
(see Refs.~\cite{Fuzfa:2007sv, Ziaeepour:2011bq, Ballesteros:2013nwa} for 
attempts to derive the vector $Q^a$ from 
physical considerations). What is more, it is not even clear how 
to provide a Lagrangian description of the interaction, as should always 
be done in fundamental physics. With the exception of Einstein frame 
scalar-tensor gravity (see Sec.~\ref{subsec:Einsteinframe}), we lack even 
an effective field 
theory Lagrangian for this purpose, in spite of substantial theoretical 
efforts to describe dark energy (and its contender,  
 modified gravity) via effective field theories 
\cite{Bloomfield:2012ff,Gubitosi:2012hu, Heisenberg:2018vsk}. 

The standard description in the large literature ({\em e.g.}, 
\cite{Capozziello:2005pa, Cruz:2006ck, Nojiri:2006zh, Nojiri:2007qi, 
Bertolami:2007gv, Sotiriou:2008it, Valiviita:2008iv, Chen:2008ft, 
Jamil:2009eb, He:2009mz, He:2010im, Clemson:2011an,  
BasteroGil:2012cm, Xu:2013jma, Costa:2013sva, Gergely:2014rna, 
Khurshudyan:2014yva,  Wetterich:2014bma, Brevik:2014lpa, Li:2014eha, 
Zimdahl:2014jsa, Amendola:2015ksp}, see \cite{Wang:2016lxa, 
SilbergleitChernin17} for 
reviews) 
assumes a spatially flat FLRW universe filled by two perfect fluids 
satisfying the equations
\be\label{fluid1} 
\dot{\rho}_1+3H \left(P_1+\rho_1 \right) =Q \,, 
\ee 
\be\label{fluid2} 
\dot{\rho}_2+3H \left(P_2+\rho_2 \right) =-Q \,. 
\ee 
where the quantity $Q(t)$ models the interaction. The explicit forms of 
$Q(t)$ adopted in the literature are arbitrary: usually this 
quantity is assumed to be a function of one or more of $a(t), \dot{a}(t), 
H(t), \rho_{1,2}(t)$ or of powers of them 
\cite{Capozziello:2005pa, Cruz:2006ck, Nojiri:2006zh, Nojiri:2007qi,  
Bertolami:2007gv, Sotiriou:2008it, Valiviita:2008iv, Chen:2008ft, 
Jamil:2009eb, He:2009mz, He:2010im, Clemson:2011an,  BasteroGil:2012cm, 
Gergely:2014rna, Khurshudyan:2014yva, Wetterich:2014bma, Brevik:2014lpa, 
Li:2014eha, Zimdahl:2014jsa,  Xu:2013jma, Costa:2013sva}. Attempts can be made to 
constrain the form of $Q(t)$ using cosmological observations 
\cite{He:2009mz, He:2010im, Xu:2013jma, Costa:2013sva}.

The total fluid with  energy density $ \rho_\text{tot}=\rho_1 +\rho_2 $ 
and pressure $ P_\text{tot}=P_1+P_2 $ obeys the conservation equation 
\be \label{totalconservation}
\dot{\rho}_\text{tot}+3H \left(P_\text{tot}+\rho_\text{tot} \right) =0 \,. 
\ee 

The extensive literature on interacting dark energy leaves the basic 
questions of a covariant and Lagrangian formulation almost completely 
unanswered, presenting models already set up in comoving coordinates in 
the  FLRW universe. A speculative covariant description of the possible 
dark energy-dark  matter interaction has been proposed in 
Ref.~\cite{Faraoni:2014vra}. There, the two 
fluids are described by the modified energy-momentum tensors
\begin{eqnarray}
T^{(1)}_{ab} &= & \left( P_1 + \rho_1 \right) u_au_b +P_1 g_{ab}
+ q_au_b +q_b u_a   \,, \label{bbb}\\
&&\nonumber\\
T^{(2)}_{ab} &= & \left( P_2 + \rho_2 \right) u_au_b +P_2 g_{ab} 
-q_au_b -q_b u_a \,,
\end{eqnarray}
where $u^a$ is a common 4-velocity pointing in the time direction of the 
observers comoving with both fluids. The four-vector  $q^c$  describes an 
energy flux density between the two fluids. In a spatially isotropic FLRW 
universe,  $q^c$ 
cannot have spatial components in comoving coordinates and points in 
the direction of comoving time, {\em i.e.}, 
\be
q^c=\alpha(t) u^c \,,
\ee
where $\alpha $ is a function of time (with $\alpha\geq 0$ to keep  
$q^c$ future-oriented).

The two fluids are imperfect fluids, but by all means not in the usual 
sense of dissipative fluid used in the relativity 
literature ({\em e.g.}, \cite{Eckart:1940te, Stephani}). Usually the 
dissipative term 
$q_au_b +q_b u_a$ in the imperfect fluid stress-energy tensor describes a 
purely  spatial (therefore, non-causal) heat flux density $q^c$ with 
$q^cu_c=0$ \cite{Eckart:1940te, Stephani}. Here, instead, $q^c$  must point 
parallel to 
$u^c$, otherwise it violates spatial 
isotropy. The traces of the fluids stress-energy tensors  $T^{(i)}_{ab}$ 
are  
\be
T^{(i)} = -\rho_i+3P_i \mp 2\alpha  \,.
\ee
While the total stress-energy tensor $ T_{ab}^\text{(tot)} = T_{ab}^{(1)} 
+ T_{ab}^{(2)} $ is covariantly conserved, each component $T_{ab}^{(i)}$   
of the mixture satisfies 
\begin{eqnarray}
&&\nabla^b T_{ab}^{(i)} = u_a u^b \nabla_b P_i + u_a u_b
\nabla^b \left( \rho_i \pm 2\alpha \right)+\nabla_a P_i  \nonumber\\
&&\nonumber\\ 
&&\ \ \ \ \ \ \ \ \ \ \ \ \ +
\left( P_i+\rho_i \pm 2\alpha \right) u^b \nabla_b u_a  \nonumber\\ 
&&\nonumber\\
&&\ \ \ \ \ \ \ \ \ \ \ \ \ +\left( P_i 
+ \rho_i  \pm 2\alpha \right) u_a \nabla^b u_b  \,.
\end{eqnarray}
Here $i=1,2$, the upper sign corresponds to fluid~1, and the lower 
one to fluid~2.  By projecting this equation on the time direction one 
obtains
\be
u^a \nabla^b T_{ab}^{(i)} = 
\left( \dot{\rho}_i \pm 2\dot{\alpha} \right) +3H\left( P_i +\rho_i 
\pm 2\alpha \right) \,,
\ee
with the usual notation $\dot{\rho}_i \equiv u^a \nabla_a \rho_i$.
 If we require that  $u^a \nabla^b T_{ab}^{(i)} =0$, the two 
fluids are covariantly conserved separately, but their perfect-fluid 
components $  \left( P_i+\rho_i \right) u_au_b +P_i g_{ab}$ are not 
because  
\be
u^a \nabla^b \left[ \left( P_i+\rho_i \right) u_au_b +P_i g_{ab} 
\right] = \pm 2 \left( \dot{\alpha} +\alpha \nabla_b u^b \right) 
\,.
\ee
In a FLRW universe this equation assumes the form
\be
\dot{\rho}_i +3H \left( P_i +\rho_i \right) =\mp 2\left( 
\dot{\alpha}+3H\alpha \right) 
\ee
and one naturally reads the right hand side as $\pm 
Q (t)$, reproducing Eqs.~(\ref{fluid1}) and~(\ref{fluid2}). Then
$\alpha(t) $ and $Q(t) $ are related by 
\be
\dot{\alpha}+3H\alpha+\frac{Q(t)}{2}=0 
\ee
or, equivalently,
\be
\label{Qa}
\frac{1}{a^3} \frac{d}{dt}\left( \alpha a^3\right) +\frac{Q(t)}{2} 
=0 \,,
\ee
yielding  
\be
\alpha(t) =-\frac{1}{2a^3(t) } \int dt \, a^3(t) Q(t) \,.
\ee

A possible physical interpretation is proposed in \cite{Faraoni:2014vra}:  
fluid~1, which is is not a perfect fluid, has effective energy 
density and pressure
\begin{eqnarray}
T_{ab}^{(1)} u^a u^b &=& \rho_1+2\alpha \neq \rho_1\,,\nonumber\\
&&\nonumber\\
P_1 &=& \frac{1}{3} \, T_{ab}^{(1)} h^{ab} \,,
\end{eqnarray}
while, for fluid~2, 
\begin{eqnarray}
T_{ab}^{(2)}u^a u^b &=& \rho_2 -2\alpha \neq \rho_2 \,,\\
&&\nonumber\\
P_2 &=& \frac{1}{3} \, T_{ab}^{(2)} h^{ab} \,.
\end{eqnarray}
The two fluids are not 
perfect fluids because their stress-enery tensors $T_{ab}^{(i)}$  contain 
the terms  $\pm \left( q_au_b +q_b u_a \right)$ describing   an energy 
transfer occurring simultaneously at all points of space without 
three-dimensional flux. 
 The  energy lost by fluid~1 per 
unit time and per unit volume is transferred to fluid~2, according to the 
splitting (\ref{fluid1})-(\ref{fluid2}).

An alternative interpretation \cite{Faraoni:2014vra} is the following:  when 
$\alpha>0$, the term $2\left( \dot{\alpha}+3H\alpha \right)$ appearing  in
fluid~1 corresponds to a dust with zero 
pressure and energy density $2\alpha$ which transfers  energy 
instantaneously to fluid~1  
taking it from fluid~2. According to fluid~2, a dust with negative energy 
density $-2\alpha$ removes energy to transfer it to fluid~1. This 
second dust obviously violates the weak energy condition, but this problem 
is not more significant than the lack of causality of usual imperfect 
fluids with spacelike heat currents \cite{Eckart:1940te, Stephani}. 

The relation~(\ref{Qa}) between the quantities $Q(t)$ and  $\alpha (t)$ 
becomes, in this picture, 
\be
Q(t)=-\frac{2\left( \alpha a^3 \right)\dot{} }{a^3} \,.
\ee
For a  three-dimensional region  with unit comoving volume and physical 
volume $a^3$,  $-2\alpha a^3$ is the energy transferred between the two 
fluids in this volume, $-(2\alpha a^3 )\dot{}$ is the energy transfer 
rate, while $Q(t) $ is the rate of energy transferred 
per unit volume.

\subsection{Interacting radiation and non-relativistic matter}

Early work on interacting fluids was motivated by the need to model the 
conversion of the cosmic fluid from radiation (with energy density and 
pressure $\rho_\text{r}$, $P_\text{r}$) to non-relativistic matter (with 
$\rho_\text{m}$ and $P_\text{m}=0$)  and is mainly due to 
Davidson \& Narlikar \cite{DavidsonNarlikar66} and McIntosh 
\cite{McIntoshNature67, McIntosh1968b, McIntosh1970} for 
spatially flat universes, May \& 
McVittie \cite{MayMcVittie1970} for $K=-1,0$ and again to May \& McVittie 
\cite{MayMcVittie1971} for $K=+1$. McIntosh assumed an {\em ad hoc} form 
of the total equation of state parameter 
\be
w_\text{tot} (t) \equiv  \frac{ P_\text{tot}}{\rho_\text{tot}} 
=\frac{P_\text{r}}{ 
\rho_\text{r}+\rho_\text{m} } 
\ee
interpolating between  $1/3$ at early 
times and $0$ at late times \cite{McIntosh1968b, McIntosh1970}, for 
example \cite{McIntosh1968b}
\be 
w_\text{tot}(t)= \frac{1}{3\left( 1+t/t_0 \right)^{\alpha}} \,,
\ee
where $\alpha$ and $t_0$ are constants. 
May \& McVittie allowed for the possibility that $P_\text{m}\neq 0$ (all 
these early works assumed $\Lambda=0$). Instead of prescribing a function 
$w_\text{tot}(t)$, they introduced a 
non-negative function $\phi$ such that  $ \dot{a}  
=g(\phi)$, where \cite{MayMcVittie1970}  
\be 
g(\phi)=  \left\{
\begin{array}{lll}
& \text{cot}\, \phi  & \quad \quad \mbox{if  } K=+1 \,,\\
&&\\
& 1/ \phi  & \quad \quad \mbox{if  } K=0 \,,\\
&&\\
& \text{coth} \, \phi  & \quad \quad \mbox{if  } K=-1 \,.
\end{array} 
\right.
\ee
$g(\phi)$ satisfies the differential equation \cite{MayMcVittie1970} 
\be
\frac{dg}{d\phi} + g^2 = -K  \label{eqforg}
\ee
and the scale factor is given by 
\be
a(t) =\frac{3w_\text{tot}(t)+1}{ 2 {\dot{\phi}  }} 
\,.\label{MayMcVittiea}
\ee
The covariant conservation equation for the fluid mixture 
$\dot{\rho}_\text{tot}+3H \left( P_\text{tot}+\rho_\text{tot}\right)=0$ is 
equivalent to
\be 
\frac{d}{dt} \left( \frac{3w_\text{tot}(t)+1}{ \dot{\phi}  }\right) 
=2g(\phi) \,,
\ee
which determines $\phi(t) $ and $a(t)$ once $w_\text{tot}(t)$ is assigned. The 
(non-)conservation equations for radiation and matter are 
\begin{eqnarray}
\dot{\rho}_\text{r}+3H\left( P_\text{r}+\rho_\text{r} \right) &=& 
Q_\text{r} 
\,,\\
&&\nonumber\\
\dot{\rho}_\text{m}+3H\left( P_\text{m}+\rho_\text{m} \right) &=& 
Q_\text{m} \,,
\end{eqnarray}
where $Q_\text{r}$ is the  radiation energy per unit 
volume converted into matter energy per unit time and 
\be
Q_\text{r}+Q_\text{m} =0 \,.
\ee
A mixture of non-interacting radiation and dust ($P_\text{m}=0$)  is 
obtained as the trivial case $Q_\text{r}=Q_\text{m}=0$, with 
\be
w_\text{tot}(t)=\frac{1}{1+ \frac{\rho_\text{m}^{(0)} 
}{\rho_\text{r}^{(0)} }\, a} 
\,.
\ee

Eq.~(\ref{eqforg}) can be rewritten as \cite{MayMcVittie1970} 
\be
\frac{d}{d\phi} \left\{ \ln \left[ (3w_\text{tot}+1) \frac{dt}{d\phi} \right] 
\right\}= \frac{2g(\phi)}{3w_\text{tot}+1} 
\ee
and integrated twice to 
\begin{eqnarray}
&& \frac{3w_\text{tot}+1}{ \dot{\phi} } = B \exp\left( 2\int 
d\phi \, \frac{g}{3w_\text{tot}+1} \right)  \,,\\
&&\nonumber\\
&& t-t_0 = B \int \frac{d\phi}{3w_\text{tot}+1} \exp \left(2\int  
\frac{d\phi' \, g}{3w_\text{tot}+1} \right) 
\end{eqnarray}
with $B$ and $t_0$ integration constants \cite{MayMcVittie1970}. Then 
Eq.~(\ref{MayMcVittiea}) gives
\be
a(t) = \frac{B}{2} \, \exp\left( 2\int d\phi \, \frac{g}{3w_\text{tot} +1} 
\right)\,.
\ee
Using
\be
F \equiv \exp\left( -2\int d\phi \, \frac{g}{3w_\text{tot}+1} \right) \,,
\ee   
one has 
\begin{eqnarray}
\rho_\text{tot}(t) &=& \frac{3F^2}{ 2\pi B^2} \left( g^2+K\right) 
\,,\\
&&\nonumber\\
P_\text{tot}(t) &=& \frac{3F^2}{2\pi B^2} \, w_\text{tot} \left( 
g^2+K\right) \,,
\end{eqnarray}
and 
\be 
\rho_\text{m}= -\rho_\text{r} +3\rho_\text{tot}^{(0)} F^2 \left( g^2+K 
\right) \,,
\ee
where
\be
\rho_\text{tot}^{(0)} \equiv \frac{1}{2\pi B^2} \,.
\ee
By further introducing \cite{MayMcVittie1970} 
\be
h(\phi) \equiv \frac{P_\text{m} }{ \rho_\text{tot}^{(0)} } \,,
\ee
the transfer rate is computed as 
\begin{eqnarray}
Q_\text{m} &=& 3\rho_\text{tot}^{(0)}  \phi^{\dot{}   } \left\{ F^4 
\frac{d}{d\phi} 
\left(  \frac{h}{F^4} \right) \right.\nonumber\\
&&\nonumber\\
&\, & \left.  - \left[ \frac{6w_\text{tot} 
(1-3w_\text{tot})}{3w_\text{tot} +1} \, 
g+  3\frac{dw_\text{tot}}{d\phi} \right] F^2 \left(g^2+K\right) \right\} 
\,.\nonumber\\
&&
\end{eqnarray}
The relation between $w_\text{tot} (t)$ and $g(\phi)$ in 
\cite{MayMcVittie1970} is not physically transparent and is dictated more 
by the mathematics than the physics. Other models of (possibly tilted) 
interacting radiation and dust fluids were studied over the years, see 
{\em e.g.} \cite{ColeyTupper86b,ColeyTupper86, Nesteruk94, 
Nesteruk:1995uu}.

\subsection{Scalar field interacting with a perfect fluid}

A scalar field can couple to a fluid or to another field. Consider a 
perfect fluid with energy density $\rho_1$ and pressure $P_1$ coupling 
explicitly to a scalar field $\phi$.  Their interaction is again described 
by the equations
\be \label{Bfluid1}
\dot{\rho}_1+3H \left(P_1+\rho_1 \right) =Q \,, 
\ee 
\be 
\dot{\rho}_{\phi}+3H \left(P_{\phi}+\rho_{\phi} \right) =-Q \,;
\label{conservationphi} 
\ee
adding them leads to the conservation equation for the total fluid with 
energy density $\rho_\text{tot}=\rho_1+\rho_{\phi}$ and pressure 
$P_\text{tot}=P_1+P_{\phi}$.

Historically, interest in a scalar field directly coupled to matter can be 
found in the Hoyle-Narlikar steady state theory \cite{HoyleNarlikar63, 
HoyleNarlikar66} in which matter is created by the so-called C-field (for 
``creation''), a scalar field tranferring its energy into matter. Although 
is usually referred to as a scalar-tensor theory, the energy transfer from 
the scalar C-field requires only a direct coupling in the context of 
general relativity \cite{HoyleNarlikar63, McIntosh1970}.  As the 
steady-state theory was abandoned by the cosmological community, the 
interest in a direct coupling between a scalar and ordinary matter waned.

In the 1980s literature devoted to reheating the universe after it 
is cooled by the 
inflationary expansion, an interaction between the inflaton scalar field 
and radiation was introduced in a phenomenological way in order to model 
the decay 
of the inflaton into ultrarelativistic particles. This process should  
happen due to the inflaton's coupling to other particles and should reheat 
the universe to allow for primordial nucleosynthesis and the formation of 
what is now the cosmic microwave background \cite{Kofman:1985aw, 
KolbTurner}.  
(Later research produced more sophisticated scenarios for ending 
inflation, most notably parametric amplification during the so-called 
preheating, see Refs.~\cite{Allahverdi:2010xz, Amin:2014eta} for 
reviews.) The phenomenological 
interaction has 
the form
\be \label{Q}
Q= \Gamma \dot{\phi}^2 \,,
\ee
where $\Gamma$ is a positive constant. The equation of 
motion (\ref{conservationphi}) for the scalar is modified to  
\be 
\dot{\phi} \left( \ddot{\phi}+3H\dot{\phi} +\Gamma 
\dot{\phi} +\frac{dV}{d\phi} \right)=0 \,.
\ee 
Since\footnote{It is $\dot{\phi}\neq 0$ because, if $\phi=\phi_0=$~const., 
the scalar field effective fluid 
reduces to a pure cosmological constant $\Lambda =V(\phi_0)$ and decouples 
from the perfect fluid.} $\dot{\phi}\neq 0$ this equation becomes a 
Klein-Gordon one with a potential augmented by a friction term of strength 
$\Gamma$ proportional to the inflaton speed $\dot{\phi}$. Taking 
this phenomenological description as paradigmatic for a scalar field-fluid 
interaction leads to 
\be 
\ddot{\phi}+3H\dot{\phi} +\Gamma 
\dot{\phi} +\frac{dV}{d\phi} =0 
\ee
for the scalar and to 
\be 
\dot{\rho}_1+3H \left(P_1+\rho_1 \right) 
=\Gamma \dot{\phi} 
\ee 
for the fluid~1 \, which has now a source $\Gamma \dot{\phi}$ on the right 
hand side. This description can be related to the covariant description of 
two interacting fluids \cite{Faraoni:2014vra} described in 
Sec.~\ref{subsec:DEDM}. Then,
\be 
\label{alphaff}
\alpha(t)= -\frac{\Gamma}{2a^3} \int dt \, 
a^3 \dot{\phi}^2 
\ee 
depends only on the kinetic energy density $\dot{\phi}^2/2$ of $\phi$. The 
decay of the field $\phi$ into fluid stops when $\phi$ becomes static.

\subsection{Einstein frame formulation of scalar-tensor gravity}
\label{subsec:Einsteinframe}

Although we confine ourselves to Einstein gravity in the rest of this 
review, here we make an exception to mention one Lagrangian and covariant 
fomulation 
of interacting dark energy. In the context of alternative theories of 
gravity, when scalar-tensor 
\cite{Brans:1961sx, Bergmann:1968ve, Wagoner:1970vr, Nordtvedt:1970uv}
or $f(R)$ \cite{Capozziello:2003tk, Carroll:2003wy, Sotiriou:2008rp, 
DeFelice:2010aj, Nojiri:2010wj} gravity is reformulated in the Einstein 
conformal frame, the scalar field degree of freedom looks like an ordinary 
scalar field in general relativity, except for the fact that it couples 
explicitly to matter. If the latter is the dark matter fluid in a FLRW 
cosmology, one then obtains a direct interaction of this dark matter with 
the extra scalar degree of freedom (which could be used to model dark 
energy) contained in the theory in addition to the two massless spin two 
modes of Einstein theory.

Let us illustrate how this works using Jordan-Brans-Dicke gravity 
\cite{Jordan38, Jordan:1959eg, Brans:1961sx} and cosmology 
\cite{FujiiMaeda, Faraoni:2004pi}. Brans-Dicke's  
is the simplest scalar-tensor framework and the prototypical alternative 
to  general relativity. The Jordan frame Brans-Dicke action is 
\cite{Brans:1961sx} 
\be 
S^{(BD)} =\frac{1}{16\pi} \, \int d^4 x \, \sqrt{-g} \left[
\phi R -\frac{\omega}{\phi} \, g^{ab} \nabla_a\phi \, \nabla_b \phi
-V( \phi) \right]  + S^\text{(m)} \,, \label{1}
\ee
where $ S^\text{(m)}=\int d^4 x \, \sqrt{-g} \, {\cal L}^\text{(m)} $ 
describes the 
matter sector and the Brans-Dicke scalar field $\phi>0$ is, roughly 
speaking, the inverse of the effective gravitational coupling strength 
$G_\text{eff}$, which becomes a field in this theory. The dimensionless 
parameter $\omega$ is the ``Brans-Dicke 
coupling'' \cite{Brans:1961sx} and the Jordan frame is the pair of 
dynamical variables 
$\left( g_{ab}, \phi \right)$.

There is a second representation of the theory, the so-called Einstein 
conformal frame, {\em i.e.},  the pair of variables $\left( 
\tilde{g}_{ab}, \tilde{\phi} \right)$  where the new metric 
$\tilde{g}_{ab}$ is obtained  from 
$g_{ab}$ by means of the  conformal transformation  
\be
g_{ab}\rightarrow \tilde{g}_{ab}=\Omega^2 \, g_{ab} \,,  \;\;\;\;\;\;\ 
\Omega=\sqrt{\phi} \,,  \label{confotransf}
\ee
while the new scalar field comes from the non-linear 
redefinition\footnote{This non-linear redefinition turns the kinetic 
energy density of $\tilde{\phi}$ into its canonical form.}  
\be \label{46} 
\tilde{\phi}( \phi)= \sqrt{ \frac{2\omega+3}{16\pi G} } \,
\ln \left( \frac{\phi}{\phi_0} \right) 
\ee 
(where $\omega > -3/2$). Under this transformation, the 
Brans-Dicke action~(\ref{1}) assumes its Einstein frame form  
\cite{Dicke:1961gz} 
\begin{eqnarray} 
S& =& \int d^4 x \, \left\{ \sqrt{ -\tilde{g}} \left[
\frac{ \tilde{R}}{16\pi G} -\frac{1}{2} \, \tilde{g}^{ab}
\tilde{\nabla}_a\tilde{\phi} \tilde{\nabla}_b\tilde{\phi} -U\left(
\tilde{\phi} \right) \right] \right.
\nonumber \\ 
& & \left. + \mbox{e}^{ -8\sqrt{ \frac{\pi G}{2\omega +3} } \,\,
\tilde{\phi} } {\cal L}^\text{(m)} 
\left[ \tilde{g} \right] \right\} \,,
\label{47} 
\end{eqnarray} 
where $\tilde{\nabla}_a$ is the covariant derivative
operator of the rescaled metric $\tilde{g}_{ab}$ and 
\be \label{47bis}
U\left( \tilde{\phi} \right) = V\left[ \phi \left( \tilde{\phi} \right)
\right] \exp \left( -8 \sqrt{\frac{\pi G}{2\omega+3} } \, \tilde{\phi}
\right) \, .
\ee
In the Einstein conformal frame the scalar field $\tilde{\phi}$ couples 
directly to matter through the exponential factor multiplying the
Lagrangian density ${\cal L}^{(\text{m})}$ in the action~(\ref{47}). 
As a consequence, in this 
frame  the matter energy-momentum tensor is not covariantly 
conserved. Mathematically, this happens because the equation $\nabla^{b} 
\, 
T_{ab}^\text{(m)} =0 $ is not conformally invariant \cite{Wald}. The 
conformally transformed stress-energy tensor $\tilde{T}_{ab}^\text{(m)} 
$ obeys the corrected equation 
\be \label{43}
\tilde{\nabla}^{b} \, \tilde{T}_{ab}^\text{(m)} =-\,\frac{d}{d\phi}
\left[  \ln \Omega (\phi) \right] \, \tilde{T}^\text{(m)} \, \tilde{\nabla}_a
\phi \,.
\ee
Only conformally invariant matter with vanishing trace $T^\text{(m)} =0 $ 
remains covariantly conserved after the conformal transformation.

Consider the Einstein frame  stress-energy tensor derived from 
the usual expression \cite{Wald} 
\be \label{43bis}
\tilde{T}_{ab}^\text{(m)}=\frac{-2}{\sqrt{ -\tilde{g} } } \, \frac{  \delta
\left( \sqrt{-\tilde{g}} \, \, {\cal L}^\text{(m)} \right) }{\delta
\tilde{g}^{ab} } \,;
\ee
it is easy to see that 
\begin{eqnarray} 
\tilde{T}_{ab}^\text{(m)} &=& \Omega^{-2} \, T_{ab}^\text{(m)} \,, 
\label{eq:43tersay}\\
&&\nonumber\\
  \widetilde{  { {T_{a}}^{b}}^\text{(m)} } &=& \Omega^{-4} \, 
{{T_a}^b}^\text{(m)}  \,,\\
&&\nonumber\\
{\tilde{T}}^{ab} &=& \Omega^{-6} \, {T^{ab}}^\text{(m)} \,, \label{43ter}
\end{eqnarray}
and
\be \label{43quater}
\tilde{T}^\text{(m)}= \Omega^{-4} \, T^\text{(m)} \,.
\ee
In particular, a perfect fluid stress-energy tensor maps to 
 \be \label{4305}
\tilde{T}_{ab}^\text{(m)}=\left( 
\tilde{P}^\text{(m)}+\tilde{\rho}^\text{(m)} 
 \right) \tilde{u}_a
\tilde{u}_b +\tilde{P}^\text{(m)} \, \tilde{g}_{ab} 
\ee
under the conformal transformation, and $ \tilde{g}_{ab} \, \tilde{u}^a 
\tilde{u}^b=-1$ yields 
\be \label{4307}
\tilde{u}^a=\Omega^{-1} \, u^a \,, \;\;\;\;\;\;\;
\tilde{u}_a=\Omega \, u_a \,.
\ee
As a consequence, 
\begin{eqnarray} 
&& \left( \tilde{P}^\text{(m)} + \tilde{\rho}^\text{(m)} \right) 
\tilde{u}_a  \tilde{u}_b    + \tilde{P}^\text{(m)} \, 
\tilde{g}_{ab} \nonumber\\
&&\nonumber\\
&& = \Omega^{-2}
\left[ \left( P^\text{(m)} +\rho^\text{(m)} \right) u_a u_b 
  + P^\text{(m)} \, g_{ab} \right]    \,, \label{4308} 
\end{eqnarray}
as follows from Eqs.~(\ref{43ter}), (\ref{4305}), 
and~(\ref{4307}). Therefore, it must be 
\be \label{4309}
\tilde{\rho}^\text{(m)}=\Omega^{-4} \, \rho^\text{(m)} \,, \;\;\;\;\;\;\;
\tilde{P}^\text{(m)}=\Omega^{-4} \, P^\text{(m)} \,.
\ee
Due to Eq.~(\ref{4309}), a Jordan frame fluid satisfying the barotropic 
equation of state~(\ref{barotropic}) is mapped into an Einstein frame 
fluid with the same equation of state.

Specializing to FLRW universes,  the conformal transformation maps 
the Jordan frame  fluid conservation equation 
\be 
\frac{d \rho^\text{(m)}}{dt} +3H \left( P^\text{(m)} +\rho^\text{(m)} \right)=0
\ee
into 
\begin{eqnarray} 
\frac{d \tilde{\rho}^\text{(m)}}{dt}+3 \tilde{H} \left( \tilde{P}^\text{(m)}
+\tilde{\rho}^\text{(m)}
\right)&=&\frac{d \left( \ln \Omega \right)}{d\phi} \,\, \dot{\phi} \left( 
3\tilde{P}^\text{(m)} - \tilde{\rho}^\text{(m)} \right) \,.\nonumber\\
&&
\end{eqnarray}
Stepping back to general spacetimes, the stress-energy tensor 
$T_{ab}^\text{(m)}$ scales  under the conformal transformation  
according to \cite{Wald}
\be 
\tilde{T}^{ab}_\text{(m)}=\Omega^s \, \, T^{ab}_\text{(m)} \,, 
\;\;\;\;\;\;\;\;\;
\tilde{T}_{ab}^\text{(m)}=\Omega^{s+4} \,\, T_{ab}^\text{(m)} \,, 
\ee
where $s$ is an appropriate conformal weight. In four spacetime 
dimensions, the Jordan frame covariant  conservation equation $\nabla^b \,  
T_{ab}^\text{(m)}=0$ corresponds to \cite{Wald,Faraoni:2004pi}
\begin{eqnarray} 
\tilde{\nabla}_a \left( \Omega^s \, T^{ab}_\text{(m)} \right) &=&\Omega^s
\, \nabla_a T^{ab}_\text{(m)} +\left( s+6 \right) \Omega^{s-1} \, 
T^{ab}_\text{(m)}\nabla_a  \Omega  \nonumber\\ 
&&\nonumber\\
&\, & - \Omega^{s-1} g^{ab} \, T^\text{(m)} \nabla_a \Omega \,. \label{x1}
\end{eqnarray}
The choice of conformal weight $s=-6$ is consistent with 
Eq.~(\ref{eq:43tersay}) and gives 
\be \label{x3}
\tilde{T}^\text{(m)} \equiv    \tilde{g}^{ab} \, 
\tilde{T}_{ab}^\text{(m)}= \Omega^{-4}  \,\, T^\text{(m)} \,,
\ee
while Eq.~(\ref{x1}) has the conformal cousin 
\be \label{x3bis}
\tilde{\nabla}_a  \tilde{T}^{ab}_\text{(m)}  =- \tilde{T}^\text{(m)}\,
\tilde{g}^{ab} \, \tilde{\nabla}_a
\left(  \ln \Omega \right)  
\ee
in the tilded world. For Brans-Dicke theory, where $\Omega=\sqrt{\phi}$,  
one has \cite{Brans:1961sx, Wagoner:1970vr, FujiiMaeda, Faraoni:2004pi} 
\be \label{x4}
\tilde{\nabla}_a  \tilde{T}^{ab}_\text{(m)}  =- \frac{1}{2\phi} \,\,
\tilde{T}^\text{(m)}
 \,  \tilde{\nabla}^b \phi  = - \sqrt{ \frac{4\pi G}{2\omega
+3} } \, \, \tilde{T}^\text{(m)} \,\, \tilde{\nabla}^b \tilde{\phi} \,.
\ee
Consider now a  dust fluid describing dark and baryonic matter in a FLRW 
universe. Setting $P^\text{(m)} =0$ yields
\begin{eqnarray} 
&& \tilde{u}_a \, \tilde{u}_b \, \tilde{\nabla}^b  \tilde{\rho}^\text{(m)} 
+\tilde{\rho}^\text{(m)} \, \tilde{u}_a \, \tilde{\nabla}^b  \tilde{u}_b
+\tilde{\rho}^\text{(m)} \, \tilde{u}_c \,\tilde{\nabla}^c \, \tilde{u}_a 
\nonumber\\
&&\nonumber\\
&\, &  - \sqrt{ \frac{4\pi G}{2\omega +3} } \, \,
\tilde{\rho}^\text{(m)} \, \tilde{\nabla}_a \tilde{\phi} =0 \,, \label{x7}
\end{eqnarray}
which shows explicitly how the scalar field interacts with ordinary matter 
through its gradient. To conclude, this picture gives a completely 
covariant and Lagrangian formulation of a scalar field interacting with a 
matter fluid. 
In the Jordan frame this scalar has gravitational nature but this 
information does not matter  in the 
Einstein frame, in which $\tilde{\phi}$ could be regarded as standard 
scalar field, except for its anomalous coupling to (non-conformal) matter.

In principle, when performing the conformal transformation one should also 
rescale all fundamental and derived units, according to Dicke's 
prescription \cite{Dicke:1961gz}, with the result that the two conformal 
frames would then be physically equivalent \cite{Dicke:1961gz}. This 
physical equivalence has been the subject of a debate filling a fairly 
large literature. This equivalence issue is immaterial here since there is 
no doubt that the transformations~(\ref{confotransf}) and (\ref{46}) 
constitute a well-defined map between the two worlds and here we are 
interested in mathematical solutions for a scalar field coupled to a fluid 
(or for two coupled effective fluids). Therefore, Jordan frame 
cosmological solutions in the Jordan frame (where $G_\text{eff} \simeq 
\phi^{-1}$ varies)  translate into mathematical solutions in the Einstein 
frame, where formally the theory can be interpreted as general relativity 
with the anomalous coupling of an (otherwise canonical) scalar field 
$\tilde{\phi}$ to the cosmological fluid(s).

One could think of using a tensor-multi-scalar theory to obtain 
multi-fluids in the Einstein conformal frame. However, in general, one 
cannot recast the transformed fields so that more than one has canonical 
kinetic energy \cite{Kaiser:2010ps}. The chiral cosmological 
models with non-canonical kinetic energies thus obtained still have 
analytical solutions. A methods to search for them has been proposed in 
\cite{Chervon:2019nwq}. This is a generalization of a previous method designed  
to obtain solutions of quintom models with potential in 
\cite{Arefeva:2005mka,Vernov:2006dm}.

There are other methods to find analytical cosmological solutions with 
nonminimally coupled models scalar fields \cite{Kamenshchik:2012pw} and to 
connect integrable cosmologies in the Jordan and the Einstein frames 
\cite{Kamenshchik:2013dga, Kamenshchik:2015cla}.

\subsection{Two interacting scalar fields}

There are cosmological scenarios in which both interacting fluids are 
actually effective fluids coming from scalar fields $\phi$ and $\psi$. In 
the previous picture, if 
the first fluid is the $\psi$-fluid with self-interaction potential $ 
U(\psi)$, then 
\begin{eqnarray}
\rho_1 &=& \frac{\dot{\psi}^2}{2}+U(\psi) \,,\\
&&\nonumber\\
P_1 &=& \frac{\dot{\psi}^2}{2}-U(\psi) \,,
\end{eqnarray}
and the equation of motion for  $\psi$ becomes
\be 
\ddot{\psi}+3H\dot{\psi} - \Gamma \,\frac{ \dot{\phi}^2}{\dot{\psi}}  
+\frac{dU}{d\psi} = 0  
\ee 
(assuming $\dot{\psi}\neq 0$). When $|\dot{\psi}|$ is large and 
increasing and $\Gamma>0$, the extra term $- 
\Gamma \, \dot{\phi}^2/ \dot{\psi} $ enhances the motion of $\psi$ 
and plays the role of antifriction for $\psi$. If $\psi$ decreases, this 
term describes an effective friction opposing the motion of $\psi$.  The 
quantity $\alpha$ given by (\ref{alphaff}) can still be introduced.

More generally, one can couple the two scalars by introducing a common 
potential $ V \left(\phi, \psi\right)$ that modifies their equations of 
motion to
\begin{eqnarray}
&& \ddot{\phi}+3H\dot{\phi} + \frac{\partial V}{\partial \phi} =0 \,, \\
&&\nonumber\\
&& \ddot{\psi}+3H\dot{\psi} + \frac{\partial V}{\partial \psi} = 0\,. 
\end{eqnarray}
This is done in preheating scenarios designed to end inflation and raise 
the temperature of the universe \cite{Allahverdi:2010xz, Amin:2014eta}.  
One of the two 
fields is the inflaton and 
the other denotes any field to which the inflaton couples (bosonic fields 
are generated in the early stages and fermionic fields near the end of 
preheating). The 
interaction potential commonly used in preheating has the form 
\be
V(\phi,\psi) = \frac{m_{\phi}^2}{2} \, \phi^2 + \frac{m_{\psi}^2}{2} \, 
\psi^2  + \frac{g}{2}\,  \phi^2 \psi^2 \,.
\ee

Quintom models of dark energy (see Ref.~\cite{Cai:2009zp} for a review) 
are two-component dark energy models in which a phantom scalar field 
interacts with a second scalar field. Sometimes, for simplicity, the two 
scalars are modelled with two fluids \cite{Zhang:2005kj}.

\section{Concluding remarks}
\setcounter{equation}{0}
\label{sec:10}

The scientist's view of multiple fluid FLRW cosmology has changed 
significantly since the discovery of the first mathematical solutions 
describing this physics in the 1960s and 1970s. Early interest was in 
radiation and dust filling a (possibly spatially curved) FLRW universe and 
was motivated by the recent discovery of the cosmic microwave background 
radiation by Penzias and Wilson \cite{Penzias:1965wn}, and expanded to 
consider any two fluids that could give exact solutions in terms of 
elementary functions.  The search for FLRW solutions quickly expanded to 
include effective fluids describing spatial curvature and the cosmological 
constant.

The modern view is quite different: the standard $\Lambda$CDM model of the 
universe includes dark energy and dark matter fluids in a spatially flat 
FLRW universe. Attempts to explain the current acceleration of the 
universe without the {\em ad hoc} dark energy by modifing gravity at large 
scales instead \cite{Sotiriou:2008rp, DeFelice:2010aj, Nojiri:2010wj} end 
up being not so different from dark energy scenarios because in a FLRW 
universe the deviations of gravity from general relativity in the field 
equations can be grouped in the right hand side, producing an effective 
perfect fluid in a wide spectrum of theories of gravity 
\cite{Gurses:2020kpv}.

Another ingredient that was hardly present in cosmology in the 1960s is 
the scalar field: since 1980 we have witnessed an enormous interest in 
inflationary scenarios of the early universe. Inflation solves the three 
problems of standard Big Bang cosmology, {\em i.e.}, the horizon, 
flatness, and monopole problems and provides a mechanism to generate 
density fluctuations which constitute the seeds for structure formation in 
the form of quantum fluctuations of the inflaton. In most inflationary 
scenarios, the inflaton driving the rapid expansion of the universe is a 
scalar field. This is true also for $R^2$ inflation 
\cite{Starobinsky:1980te, StarobinskyI} which, although 
correcting Einstein theory, can be seen as the result of the dynamics of 
the effective extra scalar degree of freedom. It is now well known that a 
scalar field $\phi$ with timelike gradient $\nabla_a\phi$ is equivalent to 
a perfect fluid which, in general, has a dynamical equation of state. It 
was realized rather early that a free scalar is equivalent to a stiff 
fluid, which appears in several integrability cases of the 
Einstein-Friedmann equations, but interesting physical scenarios for the 
scalar field were still missing.

Having learned the lesson from inflation, the scalar field was naturally 
the main ingredient for modelling dark energy after the 1998 discovery of 
the cosmic acceleration with type Ia supernovae \cite{Perlmutter:1997zf, 
Riess:1998cb, Perlmutter:1998np}. In all these situations dark and 
baryonic matter form one fluid and dark energy, or its scalar field 
version, another fluid. It is possible, and it would be even convenient 
for modellers if these dark fluids were interacting directly, which has 
been the subject of a considerable literature. In the meantime, 
scalar-tensor gravity originally proposed by Brans \& Dicke 
\cite{Brans:1961sx} and later generalized \cite{Bergmann:1968ve, 
Wagoner:1970vr, Nordtvedt:1970uv}, has been the subject of much interest 
and many other theories of gravity have been studied 
(see the reviews \cite{Willbook, Will:2014kxa, CapozzielloFaraoni, 
Clifton:2011jh, HarkoLobobook, Heisenberg:2018vsk}). After 2004, $f(R)$ 
gravity (a 
subclass of scalar-tensor gravity) was 
studied intensely to explain the present acceleration of the universe 
\cite{Sotiriou:2008rp, DeFelice:2010aj, Nojiri:2010wj}. Nowadays, testing 
gravity constitutes added value for astrophysical and cosmological 
observations and satellite missions at vastly different scales, as well as 
theoretical research programs. In the last decade or so, much interest has 
been devoted to ``second generation'' scalar-tensor theories of gravity 
(see  \cite{Kobayashi:2019hrl, Langlois:2018dxi} for reviews) known as 
Horndeski \cite{Horndeski:1974wa, Nicolis:2008in, 
Deffayet:2009wt, Kobayashi:2011nu, Deffayet:2011gz} and 
beyond-Horndeski and Degenerate 
Higher Order Scalar-Tensor (DHOST) theories \cite{Gleyzes:2014dya, 
Gleyzes:2014qga, Langlois:2015cwa, Langlois:2015skt, 
Achour:2016rkg, Crisostomi:2016czh, Motohashi:2016ftl, 
BenAchour:2016fzp}. They explore the vast 
lanscape of the most general scalar-field based theories of gravity that 
have second order equations and are compatible with theoretical and 
observational constraints \cite{Bettoni:2016mij}. The motivation for 
studying these theories is primarily cosmological and, eventually, 
multi-fluid FLRW universes enter the description of these cosmologies.

While trying to connect the old with the new literature, we have reported 
analytical multi-(effective) fluid solutions expressed by an integral in 
finite form containing only elementary functions. Either the scale factor 
is a function $a(t)$ of the comoving time $t$, or the latter is a function 
$t(a)$ (this relation can sometimes be inverted to obtain $a(t)$ 
explicitly). Alternatively, the solution is expressed in parametric form 
with the conformal time $\eta$ as the parameter.

A key element of the discussion is the treatment of the spatial curvature 
of FLRW universes and of the cosmological constant $\Lambda$ as effective 
fluids, reducing formally the Einstein-Friedmann equations to those for an 
effective $K=0, \Lambda=0$, multi-fluid situation. While this procedure is 
followed routinely for the cosmological constant and for the effective 
energy density of spatial curvature, it is not normally implemented for 
the effective pressure of the latter.

Following the key realization of Refs.~\cite{Jacobs1968,McIntosh1972, 
McIntoshFoyster1972, Chen:2014fqa} that the integral expressing the scale 
factor has an hypergeometric series representation which truncates for 
special values of its arguments (or, alternatively, using the Chebysev 
theorem of integration), the two-fluid situations in which $t(a)$ or 
$a(t)$ can be obtained analytically in terms of elementary functions can 
be determined. One can then compile a catalogue of two-fluid solutions 
expressed in terms of elementary functions and select those of physical 
interest.

To keep this review manageable, we restricted its scope by excluding from 
it certain extensions, and related topics that are of interest in modern 
cosmology. First, we did not discuss tilted fluids 
\cite{SmootGorensteinMuller77, ColeyTupper86, ColeyTupper86b, ColeyASS89, 
ColeyWainwright92, Nesteruk94, Nesteruk:1995uu, Coley:1995dpj, 
WainwrightEllisBook, Verma07} and imperfect fluids, which are interesting 
possibilities to consider. FLRW cosmologies do not necessarily represent 
perfect fluid solutions but can also describe a viscous fluid in a 
non-comoving frame.  Imperfect fluids are characterized by spacelike 
energy fluxes and anisotropic stresses \cite{Eckart:1940te, Stephani, Wald}, 
which are excluded by spatial isotropy in a FLRW universe, but bulk 
viscosity is allowed and has been studied extensively in the literature. 
Bulk viscosity could be due, for example, to particle production in the 
early universe. Likewise, we did not discuss multi-fluid or scalar field 
anisotropic Bianchi universes and cosmic no-hair 
theorems \cite{Goldwirth:1991rj}. 
Finally, multi-fluid FLRW solutions in alternative theories of gravity 
have been omitted, except for the Einstein frame description of 
scalar-tensor gravity, which is formally equivalent to a cosmological 
scalar field always present in the theory that couples explicitly to the 
matter sector and could explain interacting dark energy in a fully 
covariant and Lagrangian way. The literature on modified gravity and 
cosmology is, 
however, huge and a comprehensive review of multi-fluid or multi-effective 
fluid cosmology in all of these theories would be a book-size undertaking 
which goes well beyond the purposes of this work.

We did not mention Chaplygin gas cosmology, in which a single gas 
interpolates between different fluids during the evolution of the universe 
\cite{Kamenshchik:2001cp}. This is an interesting alternative and the 
relations between its FLRW solutions and those of multi-fluids are not 
completely explored. At the same time, we restricted to exact FLRW 
universes, ignoring their perturbations. However, the latter are the main 
source of information in modern cosmology through the imprints that they 
left in the cosmic microwave background and through large-scale structure 
surveys. Perturbations of multi-fluid FLRW universes constitute an 
important part of cosmology.

Even restricting the scope of this work, however, left many interesting 
multi-(effective) fluids to contemplate. This richness is due to the 
variety of real or effective matter sources, including real fluids with 
barotropic equations of state, scalar fields with a large variety of 
physically interesting potentials, the gravitational scalar field of 
scalar-tensor gravity in the Einstein frame, the cosmological constant 
$\Lambda$, and spatial curvature. The knowledge of analytical solutions 
must 
be supplemented by phase space analysis to determine which ones are 
attractors in phase space and, therefore, physically important. This type 
of study is fundamental in theories of inflation and dark energy. 
Nevertheless, the proliferation of inflationary and dark energy models, 
and 
the associated phase space studies has sometimes obscured the full 
covariance of a theory or its Lagrangian formulation, as we have pointed 
out in the context of interacting dark energy.

Finally, it is rather unfortunate that in recent years old 
solutions, integrability conditions, and methods have been forgotten and 
needed to be 
rediscovered, which can be partially explained by the very different 
points of view on this subject in the last half century. We set out with 
the goal of providing a starting point for the less experienced reader and 
a reference for the expert one, although a review cannot replace the 
thorough study of the relevant literature. We hope that highlighting 
multi-fluid cosmology will bring i)~a better understanding of the 
relations between the various possible components of the universe; 
ii)~more 
efficient research in the context of Einstein theory; iii)~will facilitate 
the comparison with analytical solutions describing similar universes in 
alternative theories of gravity.

\section*{Acknowledgments} 


This work is supported by the Natural Sciences \& Engineering Research 
Council of Canada (grant no. 2016-03803 to V.F.) and by a Bishop's 
University Graduate Entrance Scholarship to S.J.

\bigskip


\appendix
\section{Derivation of Eq.~(\ref{wdot}) }
\label{AppendixA}
\renewcommand{\theequation}{A.\arabic{equation}}
\setcounter{equation}{0}

Following Ref.~\cite{McIntosh1972}, we have 
\be
w_\text{tot}(a)= \frac{ \sum_{i=1}^n w_i \rho_i^{(0)} a^{-3(w_i+1)} 
}{\sum_{j=1}^n 
\rho_j a^{-3(w_j+1)}} 
\ee
and
\begin{eqnarray}
\dot{w}_\text{tot}\left( a, H\right) &=& -\frac{3}{\rho^2} \left\{
\sum_i w_i \rho_i^{(0)} (w_i+1) a^{-3w_i-4} \dot{a} \left[ \sum_j 
\rho_i^{(0)} a^{-3(w_j+1)} \right] \right.\nonumber\\
&&\nonumber\\
&\, & \left.  - 
\sum_i w_i \rho_i^{(0)} a^{-3(w_i+1)} \sum_j \rho_j^{(0)} (w_j+1) 
a^{-3w_j-4} \dot{a} \right\} \nonumber\\
&&\nonumber\\
&=& -\frac{3}{\rho^2} \left\{
\sum_{i,j} w_i\left(w_i-w_j\right) \rho_i^{(0)} \rho_j^{(0)} 
a^{-3(w_i+w_j)-6} \right\}\frac{\dot{a}}{a} \,. 
\label{stocazzointermediate}
\end{eqnarray}
Remembering that $ \rho_i = \rho_i^{(0)} a^{-3(w_i+1)}$, we have 
\begin{eqnarray}
&& \sum_{i,j} w_i\left(w_i-w_j\right) \rho_i \rho_j = 
\sum_{i,j} \left( w_i -w_j +w_j \right) \left(w_i-w_j\right) \rho_i \rho_j  
\nonumber\\
&&\nonumber\\
&& = \sum_{i,j} \left( w_i -w_j \right)^2 \rho_i \rho_j  +
\sum_{i,j}  w_j\left(w_i-w_j\right) \rho_i \rho_j \,,
\end{eqnarray}
therefore,
\begin{eqnarray}
&& \sum_{i,j} w_i \left(w_i-w_j \right) \rho_i \rho_j  
- \sum_{i,j} w_j \left(w_i-w_j \right) \rho_i \rho_j  \nonumber\\
&&\nonumber\\
&& =\sum_{i,j}\left(  
w_i-w_j\right)^2  \rho_i \rho_j  \,,
\end{eqnarray}
or
\be
\sum_{i,j} w_i \left(w_i-w_j \right) \rho_i \rho_j  = \frac{1}{2} \, 
 \sum_{i,j}\left(w_i-w_j\right)^2 \rho_i \rho_j  \,.
\ee
Equation~(\ref{stocazzointermediate}) then becomes
\begin{eqnarray}
\dot{w}_\text{tot} &=& -\frac{3H}{2\rho_\text{tot}^2}  \sum_{i,j} 
\left(w_i-w_j\right)^2 \rho_i^{(0)} \rho_j^{(0)} a^{-3(w_i+w_j) 
-6}\nonumber\\
&&\nonumber\\
&=& -\frac{3H}{\rho_\text{tot}^2}   
\sum_{i<j} \left(w_i-w_j\right)^2 \rho_i \rho_j \,, 
\end{eqnarray}
where we used the fact that 
\be
\frac{1}{2} \, \sum_{i,j=1}^n = \sum_{i<j}^n \,.
\ee

\section{Equivalence of Eqs.~(\ref{-1dust+radiation})  and 
(\ref{CohenDustRadCurvLog})}
\label{AppendixB}
\renewcommand{\theequation}{B.\arabic{equation}}
\setcounter{equation}{0}

Set 
\be
x \equiv \frac{a+ \frac{4\pi}{3}\,  \rho_\text{m}^{(0)} }{
\sqrt{ \frac{4\pi}{3}\, \rho_\text{m}^{(0)} \left(- \frac{4\pi}{3}\, 
\rho_\text{m}^{(0)}  
+\frac{2\rho_\text{m}^{(0)} }{\rho_\text{r}^{(0)} } \right) }}
\ee
and use the inverse hyperbolic function identity

\begin{equation}
  \sinh^{-1}x=\ln \left( x+\sqrt{x^2+1} \right)  
\end{equation}
to write 

\begin{eqnarray}
\sinh^{-1} x &=& \ln \left( 
\frac{a+ \frac{4\pi}{3}\,  \rho_\text{m}^{(0)} }{
\sqrt{ \frac{4\pi}{3}\, \rho_\text{m}^{(0)}  \left( -\frac{4\pi}{3}\, 
\rho_\text{m}^{(0)}  
+\frac{2\rho_\text{m}^{(0)} }{\rho_\text{r}^{(0)} } \right) } } \right. 
\nonumber\\
&&\nonumber\\ 
&\, & \left. +
\sqrt{ 
\frac{ \left( a+ \frac{4\pi}{3}\,  \rho_\text{m}^{(0)} \right)^2 }{
\frac{4\pi}{3}\,  \rho_\text{m}^{(0)} \left(-\frac{4\pi}{3}\, 
\rho_\text{m}^{(0)}  
+\frac{2\rho_\text{m}^{(0)} }{\rho_\text{r}^{(0)}  } \right) } +1 } \, 
\right) 
\nonumber\\ 
& \,&  \nonumber\\  
&=& \ln \left[ \frac{  \sqrt{a^2 + \frac{8\pi}{3} \, \rho_\text{m}^{(0)} a 
+ 
\frac{8\pi}{3} \, \rho_\text{r}^{(0)} } +a + \frac{4\pi}{3} \, 
\rho_\text{m}^{(0)} 
}{  \sqrt{ \frac{4\pi}{3}\, \rho_\text{m}^{(0)}  \left( 
-\frac{4\pi}{3}\, \rho_\text{m}^{(0)}  
+\frac{2\rho_\text{m}^{(0)} }{\rho_\text{r}^{(0)}} \right) }  }\right]  
\nonumber\\
&&\nonumber\\  
&=& \ln \left\{ C \left[ \sqrt{a^2 + \frac{8\pi}{3} \, 
\rho_\text{m}^{(0)} a 
+  \frac{8\pi}{3} \, \rho_\text{r}^{(0)} } +a \right. \right.\nonumber\\
&&\nonumber\\
&\, & \left.\left. + \frac{4\pi}{3} \, 
\rho_\text{m}^{(0)}  \right]  \right\}  
\end{eqnarray} 
where 
\be
C \equiv \frac{1}{ 
\sqrt{ \frac{4\pi}{3}\, \rho_\text{m}^{(0)}  \left( 
-\frac{4\pi}{3}\, \rho_\text{m}^{(0)}  
+\frac{2\rho_\text{m}^{(0)} }{\rho_\text{r}^{(0)}} \right) } }  \,.
\ee
Therefore, we have 
\begin{eqnarray}
t(a) &=& \sqrt{a^2 + \frac{8\pi}{3} \, \rho_\text{m}^{(0)} a + 
\frac{8\pi}{3} \, \rho_\text{r}^{(0)} } \nonumber\\
&&\nonumber\\
&\, & -\frac{4\pi}{3} \, \rho_\text{m}^{(0)} \sinh^{-1} \left[
\frac{a+ \frac{4\pi}{3}\,  \rho_\text{m}^{(0)} }{
\sqrt{ \frac{4\pi \rho_\text{m}^{(0)} }{3} \left( -\frac{4\pi 
\rho_\text{m}^{(0)} 
}{3} 
+\frac{2\rho_\text{m}^{(0)} }{\rho_\text{r}^{(0)} } \right) }} \right] 
\nonumber\\
&&\nonumber\\
&=&  \sqrt{a^2 + \frac{8\pi}{3} \, \rho_\text{m}^{(0)} a + 
\frac{8\pi}{3} \, \rho_\text{r}^{(0)} } \nonumber\\
&&\nonumber\\
&\, & - \frac{4\pi}{3} \, \rho_\text{m}^{(0)} \ln \left\{ C \left[
\sqrt{a^2 + \frac{8\pi}{3} \, \rho_\text{m}^{(0)} a + 
\frac{8\pi}{3} \, \rho_\text{r}^{(0)} } +a  \right. \right.\nonumber\\
&&\nonumber\\
&\, & \left.\left. + \frac{4\pi}{3}\, \rho_\text{m}^{(0)} \right] \right\} 
\end{eqnarray}
or, Eq.~(\ref{-1dust+radiation}) coincides with 
Eq.~(\ref{CohenDustRadCurvLog}).

\section{Single-fluid limit of the universe with dust plus radiation and 
$K=\pm1$}
\label{AppendixC}
\renewcommand{\theequation}{C.\arabic{equation}}
\setcounter{equation}{0}

Here we discuss the single-fluid limits of the 
solutions~(\ref{DustRadPositive}) and (\ref{CohenDustRadCurvLog}) of the 
Einstein-Friedmann equations~(\ref{Friedmann1})-(\ref{conservation1})  
describing spatially curved universes sourced by dust plus radiation. As 
expected, these limits generate the well known solutions for radiation 
with curvature, however the limits to dust plus curvature are not trivial 
in both cases $K =\pm 1$.

\subsection{Positive curvature}

In the positively curved case $K=+1$, the relation between comoving 
time and scale factor is given by Eq.~(\ref{DustRadPositive}),  which we 
reproduce here for convenience:
\begin{eqnarray}
t(a) &=& t_0 - \sqrt{ \frac{8\pi}{3}\,  \rho_\text{r}^{(0)} 
+\frac{8\pi}{3}\, 
\rho_\text{m}^{(0)} a -a^2 } \nonumber\\
&&\nonumber\\
& \, & 
+\frac{4\pi}{3} \, \rho_\text{m}^{(0)} \sin^{-1} \left[ 
\frac{ -\frac{4\pi}{3} \, \rho_\text{m}^{(0)} +a }{ \sqrt{
\left( \frac{4\pi}{3} \, \rho_\text{m}^{(0)} \right)^2 +\frac{8\pi}{3} \, 
\rho_\text{r}^{(0)} } } \right] \,. \nonumber\\
&& \label{forconvenience1}
\end{eqnarray}
\noindent {\em Limit to dust and $K=+1$:} Setting $\rho_\text{r}^{(0)} = 
0$, one 
obtains the limit 
\begin{eqnarray}
t(a) &=&- \sqrt{ \frac{8\pi}{3} \, \rho_\text{m}^{(0)} a-a^2} 
+\frac{4\pi}{3} \, \rho_\text{m}^{(0)} \sin^{-1} \left(  
\frac{3a}{4\pi \rho_\text{m}^{(0)} } -1  \right) \nonumber\\
&&\nonumber\\
&\, & +t_0 \,,
\end{eqnarray}
that looks different from Eq.~(\ref{DustCurvPos}), which is instead
\begin{eqnarray}
t(a)&=&- \sqrt{ \frac{8\pi}{3} \, \rho_\text{m}^{(0)} a-a^2} 
+\frac{8\pi}{3} \, \rho_\text{m}^{(0)} \sin^{-1} \left( \sqrt{ 
\frac{3a}{8\pi \rho_\text{m}^{(0)} } } \, \right) \nonumber\\
&&\nonumber\\
&\, & +t_0 \,,
\end{eqnarray}
To proceed, use the trigonometric identities

\begin{eqnarray}
&& \sin^{-1}x = \frac{1}{2} \, \cos^{-1} \left( 1-2x^2\right) \,,\\
&&\nonumber\\
&& \cos^{-1} x = \frac{\pi}{2}-\sin^{-1} x \,,\\
&&\nonumber\\
&& \sin^{-1} (-x) = - \sin^{-1}x \,,
\end{eqnarray}
for $x\equiv \sqrt{ \frac{3a}{8\pi \rho_\text{r}^{(0)} } }$ to obtain

\begin{eqnarray}
&& \frac{8\pi}{3} \, \rho_\text{m}^{(0)} \sin^{-1} \left( \sqrt{ 
\frac{3a}{8\pi 
\rho_\text{r}^{(0)} } } \right) = \frac{4\pi}{3}\, \rho_\text{m}^{(0)} 
\cos^{-1} 
\left( 1-\frac{3a}{4\pi \rho_\text{r}^{(0)} }\right) \nonumber\\
&&\nonumber\\
&&= \frac{4\pi}{3}\, \rho_\text{m}^{(0)} \left[ \frac{\pi}{2}- \sin^{-1} 
\left( 
1-\frac{3a}{4\pi \rho_\text{r}^{(0)} } \right) \right] \nonumber\\
&&\nonumber\\
&&= \frac{2\pi^2}{3} \, \rho_\text{m}^{(0)} + \frac{4\pi}{3}\, 
\rho_\text{m}^{(0)} 
\sin^{-1} \left( \frac{3a}{4\pi \rho_\text{r}^{(0)} } -1\right) \,,
\end{eqnarray}
which yields

\begin{eqnarray}
t(a)&=& - \sqrt{ \frac{8\pi}{3} \, \rho_\text{m}^{(0)} a-a^2} 
+\frac{4\pi}{3} \, \rho_\text{m}^{(0)} \sin^{-1} \left( 
\frac{3a}{4\pi \rho_\text{r}^{(0)} } -1 \, \right) \nonumber\\
&&\nonumber\\
&\, & +t_0' 
\end{eqnarray}
(where $t_0'=t_0 +\frac{2\pi^2}{3} \, \rho_\text{m}^{(0)} $): this   
coincides with the well known equation~(\ref{DustCurvPos}).

\noindent {\em Limit to radiation and $K=+1$:} By setting 
$\rho_\text{m}^{(0)}=0$, Eq.~(\ref{forconvenience1}) gives
\be
t=t_0-\sqrt{ \frac{8\pi}{3} \, \rho_\text{r}^{(0)} -a^2} \,, 
\ee
which is inverted to 
\be
a(t)= \sqrt{ \frac{8\pi}{3}\, \rho_\text{r}^{(0)} -\left(t-t_0\right)^2} 
\,;
\ee
this equation coincides coincides with Eq.~(\ref{RadCurvPos}).

\subsection{Negative curvature}

In the negatively curved universe $K=-1$, the relation between comoving 
time and scale factor is (Eq.~(\ref{CohenDustRadCurvLog}))
\begin{eqnarray}
&&t(a) = t_0 + \sqrt{
\frac{8\pi}{3} \, \rho_\text{r}^{(0)} +\frac{8\pi}{3} \, 
\rho_\text{m}^{(0)} a 
+a^2 } 
\nonumber\\
&&\nonumber\\
& & -\frac{4\pi}{3} \, \rho_\text{m}^{(0)} \ln\left\{ C \left[ \sqrt{
\frac{8\pi}{3} \, \rho_\text{r}^{(0)} +\frac{8\pi}{3} \, 
\rho_\text{m}^{(0)} a
+a^2 } +a \right.\right.\nonumber\\
&&\nonumber\\
&\, & \left. \left. + \frac{4\pi}{3}\rho_\text{m}^{(0)}  
\right]\right\}  \,.  \label{forconvenience2}
\end{eqnarray}

\noindent {\em Limit to dust and $K=-1$:} Setting 
$\rho_\text{r}^{(0)}=0$  in Eq.~(\ref{forconvenience2}), one obtains
\begin{eqnarray}
&&t(a) = t_0 + \sqrt{ \frac{8\pi}{3} \, \rho_\text{m}^{(0)} a +a^2 } 
\nonumber\\
&&\nonumber\\
& & -\frac{4\pi}{3} \, \rho_\text{m}^{(0)} \ln\left\{ C \left[ \sqrt{
\frac{8\pi}{3} \, \rho_\text{m}^{(0)} a+a^2 } 
+a + \frac{4\pi}{3}\rho_\text{m}^{(0)}  \right]\right\} \,,\nonumber\\
&&
\end{eqnarray}
which does not coincide with Eq.~(\ref{DustCurvNeg}); the latter gives 
instead
\begin{eqnarray}
t&=&t_0 +\sqrt{ \frac{8\pi}{3} \, \rho_\text{m}^{(0)} a+a^2 } 
-\frac{8\pi}{3} \, 
\rho_\text{m}^{(0)} \sinh^{-1} \left( \sqrt{\frac{3a}{8\pi 
\rho_\text{m}^{(0)}} }  
\, \right) 
\,.\nonumber\\
&&
\end{eqnarray}
However, using the inverse hyperbolic function identities

\begin{eqnarray}
&& 2\sinh^{-1} x = \cosh^{-1} \left(2x^1+1\right) \,,\\
&&\nonumber\\
&& \cosh^{-1} x = \ln \left( x+\sqrt{x^2-1}\right) \,,
\end{eqnarray}
one has
\begin{eqnarray}
\frac{8\pi}{3} \, \rho_\text{m}^{(0)} \sinh^{-1} \left( 
\sqrt{\frac{3a}{8\pi 
\rho_\text{m}^{(0)} } }\right) &=& \frac{4\pi}{3}\, \rho_\text{m}^{(0)} 
\cosh^{-1} 
\left( 
\frac{3a}{4\pi \rho_\text{m}^{(0)}}+1\right) \nonumber\\
&&\nonumber\\  
&=& \frac{4\pi}{3}\, \rho_\text{m}^{(0)} \ln \left( \frac{3a}{4\pi 
\rho_\text{m}^{(0)} }
+1 +\sqrt{ \left( \frac{3a}{4\pi\rho_\text{m}^{(0)} } +1 \right)^2 -1} 
\right) 
\nonumber\\
&&\nonumber\\ 
&=& \frac{4\pi}{3}\, \rho_\text{m}^{(0)} \ln \left[   \frac{3}{4\pi 
\rho_\text{m}^{(0)} }  \left( a+ \frac{4\pi}{3}\, \rho_\text{m}^{(0)} 
+\sqrt{ a^2 
+ 
\frac{8\pi}{3}\, \rho_\text{m}^{(0)}  a } \right) \right] \nonumber\\
&&\nonumber\\ 
&=& \frac{4\pi}{3}\, \rho_\text{m}^{(0)} \ln \left(  a+ \frac{4\pi}{3}\, 
\rho_\text{m}^{(0)} 
+\sqrt{ a^2 + \frac{8\pi}{3}\, \rho_\text{m}^{(0)} a}\, \right) 
\nonumber\\
&&\nonumber\\
&\, & + \frac{4\pi}{3}\, \rho_\text{m}^{(0)} \ln \left( 
\frac{3}{4\pi\rho_\text{m}^{(0)}} 
\right)  \,.\label{eq:lastfornow}
\end{eqnarray}
The last and constant  term is incorporated in the constant $t_0$ and 
Eq.~(\ref{eq:lastfornow}) then coincides  with Eq.~(\ref{DustCurvNeg}).

\noindent {\em Limit to radiation and $ K=-1$:} Setting 
$\rho_\text{m}^{(0)} = 0$, one obtains immediately  the radiation-only 
limit  
\be
t(a)= t_0+ \sqrt{ \frac{8\pi}{3} \, \rho_\text{r}^{(0)} + a^2 } 
\ee
that is inverted to give 
\be
a(t)=\sqrt{ \left(t-t_0\right)^2 -\frac{8\pi}{3}\, \rho_\text{r}^{(0)} } 
\,,
\ee
which coincides with Eq.~(\ref{RadCurvNeg}).


\begin{thebibliography}{99}

\bibitem{Abbott:1984fp}
L.F.~Abbott and M.B.~Wise,
Nucl. Phys. B \textbf{244}, 541-548 (1984)
doi:10.1016/0550-3213(84)90329-8

\bibitem{Abreu} J.P. Abreu, P. Crawford, and J.P. Mimoso, 
Class. Quant. Grav. {\bf 11}, 1919-1940 (1994).   
http://dx.doi.org/10.1088/0264-9381/11/8/002, arXiv:gr-qc/9401024.

\bibitem{Planck} P.A.R. Ade {\em et al.} (Planck collaboration),
Astron. Astrophys. {\bf 571}, A1 (2014).

\bibitem{Planck:2013pxb}
P.A.R.~Ade \textit{et al.} [Planck],
Astron. Astrophys. \textbf{571}, A16 (2014)
doi:10.1051/0004-6361/201321591
[arXiv:1303.5076 [astro-ph.CO]].

\bibitem{Ade:2015lrj}
P.A.R.~Ade \textit{et al.} [Planck Collaboration],
Astron. Astrophys. \textbf{594}, A20 (2016)
doi:10.1051/0004-6361/201525898
[arXiv:1502.02114 [astro-ph.CO]].

\bibitem{Aguirregabiria:2004te}
J.M.~Aguirregabiria, L.P.~Chimento and R.~Lazkoz,
Phys. Rev. D \textbf{70}, 023509 (2004)
doi:10.1103/PhysRevD.70.023509
[arXiv:astro-ph/0403157 [astro-ph]].

\bibitem{Aguirregabiria:2003uh}
J.M.~Aguirregabiria, L.P.~Chimento, A.S.~Jakubi and R.~Lazkoz,
Phys. Rev. D \textbf{67}, 083518 (2003)
doi:10.1103/PhysRevD.67.083518
[arXiv:gr-qc/0303010 [gr-qc]].

\bibitem{Alabidi:2006wa}
L.~Alabidi and D.~Lyth,
JCAP \textbf{08}, 006 (2006)
doi:10.1088/1475-7516/2006/08/006
[arXiv:astro-ph/0604569 [astro-ph]].

\bibitem{Alabidi:2006hg}
L.~Alabidi,
JCAP \textbf{10}, 015 (2006)
doi:10.1088/1475-7516/2006/10/015
[arXiv:astro-ph/0604611 [astro-ph]].

\bibitem{AliHaimoud:2010dx}
Y.~Ali-Haimoud and C.M.~Hirata,
Phys. Rev. D \textbf{83}, 043513 (2011)
doi:10.1103/PhysRevD.83.043513
[arXiv:1011.3758 [astro-ph.CO]].

\bibitem{Allahverdi:2010xz}
R.~Allahverdi, R.~Brandenberger, F.Y.~Cyr-Racine and A.~Mazumdar,
Ann. Rev. Nucl. Part. Sci. \textbf{60}, 27-51 (2010)
doi:10.1146/annurev.nucl.012809.104511
[arXiv:1001.2600 [hep-th]].

\bibitem{Alcubierre:2001ea}
M.~Alcubierre, F.S.~Guzman, T.~Matos, D.~Nunez, L.A.~Urena-Lopez and 
P.~Wiederhold,
Class. Quant. Grav. \textbf{19}, 5017 (2002)
doi:10.1088/0264-9381/19/19/314
[arXiv:gr-qc/0110102 [gr-qc]].

\bibitem{Aldrovandi:2005ya}
R.~Aldrovandi, R.R.~Cuzinatto and L.G.~Medeiros,
Found. Phys. \textbf{36}, 1736-1752 (2006)
doi:10.1007/s10701-006-9076-6
[arXiv:gr-qc/0508073 [gr-qc]].


\bibitem{Alimohammadi:2006tw}
M.~Alimohammadi and H.M.~Sadjadi,
Phys. Lett. B \textbf{648}, 113-118 (2007)
doi:10.1016/j.physletb.2007.03.014
[arXiv:gr-qc/0608016 [gr-qc]].

\bibitem{Alimohammadi:2007jj}
M.~Alimohammadi,
Gen. Rel. Grav. \textbf{40}, 107-115 (2008)
doi:10.1007/s10714-007-0514-3
[arXiv:0706.1360 [gr-qc]].

\bibitem{Alpher:1949sef}
R.A.~Alpher and R.C.~Herman,
Phys. Rev. \textbf{75}, no.7, 1089-1095 (1949)
doi:10.1103/physrev.75.1089

\bibitem{Alvarenga:2016yxh}
F.G.~Alvarenga, R.~Fracalossi, R.C.~Freitas and S.V.B.~Gon\c{c}alves,
Gen. Rel. Grav. \textbf{49}, no.11, 136 (2017)
doi:10.1007/s10714-017-2301-0
[arXiv:1607.03478 [gr-qc]].

\bibitem{Amendola:2015ksp}
L.~Amendola and S.~Tsujikawa, {\em Dark Energy: Theory and Observations}  
(Cambridge University Press, Cambridge, UK, 2010).

\bibitem{Amin:2014eta}
M.A.~Amin, M.P.~Hertzberg, D.I.~Kaiser and J.~Karouby,
Int. J. Mod. Phys. D \textbf{24}, 1530003 (2014)
doi:10.1142/S0218271815300037
[arXiv:1410.3808 [hep-ph]].

\bibitem{Ananda:2006gf}
K.N.~Ananda and M.~Bruni,
Phys. Rev. D \textbf{74}, 023524 (2006)
doi:10.1103/PhysRevD.74.023524
[arXiv:gr-qc/0603131 [gr-qc]].

\bibitem{Ananda:2005xp}
K.N.~Ananda and M.~Bruni,
Phys. Rev. D \textbf{74}, 023523 (2006)
doi:10.1103/PhysRevD.74.023523
[arXiv:astro-ph/0512224 [astro-ph]].

\bibitem{Andrianov:2011fg}
A.A.~Andrianov, F.~Cannata and A.Y.~Kamenshchik,
JCAP \textbf{10}, 004 (2011)
doi:10.1088/1475-7516/2011/10/004
[arXiv:1105.4515 [gr-qc]].

\bibitem{Arbey:2006it}
A.~Arbey,
Phys. Rev. D \textbf{74}, 043516 (2006)
doi:10.1103/PhysRevD.74.043516
[arXiv:astro-ph/0601274 [astro-ph]].

\bibitem{Arefeva:2005mka}
I.~Y.~Aref'eva, A.~S.~Koshelev and S.~Y.~Vernov,
Phys. Rev. D \textbf{72}, 064017 (2005)
doi:10.1103/PhysRevD.72.064017
[arXiv:astro-ph/0507067 [astro-ph]].

\bibitem{AssadLima88} M.J.D. Assad and J.A. Sales de Lima, 
Gen. Relativ. Gravit. {\bf 20}, 527 (1988).

\bibitem{Ballesteros:2013nwa}
G.~Ballesteros, B.~Bellazzini and L.~Mercolli,
JCAP \textbf{05}, 007 (2014)
doi:10.1088/1475-7516/2014/05/007
[arXiv:1312.2957 [hep-th]].

\bibitem{Ballesteros:2016kdx}
G.~Ballesteros, D.~Comelli and L.~Pilo,
Phys. Rev. D \textbf{94}, no.2, 025034 (2016)
doi:10.1103/PhysRevD.94.025034
[arXiv:1605.05304 [hep-th]].

\bibitem{Barnaby:2006cq}
N.~Barnaby and J.M.~Cline,
Phys. Rev. D \textbf{73}, 106012 (2006)
doi:10.1103/PhysRevD.73.106012
[arXiv:astro-ph/0601481 [astro-ph]].

\bibitem{Barnaby:2006km}
N.~Barnaby and J.~M.~Cline,
Phys. Rev. D \textbf{75}, 086004 (2007)
doi:10.1103/PhysRevD.75.086004
[arXiv:astro-ph/0611750 [astro-ph]].

\bibitem{Barrow:1987ia}
J.D.~Barrow,
Phys. Lett. B \textbf{187}, 12-16 (1987)
doi:10.1016/0370-2693(87)90063-3

\bibitem{Barrow:1990vx}
J.D.~Barrow,
Phys. Lett. B \textbf{235}, 40-43 (1990)
doi:10.1016/0370-2693(90)90093-L

\bibitem{Barrow:1993ah}
J.D. Barrow and P.~Saich,
Class. Quant. Grav. \textbf{10}, 279-283 (1993)
doi:10.1088/0264-9381/10/2/009

\bibitem{Barrow:1993hn}
J.D.~Barrow,
Phys. Rev. D \textbf{48}, 1585-1590 (1993)
doi:10.1103/PhysRevD.48.1585

\bibitem{Barrow:1993zq}
J.D.~Barrow and A.R.~Liddle,
Phys. Rev. D \textbf{47}, no.12, R5219 (1993)
doi:10.1103/PhysRevD.47.R5219
[arXiv:astro-ph/9303011 [astro-ph]].

\bibitem{Barrow:1994nt}
J.D.~Barrow,
Phys. Rev. D \textbf{49}, 3055-3058 (1994)
doi:10.1103/PhysRevD.49.3055

\bibitem{Barrow04} J.D. Barrow, 
Class. Quant. Grav. {\bf 21}, L79 (2004).

\bibitem{BarrowBook} J.D. Barrow, {\em The Book of Universes} (W.W. 
Norton \& C., New York, 2011).

\bibitem{BarrowGallowayTipler} J.D. Barrow, G. Galloway, and F.J.T. 
Tipler, 
Mon. Not. Roy. Astron. Soc. {\bf 223},  835 (1986).

\bibitem{Bartolo:2001cw}
N.~Bartolo, S.~Matarrese and A.~Riotto,
Phys. Rev. D \textbf{65}, 103505 (2002)
doi:10.1103/PhysRevD.65.103505
[arXiv:hep-ph/0112261 [hep-ph]].

\bibitem{Bartolo:2003jx}
N.~Bartolo, S.~Matarrese and A.~Riotto,
Phys. Rev. D \textbf{69}, 043503 (2004)
doi:10.1103/PhysRevD.69.043503
[arXiv:hep-ph/0309033 [hep-ph]].

\bibitem{Basilakos:2011rx}
S.~Basilakos, M.~Tsamparlis and A.~Paliathanasis,
Phys. Rev. D \textbf{83}, 103512 (2011)
doi:10.1103/PhysRevD.83.103512
[arXiv:1104.2980 [astro-ph.CO]].



\bibitem{Basilakos:2019acj}
S.~Basilakos, N.E.~Mavromatos and J.~Sol\`a Peracaula,
Phys. Rev. D \textbf{101}, no.4, 045001 (2020)
doi:10.1103/PhysRevD.101.045001
[arXiv:1907.04890 [hep-ph]].

\bibitem{Basilakos:2019mpe}
S.~Basilakos, N.E.~Mavromatos and J.~Sol\`a Peracaula,
Int. J. Mod. Phys. D \textbf{28}, no.14, 1944002 (2019)
doi:10.1142/S0218271819440024
[arXiv:1905.04685 [hep-th]].

\bibitem{Basilakos:2020qmu}
S.~Basilakos, N.~E.~Mavromatos and J.~Sol\`a Peracaula,
Phys. Lett. B \textbf{803}, 135342 (2020)
doi:10.1016/j.physletb.2020.135342
[arXiv:2001.03465 [gr-qc]].

\bibitem{BasteroGil:2012cm}
M.~Bastero-Gil, A.~Berera, R.O.~Ramos and J.G.~Rosa,
JCAP \textbf{01}, 016 (2013)
doi:10.1088/1475-7516/2013/01/016
[arXiv:1207.0445 [hep-ph]].

\bibitem{Battefeld:2006sz}
T.~Battefeld and R.~Easther,
JCAP \textbf{03}, 020 (2007)
doi:10.1088/1475-7516/2007/03/020
[arXiv:astro-ph/0610296 [astro-ph]].

\bibitem{Battefeld:2007en}
D.~Battefeld and T.~Battefeld,
JCAP \textbf{05}, 012 (2007)
doi:10.1088/1475-7516/2007/05/012
[arXiv:hep-th/0703012 [hep-th]].

\bibitem{Battye:1999eq}
R.A.~Battye, M.~Bucher and D.~Spergel,
[arXiv:astro-ph/9908047 [astro-ph]].

\bibitem{Bayin:1994nz}
S.S.~Bayin, F.I.~Cooperstock and V.~Faraoni,
Astrophys. J. \textbf{428}, 439-446 (1994)
doi:10.1086/174256
[arXiv:astro-ph/9402033 [astro-ph]].

\bibitem{Diego} J. Beltr\'an Jim\'enez, D. Rubiera-Garcia, D. 
S\'aez-G\'omez, and V. Salzano, 
Phys. Rev. D {\bf 94}, 123520 (2016).

\bibitem{Achour:2016rkg}
J.~Ben Achour, D.~Langlois and K.~Noui,
Phys. Rev. D \textbf{93}, no.12, 124005 (2016)
doi:10.1103/PhysRevD.93.124005
[arXiv:1602.08398 [gr-qc]].


\bibitem{BenAchour:2016fzp}
J.~Ben Achour, M.~Crisostomi, K.~Koyama, D.~Langlois, K.~Noui and 
G.~Tasinato,
JHEP \textbf{12}, 100 (2016)
doi:10.1007/JHEP12(2016)100
[arXiv:1608.08135 [hep-th]].

\bibitem{BenAchour:2019ufa}
J. Ben Achour and E.R.~Livine,
JHEP \textbf{12}, 031 (2019)
doi:10.1007/JHEP12(2019)031
[arXiv:1909.13390 [gr-qc]].

\bibitem{BenAchour:2020xif}
J.~Ben Achour and E.R.~Livine,
Class. Quant. Grav. \textbf{37}, no.21, 215001 (2020)
doi:10.1088/1361-6382/abb577
[arXiv:2004.05841 [gr-qc]].

\bibitem{BenAchour:2020njq}
J.~Ben Achour and E.R.~Livine,
JHEP \textbf{03}, 067 (2020)
doi:10.1007/JHEP03(2020)067
[arXiv:2001.11807 [gr-qc]].

\bibitem{Bergmann:1968ve}
P.G.~Bergmann,
Int. J. Theor. Phys. \textbf{1}, 25-36 (1968)
doi:10.1007/BF00668828

\bibitem{Bernal:2006ci}
A.~Bernal and F.~Siddhartha Guzman,
Phys. Rev. D \textbf{74}, 103002 (2006)
doi:10.1103/PhysRevD.74.103002
[arXiv:astro-ph/0610682 [astro-ph]].

\bibitem{Bernardeau:2002jy}
F.~Bernardeau and J.-P.~Uzan,
Phys. Rev. D \textbf{66}, 103506 (2002)
doi:10.1103/PhysRevD.66.103506
[arXiv:hep-ph/0207295 [hep-ph]].

\bibitem{Bernardeau:2002jf}
F.~Bernardeau and J.-P.~Uzan,
Phys. Rev. D \textbf{67}, 121301 (2003)
doi:10.1103/PhysRevD.67.121301
[arXiv:astro-ph/0209330 [astro-ph]].

\bibitem{Bertacca:2007ux}
D.~Bertacca, S.~Matarrese and M.~Pietroni,
Mod. Phys. Lett. A \textbf{22}, 2893-2907 (2007)
doi:10.1142/S0217732307025893
[arXiv:astro-ph/0703259 [astro-ph]].

\bibitem{Bertacca:2007fc}
D.~Bertacca, N.~Bartolo and S.~Matarrese,
JCAP \textbf{05}, 005 (2008)
doi:10.1088/1475-7516/2008/05/005
[arXiv:0712.0486 [astro-ph]].

\bibitem{Bertacca:2008uf}
D.~Bertacca, N.~Bartolo, A.~Diaferio and S.~Matarrese,
JCAP \textbf{10}, 023 (2008)
doi:10.1088/1475-7516/2008/10/023
[arXiv:0807.1020 [astro-ph]].

\bibitem{Bertolami:2007gv}
O.~Bertolami, C.G.~Boehmer, T.~Harko and F.S.N.~Lobo,
Phys. Rev. D \textbf{75}, 104016 (2007)
doi:10.1103/PhysRevD.75.104016
[arXiv:0704.1733 [gr-qc]].

\bibitem{Bertolami:2016ywc}
O.~Bertolami, C.~Cosme and J.G.~Rosa,
Phys. Lett. B \textbf{759}, 1-8 (2016)
doi:10.1016/j.physletb.2016.05.047
[arXiv:1603.06242 [hep-ph]].

\bibitem{Bettoni:2016mij}
D.~Bettoni, J.~M.~Ezquiaga, K.~Hinterbichler and M.~Zumalac\'arregui,
Phys. Rev. D \textbf{95}, no.8, 084029 (2017)
doi:10.1103/PhysRevD.95.084029
[arXiv:1608.01982 [gr-qc]].

\bibitem{Billyard:2000bh} 
A.P.~Billyard and A.A.~Coley,
Phys.\ Rev.\ D {\bf 61}, 083503 (2000) [astro-ph/9908224].

\bibitem{Bloomfield:2012ff}
J.K.~Bloomfield, {\'E}.{\'E}.~Flanagan, M.~Park, and S.~Watson,
  JCAP {\bf 1308},  010 (2013) [arXiv:1211.7054[astro-ph.CO]].

\bibitem{Borowiec:2014wva}
A.~Borowiec, S.~Capozziello, M.~De Laurentis, F.S.N.~Lobo, 
A.~Paliathanasis, M.~Paolella and A.~Wojnar,
Phys. Rev. D \textbf{91}, no.2, 023517 (2015)
doi:10.1103/PhysRevD.91.023517
[arXiv:1407.4313 [gr-qc]].

\bibitem{Borowiec16} M. Szydlowski, A. Stachowski, A. Borowiec, and A. 
Wojnar, 
Eur. Phys. J. C {\bf 76}, 567 (2016).

\bibitem{Bouhmadi15} M. Bouhmadi-L\'opez, A. Errahmani, P. Martin-Moruno, 
T. Ouali, and Y. Tavakoli,
Int. J. Mod. Phys. D {\bf 24}, 1550078 (2015).  

\bibitem{Brans:1961sx}
C.~Brans and R.H.~Dicke,
Phys. Rev. \textbf{124}, 925-935 (1961)
doi:10.1103/PhysRev.124.925

\bibitem{Brevik:2014lpa}
I.~Brevik, V.V.~Obukhov and A.V.~Timoshkin,
Mod. Phys. Lett. A \textbf{29}, no.15, 1450078 (2014)
doi:10.1142/S0217732314500783
[arXiv:1404.1887 [gr-qc]].

\bibitem{Bucher:1998mh}
M.~Bucher and D.N.~Spergel,
Phys. Rev. D \textbf{60}, 043505 (1999)
doi:10.1103/PhysRevD.60.043505
[arXiv:astro-ph/9812022 [astro-ph]].

\bibitem{Burd:1988ss}
A.B.~Burd and J.D.~Barrow,
Nucl. Phys. B \textbf{308}, 929-945 (1988)
[erratum: Nucl. Phys. B \textbf{324}, 276-276 (1989)]
doi:10.1016/0550-3213(88)90135-6

\bibitem{Byrnes:2008wi}
C.T.~Byrnes, K.Y.~Choi and L.M.H.~Hall,
JCAP \textbf{10}, 008 (2008)
doi:10.1088/1475-7516/2008/10/008
[arXiv:0807.1101 [astro-ph]].

\bibitem{Byrnes:2008zy}
C.T.~Byrnes, K.Y.~Choi and L.M.H.~Hall,
JCAP \textbf{02}, 017 (2009)
doi:10.1088/1475-7516/2009/02/017
[arXiv:0812.0807 [astro-ph]].

\bibitem{Cai:2009zp}
Y.F.~Cai, E.N.~Saridakis, M.R.~Setare and J.Q.~Xia,
Phys. Rept. \textbf{493}, 1-60 (2010)
doi:10.1016/j.physrep.2010.04.001
[arXiv:0909.2776 [hep-th]].

\bibitem{Cai:2012yf}
Y.F.~Cai, M.~Li and X.~Zhang,
Phys. Lett. B \textbf{718}, 248-254 (2012)
doi:10.1016/j.physletb.2012.10.065
[arXiv:1209.3437 [hep-th]].

\bibitem{Cai:2006dm}
Y.F.~Cai, H.~Li, Y.S.~Piao and X.m.~Zhang,
Phys. Lett. B \textbf{646}, 141-144 (2007)
doi:10.1016/j.physletb.2007.01.027
[arXiv:gr-qc/0609039 [gr-qc]].

\bibitem{Cai:2007zv}
Y.F.~Cai, T.~Qiu, R.~Brandenberger, Y.S.~Piao and X.~Zhang,
JCAP \textbf{03}, 013 (2008)
doi:10.1088/1475-7516/2008/03/013
[arXiv:0711.2187 [hep-th]].

\bibitem{Cai:2007gs}
Y.F.~Cai, M.Z.~Li, J.X.~Lu, Y.S.~Piao, T.t.~Qiu and X.m.~Zhang,
Phys. Lett. B \textbf{651}, 1-7 (2007)
doi:10.1016/j.physletb.2007.05.056
[arXiv:hep-th/0701016 [hep-th]].

\bibitem{Cai:2007qw}
Y.F.~Cai, T.~Qiu, Y.S.~Piao, M.~Li and X.~Zhang,
JHEP \textbf{10}, 071 (2007)
doi:10.1088/1126-6708/2007/10/071
[arXiv:0704.1090 [gr-qc]].

\bibitem{Cai:2008ed}
Y.F.~Cai and X.~Zhang,
JCAP \textbf{06}, 003 (2009)
doi:10.1088/1475-7516/2009/06/003
[arXiv:0808.2551 [astro-ph]].

\bibitem{Cai:2008gk}
Y.F.~Cai and J.~Wang,
Class. Quant. Grav. \textbf{25}, 165014 (2008)
doi:10.1088/0264-9381/25/16/165014
[arXiv:0806.3890 [hep-th]].

\bibitem{Calcagni:2004bh}
G.~Calcagni,
Phys. Rev. D \textbf{69}, 103508 (2004)
doi:10.1103/PhysRevD.69.103508
[arXiv:hep-ph/0402126 [hep-ph]].

\bibitem{Calcagni:2004wu}
G.~Calcagni,
Phys. Rev. D \textbf{71}, 023511 (2005)
doi:10.1103/PhysRevD.71.023511
[arXiv:gr-qc/0410027 [gr-qc]].

\bibitem{Capozziello:2003tk}
S.~Capozziello, S.~Carloni and A.~Troisi,
Recent Res. Dev. Astron. Astrophys. \textbf{1}, 625 (2003)
[arXiv:astro-ph/0303041 [astro-ph]].

\bibitem{Capozziello:2005pa}
S.~Capozziello, V.F.~Cardone, E.~Elizalde, S.~Nojiri and S.D.~Odintsov,
Phys. Rev. D \textbf{73}, 043512 (2006)
doi:10.1103/PhysRevD.73.043512
[arXiv:astro-ph/0508350 [astro-ph]].

\bibitem{Capozzielloetal06} S. Capozziello, V.F. Cardone, E. Elizalde, S. 
Nojiri, and S.D. Odintsov,
Phys. Rev. D {\bf 73}, 043512 (2006). 

\bibitem{Capozziello:2009te}
S.~Capozziello, E.~Piedipalumbo, C.~Rubano and P.~Scudellaro,
Phys. Rev. D \textbf{80}, 104030 (2009)
doi:10.1103/PhysRevD.80.104030
[arXiv:0908.2362 [astro-ph.CO]].

\bibitem{CapozzielloFaraoni} S. Capozziello, V. Faraoni, {\em Beyond 
Einstein Gravity: A Survey of Gravitational Theories For Cosmology and 
Astrophysics} (Springer, New York, 2010).

\bibitem{Capozziello:2013bma}
S.~Capozziello and M.~Roshan,
Phys. Lett. B \textbf{726}, 471-480 (2013)
doi:10.1016/j.physletb.2013.08.047
[arXiv:1308.3910 [gr-qc]].
 
\bibitem{Capozziello:2014bna}
S.~Capozziello, M.~De Laurentis and R.~Myrzakulov,
Int. J. Geom. Meth. Mod. Phys. \textbf{12}, no.09, 1550095 (2015)
doi:10.1142/S0219887815500954
[arXiv:1412.1471 [gr-qc]].

\bibitem{Capozziello:2016eaz}
S.~Capozziello, M.~De Laurentis and K.F.~Dialektopoulos,
Eur. Phys. J. C \textbf{76}, no.11, 629 (2016)
doi:10.1140/epjc/s10052-016-4491-0
[arXiv:1609.09289 [gr-qc]].

\bibitem{Carr:1993aq}
B.J.~Carr and J.E.~Lidsey,
Phys. Rev. D \textbf{48}, 543-553 (1993)
doi:10.1103/PhysRevD.48.543

\bibitem{Carroll} S.M. Carroll, {\em Spacetime and Geometry: An 
Introduction to General Relativity} (Addison Wesley, San Francisco, 2004).

\bibitem{Carroll:2003wy}
S.M.~Carroll, V.~Duvvuri, M.~Trodden and M.S.~Turner,
Phys. Rev. D \textbf{70}, 043528 (2004)
doi:10.1103/PhysRevD.70.043528
[arXiv:astro-ph/0306438 [astro-ph]].

\bibitem{Carter:1995mj}
B.~Carter and D.~Langlois,
Nucl. Phys. B \textbf{454}, 402-424 (1995)
doi:10.1016/0550-3213(95)00425-R
[arXiv:hep-th/9611082 [hep-th]].

\bibitem{Carter:1998rn}
B.~Carter and D.~Langlois,
Nucl. Phys. B \textbf{531}, 478-504 (1998)
doi:10.1016/S0550-3213(98)00430-1
[arXiv:gr-qc/9806024 [gr-qc]].

\bibitem{Cataldo:2005gb}
M.~Cataldo and L.P.~Chimento,
Int. J. Mod. Phys. D \textbf{17}, 1981-1989 (2008)
doi:10.1142/S0218271808013790
[arXiv:gr-qc/0506090 [gr-qc]].

\bibitem{Charters:2009ku}
T.~Charters and J.P.~Mimoso,
JCAP \textbf{08}, 022 (2010)
doi:10.1088/1475-7516/2010/08/022
[arXiv:0909.2282 [hep-ph]].

\bibitem{Chavanis2015}  P.H. Chavanis, Phys. Rev. D {\bf 92}, 103004 
(2015).

\bibitem{Chebysev} P.L. Chebyshev, ``Sur l'integration des 
diff\'erentielles irrationnelles'',  Journal de Mathematiques 
(series~1) {\bf 18}, 87-111 (1853). 

\bibitem{Chen:2008ft}
X.m.~Chen, Y.g.~Gong and E.N.~Saridakis,
JCAP \textbf{04}, 001 (2009)
doi:10.1088/1475-7516/2009/04/001
[arXiv:0812.1117 [gr-qc]].

\bibitem{Chen:2014fqa}
S.~Chen, G.~W.~Gibbons, Y.~Li and Y.~Yang,
JCAP \textbf{12}, 035 (2014)
doi:10.1088/1475-7516/2014/12/035
[arXiv:1409.3352 [astro-ph.CO]].

\bibitem{Chernin1966} A.D. Chernin, Sov. Astron. {\bf 9}, 
871 (1966).

\bibitem{Chervon:2017kgn}
S.V.~Chervon, I.V.~Fomin and A.~Beesham,
Eur. Phys. J. C \textbf{78}, no.4, 301 (2018)
doi:10.1140/epjc/s10052-018-5795-z
[arXiv:1704.08712 [gr-qc]].

\bibitem{Chervon:2019nwq}
S.~V.~Chervon, I.~V.~Fomin, E.~O.~Pozdeeva, M.~Sami and S.~Y.~Vernov,
Phys. Rev. D \textbf{100}, no.6, 063522 (2019)
doi:10.1103/PhysRevD.100.063522
[arXiv:1904.11264 [gr-qc]].

\bibitem{Chimento:1995da}
L.P.~Chimento and A.S.~Jakubi,
Int. J. Mod. Phys. D \textbf{5}, 71-84 (1996)
doi:10.1142/S0218271896000084
[arXiv:gr-qc/9506015 [gr-qc]].

\bibitem{Chimento:2002gb}
L.P.~Chimento,
Phys. Rev. D \textbf{65}, 063517 (2002)
doi:10.1103/PhysRevD.65.063517

\bibitem{Chimento:2003qy}
L.P.~Chimento and R.~Lazkoz,
Phys. Rev. Lett. \textbf{91}, 211301 (2003)
doi:10.1103/PhysRevLett.91.211301
[arXiv:gr-qc/0307111 [gr-qc]].

\bibitem{Chimento:2005au}
L.P.~Chimento and R.~Lazkoz,
Class. Quant. Grav. \textbf{23}, 3195-3204 (2006)
doi:10.1088/0264-9381/23/9/027
[arXiv:astro-ph/0505254 [astro-ph]].

\bibitem{Chimento:2005xa}
L.P.~Chimento and D.~Pavon,
Phys. Rev. D \textbf{73}, 063511 (2006)
doi:10.1103/PhysRevD.73.063511
[arXiv:gr-qc/0505096 [gr-qc]].

\bibitem{Chimento:2006gk}
L.P.~Chimento and W.~Zimdahl,
Int. J. Mod. Phys. D \textbf{17}, 2229-2254 (2008)
doi:10.1142/S0218271808013820
[arXiv:gr-qc/0609104 [gr-qc]].

\bibitem{Chimento:2007fx}
L.P.~Chimento, F.P.~Devecchi, M.I.~Forte and G.M.~Kremer,
Class. Quant. Grav. \textbf{25}, 085007 (2008)
doi:10.1088/0264-9381/25/8/085007
[arXiv:0707.4455 [gr-qc]].

\bibitem{Chimento:2008ws}
L.P.~Chimento, M.I.~Forte, R.~Lazkoz and M.G.~Richarte,
Phys. Rev. D \textbf{79}, 043502 (2009)
doi:10.1103/PhysRevD.79.043502
[arXiv:0811.3643 [astro-ph]].

\bibitem{Chimento:2010un}
L.P.~Chimento, R.~Lazkoz and M.G.~Richarte,
Phys. Rev. D \textbf{83}, 063505 (2011)
doi:10.1103/PhysRevD.83.063505
[arXiv:1011.2345 [astro-ph.CO]].

\bibitem{Chimento:2011dw}
L.P.~Chimento, M.I.~Forte and M.G.~Richarte,
Mod. Phys. Lett. A \textbf{28}, 1250235 (2013)
doi:10.1142/S0217732312502355
[arXiv:1106.0781 [astro-ph.CO]].

\bibitem{Chluba:2010ca}
J.~Chluba and R.M.~Thomas,
Mon. Not. Roy. Astron. Soc. \textbf{412}, 748 (2011)
doi:10.1111/j.1365-2966.2010.17940.x
[arXiv:1010.3631 [astro-ph.CO]].

\bibitem{Chluba:2015gta}
J.~Chluba and Y.~Ali-Haimoud,
Mon. Not. Roy. Astron. Soc. \textbf{456}, no.4, 3494-3508 (2016)
doi:10.1093/mnras/stv2691
[arXiv:1510.03877 [astro-ph.CO]].

\bibitem{Choi:2007su}
K.~Y.~Choi, L.~M.~H.~Hall and C.~van de Bruck,
JCAP \textbf{02}, 029 (2007)
doi:10.1088/1475-7516/2007/02/029
[arXiv:astro-ph/0701247 [astro-ph]].

\bibitem{Christodoulakis:2003pg}
T.~Christodoulakis, C.~Helias, P.G. Kevrekidis, I.G. Kevrekidis and 
G.O. Papadopoulos,
[arXiv:gr-qc/0302120 [gr-qc]].

\bibitem{Christodoulakis:2013xha}
T.~Christodoulakis, N.~Dimakis and P.A.~Terzis,
J. Phys. A \textbf{47}, 095202 (2014)
doi:10.1088/1751-8113/47/9/095202
[arXiv:1304.4359 [gr-qc]].

\bibitem{Christodoulakis:2018swq}
T.~Christodoulakis, A.~Karagiorgos and A.~Zampeli,
Symmetry \textbf{10}, no.3, 70 (2018)
doi:10.3390/sym10030070

\bibitem{Clemson:2011an}
T.~Clemson, K.~Koyama, G.B.~Zhao, R.~Maartens and J.~Valiviita,
Phys. Rev. D \textbf{85}, 043007 (2012)
doi:10.1103/PhysRevD.85.043007
[arXiv:1109.6234 [astro-ph.CO]].

\bibitem{Clifton:2011jh}
T.~Clifton, P.G.~Ferreira, A.~Padilla and C.~Skordis,
Phys. Rept. \textbf{513}, 1-189 (2012)
doi:10.1016/j.physrep.2012.01.001
[arXiv:1106.2476 [astro-ph.CO]].

\bibitem{CoquereauxGrossman1982} R. Coquereaux and A. Grossmann, 
Ann. Phys. (USA) {\bf 143}, 296 (1982).

\bibitem{CohenNature1967} J.M. Cohen, Nature {\bf 216}, 249 (1967).

\bibitem{ColeyASS89} A.A. Coley, 
Astrophys. Space Sci. {\bf 155}, 193-201 (1989).

\bibitem{Coleybook} A.A. Coley, {\em Dynamical Systems and 
Cosmology} (Kluwer, Dordrecht, 2003).

\bibitem{ColeyTupper86} A.A. Coley and B.O.J. Tupper,   
J. Math. Phys. {\bf 27}, 406 (1986).

\bibitem{ColeyTupper86b} A.A. Coley and B.O.J. Tupper, 
Can. J. Phys. {\bf 64}, 204 (1986).

\bibitem{ColeyWainwright92} A.A. Coley and J. Wainwright, Class. 
Quant. Grav. {\bf 9}, 651-665 (1992). 

\bibitem{Coley:1995dpj}
A.A.~Coley and R.J.~van den Hoogen,
Class. Quant. Grav. \textbf{12}, 1977-1994 (1995)
doi:10.1088/0264-9381/12/8/015
[arXiv:gr-qc/9605061 [gr-qc]].

\bibitem{Conversi:2004pi}
L.~Conversi, A.~Melchiorri, L.~Mersini-Houghton and J.~Silk,
Astropart. Phys. \textbf{21}, 443-449 (2004)
doi:10.1016/j.astropartphys.2004.02.006
[arXiv:astro-ph/0402529 [astro-ph]].

\bibitem{Cosme:2018nly}
C.~Cosme, J.G.~Rosa and O.~Bertolami,
JHEP \textbf{05}, 129 (2018)
doi:10.1007/JHEP05(2018)129
[arXiv:1802.09434 [hep-ph]].

\bibitem{Costa:2013sva}
A.A.~Costa, X.D.~Xu, B.~Wang, E.G.M.~Ferreira and E.~Abdalla,
Phys. Rev. D \textbf{89}, no.10, 103531 (2014)
doi:10.1103/PhysRevD.89.103531
[arXiv:1311.7380 [astro-ph.CO]].

\bibitem{Crisostomi:2016czh}
M.~Crisostomi, K.~Koyama and G.~Tasinato,
JCAP \textbf{04}, 044 (2016)
doi:10.1088/1475-7516/2016/04/044
[arXiv:1602.03119 [hep-th]].

\bibitem{Crisostomi:2017aim}
M.~Crisostomi, R.~Klein and D.~Roest,
JHEP \textbf{06}, 124 (2017)
doi:10.1007/JHEP06(2017)124
[arXiv:1703.01623 [hep-th]].

\bibitem{Cruz:2006ck}
N.~Cruz, S.~Lepe and F.~Pena,
Phys. Lett. B \textbf{646}, 177-182 (2007)
doi:10.1016/j.physletb.2006.12.070
[arXiv:gr-qc/0609013 [gr-qc]].

\bibitem{Dabrowski:2003jm}
M.P.~Dabrowski, T.~Stachowiak and M.~Szydlowski,
Phys. Rev. D \textbf{68}, 103519 (2003)
doi:10.1103/PhysRevD.68.103519
[arXiv:hep-th/0307128 [hep-th]].

\bibitem{DabrowskiStelmach1986} 
M. Dabrowski, and J. Stelmach, 
Ann. Phys. (USA) {\bf 166}, 422 (1986).

\bibitem{Dabrowski:2006dd}
M.P.~Dabrowski, C.~Kiefer and B.~Sandhofer,
Phys. Rev. D \textbf{74}, 044022 (2006)
doi:10.1103/PhysRevD.74.044022
[arXiv:hep-th/0605229 [hep-th]].

\bibitem{DavidsonNarlikar66}W. Davidson and J.V. Narlikar
Progr. Phys. 29, 539 (1966).

\bibitem{DAmbroise:2007zhp}
J.~D'Ambroise,
[arXiv:0711.3916 [hep-th]].

\bibitem{DAmbroise:2010dgl}
J.~D'Ambroise,
[arXiv:1005.1410 [gr-qc]].

\bibitem{DAmbroiseWilliams} J. D'Ambroise and F.L. Williams, Int. J. Pure 
Appl. Maths. {\bf 34}, 117 (2007)

\bibitem{Dariescu2017} M.-A. Dariescu,  D.-A. Mihu, and C. Dariescu,  
Romanian J. Phys. {\bf 62}, 101 (2017).

\bibitem{Deffayet:2009wt}
C.~Deffayet, G.~Esposito-Farese and A.~Vikman,
Phys. Rev. D \textbf{79}, 084003 (2009)
doi:10.1103/PhysRevD.79.084003
[arXiv:0901.1314 [hep-th]].

\bibitem{Deffayet:2011gz}
C.~Deffayet, X.~Gao, D.A.~Steer and G.~Zahariade,
Phys. Rev. D \textbf{84}, 064039 (2011)
doi:10.1103/PhysRevD.84.064039
[arXiv:1103.3260 [hep-th]].

\bibitem{DeFelice:2010aj}
A.~De Felice and S.~Tsujikawa,
Living Rev. Rel. \textbf{13} (2010) 3 
doi:10.12942/lrr-2010-3
[arXiv:1002.4928 [gr-qc]].

\bibitem{deRitis:1990ba}
R.~de Ritis, G.~Marmo, G.~Platania, C.~Rubano, P.~Scudellaro and 
C.~Stornaiolo,
Phys. Rev. D \textbf{42}, 1091-1097 (1990)
doi:10.1103/PhysRevD.42.1091

\bibitem{deRitis2} R. de Ritis, G. Marmo, G. Platania, C. Rubano, P. 
Scudellaro, and C. Stornaiolo, Phys. Lett. 149A 79-83 (1990).

\bibitem{deRitis3} R. de Ritis. G. Platania, C. Rubano, and R. Sabatino,  
Phys. Lett. 16lA 230 (1991).

\bibitem{deSitter:1917zz}
W.~de Sitter,
Mon. Not. Roy. Astron. Soc. \textbf{78}, 3-28 (1917)

\bibitem{Dicke:1961gz}
R.H.~Dicke,
Phys. Rev. \textbf{125}, 2163-2167 (1962)
doi:10.1103/PhysRev.125.2163

\bibitem{Dimakis:2013oza}
N.~Dimakis, T.~Christodoulakis and P.A.~Terzis,
J. Geom. Phys. \textbf{77}, 97-112 (2014)
doi:10.1016/j.geomphys.2013.12.001
[arXiv:1311.4358 [gr-qc]].

\bibitem{Dimakis:2016mip}
N.~Dimakis, A.~Karagiorgos, A.~Zampeli, A.~Paliathanasis, 
T.~Christodoulakis and P.~A.~Terzis,
Phys. Rev. D \textbf{93}, no.12, 123518 (2016)
doi:10.1103/PhysRevD.93.123518
[arXiv:1604.05168 [gr-qc]].

\bibitem{Inverno} R. d'Inverno, {\em Introducing Einstein's Relativity} 
(Clarendon Press, Oxford, 1992).

\bibitem{Dussault:2020uvj}
S.~Dussault and V.~Faraoni,
Eur. Phys. J. C \textbf{80}, no.11, 1002 (2020)
doi:10.1140/epjc/s10052-020-08590-8
[arXiv:2009.03235 [gr-qc]].

\bibitem{Dutta:2016exd}
S.~Dutta, M.~Lakshmanan and S.~Chakraborty,
Int. J. Mod. Phys. D \textbf{25}, no.14, 1650110 (2016)
doi:10.1142/S0218271816501108
[arXiv:1607.03396 [gr-qc]].

\bibitem{Easther:1993qg}
R.~Easther,
Class. Quant. Grav. \textbf{10}, 2203-2216 (1993)
doi:10.1088/0264-9381/10/11/005
[arXiv:gr-qc/9308010 [gr-qc]].

\bibitem{Eckart:1940te}
C.~Eckart,
Phys. Rev. \textbf{58}, 919-924 (1940)
doi:10.1103/PhysRev.58.919

\bibitem{Edwards1972} D. Edwards, 
Mon. Not. R. Astron. Soc. {\bf 159}, 51 (1972).

\bibitem{Einstein1917} A. Einstein, 
Sitzungsb. K\"onig. Preuss. Akad. 142-152 (1917). 

\bibitem{Ellis:1988jw}
G.F.R.~Ellis,
Class. Quant. Grav. \textbf{5}, 891-901 (1988)
doi:10.1088/0264-9381/5/6/010

\bibitem{EllisMadsen1991}  G.F.R. Ellis and M. Madsen,  
Class. Quant. Grav. {\bf 8}, 667 (1991).

\bibitem{EllisMaartensMacCallum} G.F.R. Ellis, R. Maartens, and M.A.H. 
MacCallum, {\em Relativistic Cosmology} (Cambridge University Press, 
Cambridge, UK, 2012).

\bibitem{Enqvist:2004bk}
K.~Enqvist and A.~Vaihkonen,
JCAP \textbf{09}, 006 (2004)
doi:10.1088/1475-7516/2004/09/006
[arXiv:hep-ph/0405103 [hep-ph]].

\bibitem{Ermakov1880} V.P. Ermakov, Univ. Izv. Kiev {\bf 20}, 1 (1880).

\bibitem{Faraoni:1999qu}
V.~Faraoni,
Am. J. Phys. \textbf{67}, 732 (1999)
doi:10.1119/1.19361
[arXiv:physics/9901006 [physics]].

\bibitem{Faraoni:2000vg}
V.~Faraoni,
Am. J. Phys. \textbf{69}, 372-376 (2001)
doi:10.1119/1.1290250
[arXiv:physics/0006030 [physics]].

\bibitem{Faraoni:2004bb}
V.~Faraoni,
Phys. Rev. D \textbf{69}, 123520 (2004)
doi:10.1103/PhysRevD.69.123520
[arXiv:gr-qc/0404078 [gr-qc]].

\bibitem{Faraoni:2004pi} V.~Faraoni, {\em Cosmology 
in Scalar-Tensor Gravity} (Kluwer Academic, Dordrecht, 2004), 
doi:10.1007/978-1-4020-1989-0

\bibitem{Faraoni:2011ut}
V.~Faraoni,
Phys. Lett. B \textbf{703}, 228-231 (2011)
doi:10.1016/j.physletb.2011.08.018
[arXiv:1108.2102 [gr-qc]].

\bibitem{Faraoni:2012hn}
V.~Faraoni,
Phys. Rev. D \textbf{85}, 024040 (2012)
doi:10.1103/PhysRevD.85.024040
[arXiv:1201.1448 [gr-qc]].

\bibitem{Faraoni:2012bf}
V.~Faraoni and C.S.~Protheroe,
Gen. Rel. Grav. \textbf{45}, 103-123 (2013)
doi:10.1007/s10714-012-1462-0
[arXiv:1209.3726 [gr-qc]].

\bibitem{Faraoni:2014vra}
V.~Faraoni, J.B.~Dent and E.N.~Saridakis,
Phys. Rev. D \textbf{90}, no.6, 063510 (2014)
doi:10.1103/PhysRevD.90.063510 [arXiv:1405.7288 [gr-qc]].

\bibitem{Faraoni:2018qdr}
V.~Faraoni and J.~Cot\'e,
Phys. Rev. D \textbf{98}, no.8, 084019 (2018)
doi:10.1103/PhysRevD.98.084019
[arXiv:1808.02427 [gr-qc]].

\bibitem{Faraoni:2020tpe}
V.~Faraoni,
Symmetry \textbf{12}, no.1, 147 (2020)
doi:10.3390/sym12010147
[arXiv:2001.02126 [gr-qc]]. 

\bibitem{Felten:1986zz}
J.~E.~Felten and R.~Isaacman,
Rev. Mod. Phys. \textbf{58}, 689-698 (1986)
doi:10.1103/RevModPhys.58.689

\bibitem{Feng:2004ff}
B.~Feng, M.~Li, Y.S.~Piao and X.~Zhang,
Phys. Lett. B \textbf{634}, 101-105 (2006)
doi:10.1016/j.physletb.2006.01.066
[arXiv:astro-ph/0407432 [astro-ph]].

\bibitem{Feng:2006ya}
B.~Feng,
[arXiv:astro-ph/0602156 [astro-ph]].

\bibitem{Fomin18} I.V.  Fomin, 
Russ Phys J 61, 843-851 (2018). 
https://doi.org/10.1007/s11182-018-1468-5

\bibitem{Frampton:2011sp}
P.H.~Frampton, K.J.~Ludwick and R.J.~Scherrer,
Phys. Rev. D \textbf{84}, 063003 (2011)
doi:10.1103/PhysRevD.84.063003
[arXiv:1106.4996 [astro-ph.CO]].

\bibitem{Friedmann1922} A. Friedmann,   
Zeit. Physik A {\bf 10}, 377-386 (1922)
doi:10.1007/BF01332580

\bibitem{Friedmann1924} A. Friedmann,   
Zeit. Physik A {\bf 21}, 326-322 (1924, 
doi:10.1007/BF01328280  
Translated in Gen. Rel. Grav. {\bf 31}, 31 (1999).

\bibitem{Fuchs:2004xe}
B.~Fuchs and E.W.~Mielke,
Mon. Not. Roy. Astron. Soc. \textbf{350}, 707 (2004)
doi:10.1111/j.1365-2966.2004.07679.x
[arXiv:astro-ph/0401575 [astro-ph]].

\bibitem{FujiiMaeda} Y. Fujii and K. Maeda, {\em The Scalar--Tensor 
Theory of Gravity} (Cambridge University Press, Cambridge, UK, 2003).

\bibitem{Fuzfa:2007sv}
A.~F\"{u}zfa and J.M.~Alimi,
Phys. Rev. D \textbf{75}, 123007 (2007)
doi:10.1103/PhysRevD.75.123007
[arXiv:astro-ph/0702478 [astro-ph]].

\bibitem{Gergely:2014rna} 
  L.A. Gergely and S.~Tsujikawa,
  Phys.\ Rev.\ D {\bf 89}, 064059 (2014) [arXiv:1402.0553 [hep-th]].

\bibitem{Gilman:1970zv}
R.C.~Gilman,
Phys. Rev. D \textbf{2}, 1400-1410 (1970)
doi:10.1103/PhysRevD.2.1400

\bibitem{Gionti:2017ffe}
G.~Gionti, S.J. and A.~Paliathanasis,
Mod. Phys. Lett. A \textbf{33}, no.16, 1850093 (2018)
doi:10.1142/S0217732318500931
[arXiv:1711.11106 [gr-qc]].

\bibitem{Gleyzes:2014dya}
J.~Gleyzes, D.~Langlois, F.~Piazza and F.~Vernizzi,
Phys. Rev. Lett. \textbf{114}, no.21, 211101 (2015)
doi:10.1103/PhysRevLett.114.211101
[arXiv:1404.6495 [hep-th]].

\bibitem{Gleyzes:2014qga}
J.~Gleyzes, D.~Langlois, F.~Piazza and F.~Vernizzi,
JCAP \textbf{02}, 018 (2015)
doi:10.1088/1475-7516/2015/02/018
[arXiv:1408.1952 [astro-ph.CO]].

\bibitem{Goldwirth:1991rj}
D.S.~Goldwirth and T.~Piran,
Phys. Rept. \textbf{214}, 223-291 (1992)
doi:10.1016/0370-1573(92)90073-9

\bibitem{Gong:2007se}
Y.~Gong and X.~Chen,
Phys. Rev. D \textbf{76}, 123007 (2007)
doi:10.1103/PhysRevD.76.123007
[arXiv:0708.2977 [astro-ph]].

\bibitem{Gorini04} V. Gorini, A. Kamenshchik, U. Moschella, and V. 
Pasquier,  
Phys. Rev. D {\bf 69}, 123512 (2004).

\bibitem{Gubitosi:2012hu}
  G.~Gubitosi, F.~Piazza, and F.~Vernizzi,
  JCAP {\bf 1302},  032 (2013) [arXiv:1210.0201[hep-th]].

\bibitem{Gumjudpai:2007bx}
B.~Gumjudpai,
Gen. Rel. Grav. \textbf{41}, 249-265 (2009)
doi:10.1007/s10714-008-0665-x
[arXiv:0710.3598 [gr-qc]].

\bibitem{Gumjudpai:2007qq}
B.~Gumjudpai,
Astropart. Phys. \textbf{30}, 186-191 (2008)
doi:10.1016/j.astropartphys.2008.09.006
[arXiv:0708.3674 [gr-qc]].

\bibitem{Gumjudpai:2008mg}
B.~Gumjudpai,
JCAP \textbf{09}, 028 (2008)
doi:10.1088/1475-7516/2008/09/028
[arXiv:0805.3796 [gr-qc]].

\bibitem{Gumjudpai:2009ws}
B.~Gumjudpai,
[arXiv:0904.2746 [gr-qc]].

\bibitem{Guo:2004fq}
Z.K.~Guo, Y.S.~Piao, X.M.~Zhang and Y.Z.~Zhang,
Phys. Lett. B \textbf{608}, 177-182 (2005)
doi:10.1016/j.physletb.2005.01.017
[arXiv:astro-ph/0410654 [astro-ph]].

\bibitem{Guo:2006pc}
Z.K.~Guo, Y.S.~Piao, X.~Zhang and Y.Z.~Zhang,
Phys. Rev. D \textbf{74}, 127304 (2006)
doi:10.1103/PhysRevD.74.127304
[arXiv:astro-ph/0608165 [astro-ph]].

\bibitem{Gurses:2020kpv}
M.~Gurses and Y.~Heydarzade,
Eur. Phys. J. C \textbf{80}, no.11, 1061 (2020)
doi:10.1140/epjc/s10052-020-08641-0
[arXiv:2009.12825 [gr-qc]].

\bibitem{Guth:1980zm}
A.H.~Guth,
Phys. Rev. D \textbf{23}, 347-356 (1981)
doi:10.1103/PhysRevD.23.347

\bibitem{Guzman:2003kt}
F.S.~Guzman and L.A.~Urena-Lopez,
Phys. Rev. D \textbf{68}, 024023 (2003)
doi:10.1103/PhysRevD.68.024023
[arXiv:astro-ph/0303440 [astro-ph]].

\bibitem{Harko:2013gha}
T.~Harko, F.S.N.~Lobo and M.K.~Mak,
Eur. Phys. J. C \textbf{74}, 2784 (2014)
doi:10.1140/epjc/s10052-014-2784-8
[arXiv:1310.7167 [gr-qc]].

\bibitem{HarkoLobobook} T. Harko and F.S.N. Lobo, {\em Extensions of f(R) 
Gravity} (Cambridge University Press, Cambridge, 2018)

\bibitem{Harrison1967} E.R. Harrison,  Mon. Not. Roy. Astron. Soc.  
{\bf 137}, 69 (1967).

\bibitem{Hawkins:2001zx}
R.M.~Hawkins and J.E.~Lidsey,
Phys. Rev. D \textbf{66}, 023523 (2002)
doi:10.1103/PhysRevD.66.023523
[arXiv:astro-ph/0112139 [astro-ph]].

\bibitem{He:2009mz}
J.H.~He, B.~Wang and Y.P.~Jing,
JCAP \textbf{07}, 030 (2009)
doi:10.1088/1475-7516/2009/07/030
[arXiv:0902.0660 [gr-qc]].

\bibitem{He:2010im}
J.H.~He, B.~Wang and E.~Abdalla,
Phys. Rev. D \textbf{83}, 063515 (2011)
doi:10.1103/PhysRevD.83.063515
[arXiv:1012.3904 [astro-ph.CO]].

\bibitem{Heisenberg:2018vsk}
L.~Heisenberg,
Phys. Rept. \textbf{796}, 1-113 (2019)
doi:10.1016/j.physrep.2018.11.006
[arXiv:1807.01725 [gr-qc]].

\bibitem{Horndeski:1974wa}
G.W.~Horndeski,
Int. J. Theor. Phys. \textbf{10}, 363-384 (1974)
doi:10.1007/BF01807638

\bibitem{HoyleNarlikar63} F. Hoyle and J.V.  Narlikar, 
 Proc. Roy. Soc. (London) A273, 1 (1963).

\bibitem{HoyleNarlikar66} F. Hoyle and J.V. Narlikar,
Proc. Roy. Soc. (London) A294, 138 (1966).

\bibitem{Hu:1992dc}
W.~Hu and J.~Silk,
Phys. Rev. D \textbf{48}, 485-502 (1993)
doi:10.1103/PhysRevD.48.485

\bibitem{Huang:2009xa}
Q.G.~Huang,
JCAP \textbf{05}, 005 (2009)
doi:10.1088/1475-7516/2009/05/005
[arXiv:0903.1542 [hep-th]].

\bibitem{Hughston69} L.P. Hughston, 
Astrophys. J. {\bf 158}, 98 (1969).

\bibitem{HughstonJacobs70} L.P. Hughston and K.C. Jacobs,  
Astrophys. J. {\bf 160}, 147 (1970).


\bibitem{HughstonShepley1970} L.P. Hughston and L.C. Shepley,  
Astrophys. J. {\bf 160}, 333 (1970). 

\bibitem{JacobsNature1967} K.C. Jacobs,  Nature {\bf 215}, 1156 
(1967).

\bibitem{Jacobs1968} K.C. Jacobs, Astrophys. J. {\bf 153}, 661 
(1968).

\bibitem{Jamil:2009eb}
M.~Jamil, E.N.~Saridakis and M.R.~Setare,
Phys. Rev. D \textbf{81}, 023007 (2010)
doi:10.1103/PhysRevD.81.023007
[arXiv:0910.0822 [hep-th]].

\bibitem{Jordan38} P. Jordan, Naturwiss. 26, 417 (1938).

\bibitem{Jordan:1959eg}
P.~Jordan,
Z. Phys. \textbf{157}, 112-121 (1959)
doi:10.1007/BF01375155

\bibitem{Joseph:2019icj}
A.~Joseph and R.~Saha,
[arXiv:1912.06782 [gr-qc]].

\bibitem{Kaiser:2010ps}
D.~I.~Kaiser,
Phys. Rev. D \textbf{81}, 084044 (2010)
doi:10.1103/PhysRevD.81.084044
[arXiv:1003.1159 [gr-qc]].

\bibitem{Kamenshchik:2001cp} 
  A.Y. Kamenshchik, U. Moschella and V. Pasquier,
  Phys.\ Lett.\ B {\bf 511}, 265 (2001)  [gr-qc/0103004].

\bibitem{Kamenshchik:2012pw}
A.Y.~Kamenshchik, A.~Tronconi, G.~Venturi and S.Y.~Vernov,
Phys. Rev. D \textbf{87}, no.6, 063503 (2013) 
doi:10.1103/PhysRevD.87.063503
[arXiv:1211.6272 [gr-qc]].

\bibitem{Kamenshchik:2013dga}
A.Y.~Kamenshchik, E.O.~Pozdeeva, A.~Tronconi, G.~Venturi and 
S.Y.~Vernov,
Class. Quant. Grav. \textbf{31}, 105003 (2014)
doi:10.1088/0264-9381/31/10/105003
[arXiv:1312.3540 [hep-th]].

\bibitem{Kamenshchik:2015cla}
A.Y.~Kamenshchik, E.O.~Pozdeeva, A.~Tronconi, G.~Venturi and 
S.Y.~Vernov,
Class. Quant. Grav. \textbf{33}, no.1, 015004 (2016)
doi:10.1088/0264-9381/33/1/015004
[arXiv:1509.00590 [gr-qc]].

\bibitem{Kharbedya1976} L.I. Kharbediya,
Astron. Zh. (USSR) {\bf 53}, 1145 (1976) (in Russian).

\bibitem{Khurshudyan:2014yva}
M.~Khurshudyan, E.~Chubaryan and B.~Pourhassan,
Int. J. Theor. Phys. \textbf{53}, 2370 (2014)
doi:10.1007/s10773-014-2036-6
[arXiv:1402.2385 [gr-qc]].

\bibitem{Kim:2006te}
S.A.~Kim and A.R.~Liddle,
Phys. Rev. D \textbf{74}, 063522 (2006)
doi:10.1103/PhysRevD.74.063522
[arXiv:astro-ph/0608186 [astro-ph]].

\bibitem{Kim:2010ud}
S.A.~Kim, A.R.~Liddle and D.~Seery,
Phys. Rev. Lett. \textbf{105}, 181302 (2010)
doi:10.1103/PhysRevLett.105.181302
[arXiv:1005.4410 [astro-ph.CO]].

\bibitem{Kobayashi:2011nu}
T.~Kobayashi, M.~Yamaguchi and J.~Yokoyama,
Prog. Theor. Phys. \textbf{126}, 511-529 (2011)
doi:10.1143/PTP.126.511
[arXiv:1105.5723 [hep-th]].

\bibitem{Kobayashi:2019hrl}
T.~Kobayashi,
Rept. Prog. Phys. \textbf{82}, no.8, 086901 (2019)
doi:10.1088/1361-6633/ab2429
[arXiv:1901.07183 [gr-qc]].

\bibitem{Kofinas03} G. Kofinas, R. Maartens, and E. Papantonopoulos,  
J. High Energy Phys. {\bf 2003}, 066 (2003).

\bibitem{Kofman:1985aw}
L.A.~Kofman, A.D.~Linde and A.A.~Starobinsky,
Phys. Lett. B \textbf{157}, 361-367 (1985)
doi:10.1016/0370-2693(85)90381-8

\bibitem{KolbTurner} E.W. Kolb and M.S. Turner, {\em The Early 
Universe} (Addison-Wesley, Redwood 
City, CA, 1990).

\bibitem{Kritpetch:2019eof}
C.~Kritpetch, J.~Sanongkhun, P.~Vanichchapongjaroen and B.~Gumjudpai,
Mod. Phys. Lett. A \textbf{35}, no.19, 2050157 (2020)
doi:10.1142/S0217732320501576
[arXiv:1908.11265 [gr-qc]].

\bibitem{Kruger:2000nra}
A.T.~Kruger and J.W.~Norbury,
Phys. Rev. D \textbf{61}, 087303 (2000)
doi:10.1103/PhysRevD.61.087303
[arXiv:gr-qc/0004039 [gr-qc]].

\bibitem{LandauLifschitz} L.D. Landau and E.M. Lifschitz, {\em The 
Classical Theory of Fields} (Pergamon Press, Oxford, 1989), pp.~363-367.

\bibitem{Langlois:2015cwa}
D.~Langlois and K.~Noui,
JCAP \textbf{02}, 034 (2016)
doi:10.1088/1475-7516/2016/02/034
[arXiv:1510.06930 [gr-qc]].

\bibitem{Langlois:2015skt}
D.~Langlois and K.~Noui,
JCAP \textbf{07}, 016 (2016)
doi:10.1088/1475-7516/2016/07/016
[arXiv:1512.06820 [gr-qc]].

\bibitem{Langlois:2018dxi}
D.~Langlois,
Int. J. Mod. Phys. D \textbf{28}, no.05, 1942006 (2019)
doi:10.1142/S0218271819420069
[arXiv:1811.06271 [gr-qc]].

\bibitem{Lazkoz:2006pa}
R.~Lazkoz and G.~Leon,
Phys. Lett. B \textbf{638}, 303-309 (2006)
doi:10.1016/j.physletb.2006.05.075
[arXiv:astro-ph/0602590 [astro-ph]].

\bibitem{Lazkoz:2007mx}
R.~Lazkoz, G.~Leon and I.~Quiros,
Phys. Lett. B \textbf{649}, 103-110 (2007)
doi:10.1016/j.physletb.2007.03.060
[arXiv:astro-ph/0701353 [astro-ph]].

\bibitem{Lee:2008jp}
J.W.~Lee and S.~Lim,
JCAP \textbf{01}, 007 (2010)
doi:10.1088/1475-7516/2010/01/007
[arXiv:0812.1342 [astro-ph]].

\bibitem{Lee:2008ab}
J.W.~Lee,
J. Korean Phys. Soc. \textbf{54}, 2622 (2009)
doi:10.3938/jkps.54.2622
[arXiv:0801.1442 [astro-ph]].

\bibitem{Lemaitre1927}  G. Lema\^itre, 
Ann. Soc. Sci. Bruxelles~A {\bf 47}, 49 (1927).

\bibitem{Lemaitre1930} G. Lema\^itre, Bull. Astron. Inst. Neth. {\bf 
5}, No. 200 (1930).

\bibitem{Lemaitre1931} G. Lema\^itre, Mon. Not. Roy. Astron. Soc.  
{\bf 95}, 483 (1931).

\bibitem{Lemaitre1933} G. Lema\^itre, Ann. Soc. Sci. Bruxelles IA  
{\bf 53}, 51 (1933).

\bibitem{Leon:2012vt}
G.~Leon, Y.~Leyva and J.~Socorro,
Phys. Lett. B \textbf{732}, 285-297 (2014)
doi:10.1016/j.physletb.2014.03.053
[arXiv:1208.0061 [gr-qc]].

\bibitem{Leon:2018lnd}
G.~Leon, A.~Paliathanasis and J.L.~Morales-Mart\'\i{}nez,
Eur. Phys. J. C \textbf{78}, no.9, 753 (2018)
doi:10.1140/epjc/s10052-018-6225-y
[arXiv:1808.05634 [gr-qc]].

\bibitem{Li:2013nal}
B.~Li, T.~Rindler-Daller and P.R.~Shapiro,
Phys. Rev. D \textbf{89}, no.8, 083536 (2014)
doi:10.1103/PhysRevD.89.083536
[arXiv:1310.6061 [astro-ph.CO]].

\bibitem{Li:2014eha}
Y.H.~Li, J.F.~Zhang and X.~Zhang,
Phys. Rev. D \textbf{90}, no.6, 063005 (2014)
doi:10.1103/PhysRevD.90.063005
[arXiv:1404.5220 [astro-ph.CO]].

\bibitem{Liddle:1988tb}
A.R.~Liddle,
Phys. Lett. B \textbf{220}, 502-508 (1989)
doi:10.1016/0370-2693(89)90776-4

\bibitem{Liddle:1998xm}
A.R.~Liddle and R.J.~Scherrer,
Phys. Rev. D \textbf{59}, 023509 (1999)
doi:10.1103/PhysRevD.59.023509
[arXiv:astro-ph/9809272 [astro-ph]].

\bibitem{Liddle} A. Liddle, {\em An Introduction to Modern Cosmology}  
(Wiley, New York, 2015).

\bibitem{LiddleLyth} A.R. Liddle and D.H. Lyth, {\em Cosmological 
Inflation and Large-Scale Structure} (Cambridge University Press,
Cambridge, UK, 2000).

\bibitem{Lidsey:1991zp}
J.E.~Lidsey,
Phys. Lett. B \textbf{273}, 42-46 (1991)
doi:10.1016/0370-2693(91)90550-A

\bibitem{Lidsey:1991dz}
J.E.~Lidsey,
Class. Quant. Grav. \textbf{8}, 923-933 (1991)
doi:10.1088/0264-9381/8/5/016

\bibitem{Lidsey:1992wk}
J.E.~Lidsey,
Gen. Rel. Grav. \textbf{25}, 399-407 (1993)
doi:10.1007/BF00757120

\bibitem{Lidsey:1994qa}
J.E. Lidsey and I.~Waga,
Phys. Rev. D \textbf{51}, 444-449 (1995)
doi:10.1103/PhysRevD.51.444
[arXiv:astro-ph/9408016 [astro-ph]].

\bibitem{Lidsey:1995np}
J.E.~Lidsey, A.R.~Liddle, E.W.~Kolb, E.J.~Copeland, T.~Barreiro and 
M.~Abney,
Rev. Mod. Phys. \textbf{69}, 373-410 (1997)
doi:10.1103/RevModPhys.69.373
[arXiv:astro-ph/9508078 [astro-ph]].

\bibitem{Lidsey:2003ze}
J.E.~Lidsey,
Class. Quant. Grav. \textbf{21}, 777-786 (2004)
doi:10.1088/0264-9381/21/4/002
[arXiv:gr-qc/0307037 [gr-qc]].

\bibitem{Lima:2001fi}
J.A.S.~Lima,
Am. J. Phys. \textbf{69}, 1245-1247 (2001)
doi:10.1119/1.1405506
[arXiv:astro-ph/0109215 [astro-ph]].

\bibitem{Linde:1996gt}
A.D.~Linde and V.F.~Mukhanov,
Phys. Rev. D \textbf{56}, R535-R539 (1997)
doi:10.1103/PhysRevD.56.R535
[arXiv:astro-ph/9610219 [astro-ph]].

\bibitem{Linde:2005yw}
A.D.~Linde and V.~Mukhanov,
JCAP \textbf{04}, 009 (2006)
doi:10.1088/1475-7516/2006/04/009
[arXiv:astro-ph/0511736 [astro-ph]].

\bibitem{Liouville1853} J.J. Liouville, Math. Pures Appl. Paris 18 (1), 71 
(1853).

\bibitem{Lucchin:1984yf}
F.~Lucchin and S.~Matarrese,
Phys. Rev. D \textbf{32}, 1316 (1985)
doi:10.1103/PhysRevD.32.1316

\bibitem{Lyth:2002my}
D.H.~Lyth, C.~Ungarelli and D.~Wands,
Phys. Rev. D \textbf{67}, 023503 (2003)
doi:10.1103/PhysRevD.67.023503
[arXiv:astro-ph/0208055 [astro-ph]].

\bibitem{Maartens:1995uz}
R.~Maartens, D.R.~Taylor and N.~Roussos,
Phys. Rev. D \textbf{52}, 3358-3364 (1995)
doi:10.1103/PhysRevD.52.3358

\bibitem{Madsen2} M.S. Madsen, Astrophys. Space Sci. {\bf 113}, 205 
(1985).

\bibitem{Madsen:1988ph}
M.S.~Madsen,
Class. Quant. Grav. \textbf{5}, 627-639 (1988)
doi:10.1088/0264-9381/5/4/010

\bibitem{MarchisottoZakeri} E.A. Marchisotto and G.-A. Zakeri  
Coll. Math. J. {\bf 25}, 295-308 (1994).

\bibitem{Martin:2013tda}
J.~Martin, C.~Ringeval and V.~Vennin,
Phys. Dark Univ. \textbf{5-6}, 75-235 (2014)
doi:10.1016/j.dark.2014.01.003
[arXiv:1303.3787 [astro-ph.CO]].

\bibitem{Matos:1998vk}
T.~Matos and F.S.~Guzman,
Class. Quant. Grav. \textbf{17}, L9-L16 (2000)
doi:10.1088/0264-9381/17/1/102
[arXiv:gr-qc/9810028 [gr-qc]].

\bibitem{Matos:1999et}
T.~Matos, F.S.~Guzman and L.A.~Urena-Lopez,
Class. Quant. Grav. \textbf{17}, 1707-1712 (2000)
doi:10.1088/0264-9381/17/7/309
[arXiv:astro-ph/9908152 [astro-ph]].

\bibitem{Matos:2001ps}
T.~Matos, F.S.~Guzman, L.A.~Urena-Lopez and D.~Nunez,
doi:10.1007/0-306-47115-9\_16
[arXiv:astro-ph/0102419 [astro-ph]].

\bibitem{Matos:2008ag}
T.~Matos, A.~Vazquez-Gonzalez and J.~Magana,
Mon. Not. Roy. Astron. Soc. \textbf{393}, 1359-1369 (2009)
doi:10.1111/j.1365-2966.2008.13957.x
[arXiv:0806.0683 [astro-ph]].

\bibitem{May1975} T.L. May, 
Astrophys. J. {\bf 199}, 322 (1975).

\bibitem{Mavromatos:2020crd}
N.E.~Mavromatos, J.~Sol\`a Peracaula and S.~Basilakos,
Universe \textbf{6}, no.11, 218 (2020)
doi:10.3390/universe6110218
[arXiv:2008.00523 [gr-qc]].

\bibitem{Mavromatos:2020kzj}
N.E.~Mavromatos and J.~Sol\`a Peracaula,
doi:10.1140/epjs/s11734-021-00197-8 
[arXiv:2012.07971 [hep-ph]].

\bibitem{MayMcVittie1970} T.K. May and G.C. McVittie, 
Mon. Not. Roy. Astron. Soc. {\bf 148}, 407 (1070).

\bibitem{MayMcVittie1971} T.L. May and G.C. McVittie, 
Mon. Not Roy. Astron. Soc. {\bf 153}, 491-500 (1971).

\bibitem{McIntoshNature67} C.B.G. Mclntosh,
Nature {\bf 216}, 1297-1298 (1967). 

\bibitem{McIntosh1968} C.G.B. McIntosh, Mon. Not. Roy. Astron. Soc. 
{\bf 138}, 423 (1968).

\bibitem{McIntosh1968b} C.G.B. McIntosh, Mon. Not. Roy. Astron. Soc.  
{\bf 140}, 461 (1968).

\bibitem{McIntosh1970} C.B.G. McIntosh
J. Math. Phys. {\bf 11}, 250-252 (1970).

\bibitem{McIntosh1972} C.B.G. McIntosh, Austral. J. Phys. {\bf 25}, 
75-82 (1972).

\bibitem{McIntoshFoyster1972} C.B.G. McIntosh and J.M. Foyster,
Austral. J. Phys. {\bf 25}, 83-89 (1972).

\bibitem{Mendez:1996ug}
V.~M\'endez,
Class. Quant. Grav. \textbf{13}, 3229-3239 (1996)
doi:10.1088/0264-9381/13/12/013

\bibitem{Meyers:2010rg}
J.~Meyers and N.~Sivanandam,
Phys. Rev. D \textbf{83}, 103517 (2011)
doi:10.1103/PhysRevD.83.103517
[arXiv:1011.4934 [astro-ph.CO]].

\bibitem{Mishra:2018dzq}
S.~Mishra and S.~Chakraborty,
Eur. Phys. J. C \textbf{78}, no.11, 917 (2018)
doi:10.1140/epjc/s10052-018-6405-9
[arXiv:1811.08279 [gr-qc]].

\bibitem{Misner:1967uu}
C.W.~Misner,
Astrophys. J. \textbf{151}, 431-457 (1968)
doi:10.1086/149448

\bibitem{MohseniSadjadi:2006hb}
H.~Mohseni Sadjadi and M.~Alimohammadi,
Phys. Rev. D \textbf{74}, 043506 (2006)
doi:10.1103/PhysRevD.74.043506
[arXiv:gr-qc/0605143 [gr-qc]].

\bibitem{Motohashi:2016ftl}
H.~Motohashi, K.~Noui, T.~Suyama, M.~Yamaguchi and D.~Langlois,
JCAP \textbf{07}, 033 (2016)
doi:10.1088/1475-7516/2016/07/033
[arXiv:1603.09355 [hep-th]].

\bibitem{Slava} V. Mukhanov, {\em Physical Foundations of Cosmology} 
(Cambridge University Press, Cambridge, UK, 2005).

\bibitem{Mukhanov:1990me}
V.~F.~Mukhanov, H.~A.~Feldman and R.~H.~Brandenberger,
Phys. Rept. \textbf{215}, 203-333 (1992)
doi:10.1016/0370-1573(92)90044-Z

\bibitem{Muslimov:1990be}
A.G.~Muslimov,
Class. Quant. Grav. \textbf{7}, 231-237 (1990)
doi:10.1088/0264-9381/7/2/015

\bibitem{Naidu:2021nwh}
N.~F.~Naidu, S.~Carloni and P.~Dunsby, 
Phys. Rev. D \textbf{104}, no.4, 044014 (2021)
doi:10.1103/PhysRevD.104.044014
[arXiv:2102.05693 [gr-qc]].

\bibitem{Naruko:2008sq}
A.~Naruko and M.~Sasaki,
Prog. Theor. Phys. \textbf{121}, 193-210 (2009)
doi:10.1143/PTP.121.193
[arXiv:0807.0180 [astro-ph]].

\bibitem{Nesteruk94} A.V. Nesteruk, Class. Quant. Grav. 11, L15 (1994).

\bibitem{Nesteruk:1995uu}
A.V.~Nesteruk and A.C.~Ottewill,
Class. Quant. Grav. \textbf{12}, 51-57 (1995)
doi:10.1088/0264-9381/12/1/005

\bibitem{Nicolis:2008in}
A.~Nicolis, R.~Rattazzi and E.~Trincherini,
Phys. Rev. D \textbf{79}, 064036 (2009)
doi:10.1103/PhysRevD.79.064036
[arXiv:0811.2197 [hep-th]].

\bibitem{Nojiri:2004pf}
S.~Nojiri and S.D. Odintsov,
Phys. Rev. D \textbf{70}, 103522 (2004)
doi:10.1103/PhysRevD.70.103522
[arXiv:hep-th/0408170 [hep-th]].

\bibitem{Nojirietal05} S. Nojiri, S.D. Odintsov, and S. Tsujikawa,  
Phys. Rev. D {\bf 71}, 063004 (2005).

\bibitem{NojiriOdintsov05}
S. Nojiri and S.D. Odintsov,
Phys. Rev. D {\bf 72}, 023003 (2005).

\bibitem{Nojiri:2006zh}
S.~Nojiri and S.D.~Odintsov,
Phys. Lett. B \textbf{639}, 144-150 (2006)
doi:10.1016/j.physletb.2006.06.065
[arXiv:hep-th/0606025 [hep-th]].

\bibitem{Nojiri:2007qi}
S.~Nojiri and S.D.~Odintsov,
Phys. Lett. B \textbf{649}, 440-444 (2007)
doi:10.1016/j.physletb.2007.02.042
[arXiv:hep-th/0702031 [hep-th]].

\bibitem{Nojiri:2010wj}
S.~Nojiri and S.~D.~Odintsov,
Phys. Rept. \textbf{505} (2011) 59
doi:10.1016/j.physrep.2011.04.001
[arXiv:1011.0544 [gr-qc]].

\bibitem{Nordtvedt:1970uv}
K.~Nordtvedt, Jr.,
Astrophys. J. \textbf{161}, 1059-1067 (1970)
doi:10.1086/150607

\bibitem{Nowakowski:2002kh}
M.~Nowakowski and H.~Rosu,
Phys. Rev. E \textbf{65}, 047602 (2002)
doi:10.1103/PhysRevE.65.047602
[arXiv:physics/0110066 [physics]].

\bibitem{Ozer:1992wh}
M.~\"{O}zer and M.O.~Taha,
Phys. Rev. D \textbf{45}, R997-R999 (1992)
doi:10.1103/PhysRevD.45.R997

\bibitem{Padmanabhan:2002sh}
T.~Padmanabhan and T.R.~Choudhury,
Phys. Rev. D \textbf{66}, 081301 (2002)
doi:10.1103/PhysRevD.66.081301
[arXiv:hep-th/0205055 [hep-th]].

\bibitem{Pailas:2020xhh}
T.~Pailas, N.~Dimakis, A.~Paliathanasis, P.A.~Terzis and 
T.~Christodoulakis,
Phys. Rev. D \textbf{102}, no.6, 063524 (2020)
doi:10.1103/PhysRevD.102.063524
[arXiv:2005.11726 [gr-qc]].

\bibitem{Paliathanasis:2011jq}
A.~Paliathanasis, M.~Tsamparlis and S.~Basilakos,
Phys. Rev. D \textbf{84}, 123514 (2011)
doi:10.1103/PhysRevD.84.123514
[arXiv:1111.4547 [astro-ph.CO]].

\bibitem{Paliathanasis:2014zxa}
A.~Paliathanasis, M.~Tsamparlis and S.~Basilakos,
Phys. Rev. D \textbf{90}, no.10, 103524 (2014)
doi:10.1103/PhysRevD.90.103524
[arXiv:1410.4930 [gr-qc]].

\bibitem{Paliathanasis:2014yfa}
A.~Paliathanasis and M.~Tsamparlis,
Phys. Rev. D \textbf{90}, no.4, 043529 (2014)
doi:10.1103/PhysRevD.90.043529
[arXiv:1408.1798 [gr-qc]].

\bibitem{Paliathanasis:2015arj}
A.~Paliathanasis, M.~Tsamparlis, S.~Basilakos and J.D.~Barrow,
Phys. Rev. D \textbf{93}, no.4, 043528 (2016)
doi:10.1103/PhysRevD.93.043528
[arXiv:1511.00439 [gr-qc]].

\bibitem{Paliathanasis:2016heb}
A.~Paliathanasis and S.~Capozziello,
Mod. Phys. Lett. A \textbf{31}, no.32, 1650183 (2016)
doi:10.1142/S0217732316501832
[arXiv:1602.08914 [gr-qc]].

\bibitem{Payne70} A.D. Payne,
Austral. J. Phys. {\bf 23}, 177-185 (1970).

\bibitem{Peebles:1968ja}
P.J.E.~Peebles,
Astrophys. J. \textbf{153}, 1 (1968)
doi:10.1086/149628

\bibitem{Peeblesbook} P.J.E. Peebles, {\em Principles of Physical 
Cosmology} (Princeton University Press, Princeton, 1993).

\bibitem{Peebles:1987ek}
P.J.E.~Peebles and B.~Ratra,
Astrophys. J. Lett. \textbf{325}, L17 (1988)
doi:10.1086/185100

\bibitem{Penzias:1965wn}
A.A.~Penzias and R.W.~Wilson,
Astrophys. J. \textbf{142}, 419-421 (1965)
doi:10.1086/148307

\bibitem{Perlmutter:1997zf}
S.~Perlmutter \textit{et al.} [Supernova Cosmology Project],
Nature \textbf{391}, 51-54 (1998)
doi:10.1038/34124
[arXiv:astro-ph/9712212 [astro-ph]].

\bibitem{Perlmutter:1998np}
S.~Perlmutter \textit{et al.} [Supernova Cosmology Project],
Astrophys. J. \textbf{517}, 565-586 (1999)
doi:10.1086/307221
[arXiv:astro-ph/9812133 [astro-ph]].

\bibitem{Peterson:2010mv}
C.M.~Peterson and M.~Tegmark,
Phys. Rev. D \textbf{84}, 023520 (2011)
doi:10.1103/PhysRevD.84.023520
[arXiv:1011.6675 [astro-ph.CO]].

\bibitem{Phetnora:2008mf}
T.~Phetnora, R.~Sooksan and B.~Gumjudpai,
Gen. Rel. Grav. \textbf{42}, 225-240 (2010)
doi:10.1007/s10714-009-0831-9
[arXiv:0805.3794 [gr-qc]].

\bibitem{Piedipalumbo:2019snr}
E.~Piedipalumbo, M.~De Laurentis and S.~Capozziello,
Phys. Dark Univ. \textbf{27}, 100444 (2020)
doi:10.1016/j.dark.2019.100444
[arXiv:1912.08089 [gr-qc]].

\bibitem{Pimentel} L.O. Pimentel, Class. Quant. Grav. {\bf 6}, L263 
(1989).

\bibitem{Pinney50} E. Pinney, Proc. Amer. Math. Soc. {\bf 1}, 681 (1950).

\bibitem{Pradhan13} A. Pradhan, Indian J. Phys. {\bf 88}, 215-223 
(2014), arXiv:1211.1882 [physics.gen-ph].

\bibitem{Pucheu:2009jw}
M.L.~Pucheu and M.~Bellini,
Nuovo Cim. B \textbf{125}, 851-859 (2010)
doi:10.1393/ncb/i2010-10888-0
[arXiv:0906.1824 [gr-qc]].

\bibitem{Pujolas:2011he}
O.~Pujolas, I.~Sawicki and A.~Vikman,
JHEP \textbf{11}, 156 (2011)
doi:10.1007/JHEP11(2011)156
[arXiv:1103.5360 [hep-th]].

\bibitem{Qiu:2010ux}
T.~Qiu,
Mod. Phys. Lett. A \textbf{25}, 909-921 (2010)
doi:10.1142/S021773231000006X
[arXiv:1002.3971 [hep-th]].

\bibitem{Riess:1998cb}
A.G.~Riess \textit{et al.} [Supernova Search Team],
Astron. J. \textbf{116}, 1009-1038 (1998)
doi:10.1086/300499
[arXiv:astro-ph/9805201 [astro-ph]].

\bibitem{Rigopoulos:2005ae}
G.I.~Rigopoulos, E.P.S.~Shellard and B.J.W.~van Tent,
Phys. Rev. D \textbf{73}, 083522 (2006)
doi:10.1103/PhysRevD.73.083522
[arXiv:astro-ph/0506704 [astro-ph]].

\bibitem{Rigopoulos:2005us}
G.I.~Rigopoulos, E.P.S.~Shellard and B.J.W.~van Tent,
Phys. Rev. D \textbf{76}, 083512 (2007)
doi:10.1103/PhysRevD.76.083512
[arXiv:astro-ph/0511041 [astro-ph]].

\bibitem{Rindler-Daller:2013zxa}
T.~Rindler-Daller and P.R.~Shapiro,
Mod. Phys. Lett. A \textbf{29}, no.2, 1430002 (2014)
doi:10.1142/S021773231430002X
[arXiv:1312.1734 [astro-ph.CO]].

\bibitem{Robles:2012uy}
V.H.~Robles and T.~Matos,
Mon. Not. Roy. Astron. Soc. \textbf{422}, 282-289 (2012)
doi:10.1111/j.1365-2966.2012.20603.x
[arXiv:1201.3032 [astro-ph.CO]].

\bibitem{Robles:2018fur}
V.H.~Robles, J.S.~Bullock and M.~Boylan-Kolchin,
Mon. Not. Roy. Astron. Soc. \textbf{483}, no.1, 289-298 (2019)
doi:10.1093/mnras/sty3190
[arXiv:1807.06018 [astro-ph.CO]].

\bibitem{Roeder1967} R.C. Roeder, Nature {\bf 216}, 774 (1967).

\bibitem{Rosu:1999ud}
H.C.~Rosu, P.~Espinoza and M.~Reyes,
Nuovo Cim. B \textbf{114}, 1439-1444 (1999)
[arXiv:gr-qc/9910070 [gr-qc]].

\bibitem{Rosu:2005rz}
H.C.~Rosu and P.~Ojeda-May,
Int. J. Theor. Phys. \textbf{45}, 1191-1196 (2006)
doi:10.1007/s10773-006-9123-2
[arXiv:gr-qc/0510004 [gr-qc]].

\bibitem{Rosu:2010if}
H.C.~Rosu and K.~V.~Khmelnytskaya,
SIGMA \textbf{7}, 013 (2011)
doi:10.3842/SIGMA.2011.013
[arXiv:1012.1920 [gr-qc]].

\bibitem{Sadeghi:2008qp}
J.~Sadeghi, M.R.~Setare, A.~Banijamali and F.~Milani,
Phys. Lett. B \textbf{662}, 92-96 (2008)
doi:10.1016/j.physletb.2008.02.062
[arXiv:0804.0553 [hep-th]].

\bibitem{Salopek:1990jq}
D.S.~Salopek and J.R.~Bond,
Phys. Rev. D \textbf{42}, 3936-3962 (1990)
doi:10.1103/PhysRevD.42.3936

\bibitem{Sapar1970} A. Sapar, Astron. Zh. (USSR) {\bf 47}, 503 
(1970) (in Russian).

\bibitem{Saridakis:2009jq}
E.N.~Saridakis and J.M.~Weller,
Phys. Rev. D \textbf{81}, 123523 (2010)
doi:10.1103/PhysRevD.81.123523
[arXiv:0912.5304 [hep-th]].

\bibitem{Saridakis:2009ej}
E.N.~Saridakis,
Nucl. Phys. B \textbf{830}, 374-389 (2010)
doi:10.1016/j.nuclphysb.2010.01.005
[arXiv:0903.3840 [astro-ph.CO]].

\bibitem{Sasaki:2006kq}
M.~Sasaki, J.~Valiviita and D.~Wands,
Phys. Rev. D \textbf{74}, 103003 (2006)
doi:10.1103/PhysRevD.74.103003
[arXiv:astro-ph/0607627 [astro-ph]].

\bibitem{Sasaki:2008uc}
M.~Sasaki,
Prog. Theor. Phys. \textbf{120}, 159-174 (2008)
doi:10.1143/PTP.120.159
[arXiv:0805.0974 [astro-ph]].

\bibitem{Schunck:1994yd}
F.E.~Schunck and E.W.~Mielke,
Phys. Rev. D \textbf{50}, 4794-4806 (1994)
doi:10.1103/PhysRevD.50.4794
[arXiv:gr-qc/9407041 [gr-qc]].

\bibitem{Semiz:2012zz}
I.~Semiz,
Phys. Rev. D \textbf{85}, 068501 (2012)
doi:10.1103/PhysRevD.85.068501

\bibitem{Setare:2006rf}
M.R.~Setare,
Phys. Lett. B \textbf{641}, 130-133 (2006)
doi:10.1016/j.physletb.2006.08.039
[arXiv:hep-th/0611165 [hep-th]].

\bibitem{Setare:2008pz}
M.R.~Setare and E.N.~Saridakis,
Phys. Lett. B \textbf{668}, 177-181 (2008)
doi:10.1016/j.physletb.2008.08.033
[arXiv:0802.2595 [hep-th]].

\bibitem{Setare:2008si}
M.R.~Setare and E.N.~Saridakis,
JCAP \textbf{09}, 026 (2008)
doi:10.1088/1475-7516/2008/09/026
[arXiv:0809.0114 [hep-th]].

\bibitem{Setare:2008sf}
M.R.~Setare and E.N.~Saridakis,
Phys. Rev. D \textbf{79}, 043005 (2009)
doi:10.1103/PhysRevD.79.043005
[arXiv:0810.4775 [astro-ph]].

\bibitem{Setare:2008dw}
M.R.~Setare and E.N.~Saridakis,
Int. J. Mod. Phys. D \textbf{18}, 549-557 (2009)
doi:10.1142/S0218271809014625
[arXiv:0807.3807 [hep-th]].

\bibitem{Setare:2008ci}
M.R.~Setare and J.~Sadeghi,
Int. J. Theor. Phys. \textbf{47}, 3219-3225 (2008)
doi:10.1007/s10773-008-9757-3
[arXiv:0805.1117 [gr-qc]].

\bibitem{Shi:2008df}
S.G.~Shi, Y.~S.~Piao and C.~F.~Qiao,
JCAP \textbf{04}, 027 (2009)
doi:10.1088/1475-7516/2009/04/027
[arXiv:0812.4022 [astro-ph]].

\bibitem{Shikin1968} I.S. Shikin, Sov. Phys. (Doklady) {\bf 13}, 320 
(1968).

\bibitem{ShtanovSahni02} Y. Shtanov and V. Sahni,
Class. Quant. Grav. {\bf 19}, L101-L107 (2002).

\bibitem{SilbergleitChernin17} A.S. Silbergleit and A.D. Chernin, {\em  
Interacting Dark Energy and the Expansion of the Universe}, Springer 
Briefs in Physics (Springer, New York, 2017).

\bibitem{SilvaCosta09}  S. Silva e Costa, 
Mod. Phys. Lett. A {\bf 24}, 531-540 (2009) doi:10.1142/S021773230902845X.

\bibitem{Sistero1972} R.F. Sistero, 
Astrophys. Space Sci. {\bf 17}, 150 (1972). 

\bibitem{SmootGorensteinMuller77} G.F. Smoot, M.V. Gorenstein, and R.A.  
Muller, Phys. Rev. Lett. {\bf 39}, 898-901 (1977).

\bibitem{Sonego:2011rb}
S.~Sonego and V.~Talamini,
Am. J. Phys. \textbf{80}, 670-679 (2012)
doi:10.1119/1.4731258
[arXiv:1112.4319 [physics.ed-ph]].

\bibitem{Sotiriou:2008it}
T.P.~Sotiriou and V.~Faraoni,
Class. Quant. Grav. \textbf{25}, 205002 (2008)
doi:10.1088/0264-9381/25/20/205002
[arXiv:0805.1249 [gr-qc]].

\bibitem{Sotiriou:2008rp}
T.P.~Sotiriou and V.~Faraoni,
Rev. Mod. Phys. \textbf{82}, 451-497 (2010)
doi:10.1103/RevModPhys.82.451
[arXiv:0805.1726 [gr-qc]].

\bibitem{Spergel:1996ai}
D.~Spergel and U.L.~Pen,
Astrophys. J. Lett. \textbf{491}, L67-L71 (1997)
doi:10.1086/311074
[arXiv:astro-ph/9611198 [astro-ph]].

\bibitem{StabellRefsdal} R. Stabell and S. Refsdal, Mon. Not. Roy. 
Astron. Soc. {\bf 132}, 379 (1966).

\bibitem{StarkovichCooperstock92} S.P. Starkovich and F.I. Cooperstock, 
Astrophys. J. 398, 1 (1992). 

\bibitem{Starobinsky:1980te}
A.A.~Starobinsky,
Phys. Lett. B \textbf{91}, 99-102 (1980)
doi:10.1016/0370-2693(80)90670-X

\bibitem{StarobinskyI}   
A.A.~Starobinsky,
JETP Lett.\  {\bf 30} (1979) 682
[Pisma Zh.\ Eksp.\ Teor.\ Fiz.\  {\bf 30} (1979) 719].

\bibitem{Stefancic:2004kb}
H.~\u{S}tefan\u{c}i\u{c},
Phys. Rev. D \textbf{71}, 084024 (2005)
doi:10.1103/PhysRevD.71.084024
[arXiv:astro-ph/0411630 [astro-ph]].

\bibitem{Stephani} H. Stephani, {\em General Relativity: An 
Introduction to the Theory of the Gravitational Field} (Cambridge 
University Press, Cambridge, UK, 1982).

\bibitem{StephaniExact} H. Stephani, D. Kramer, M. MacCallum, C. 
Hoenselaers, E.  Herlt, {\em Exact Solutions of Einstein's Field 
Equations}, 2nd edition (Cambridge University Press, Cambridge, UK, 2003).

\bibitem{Suarez:2011yf}
A.~Suarez and T.~Matos,
Mon. Not. Roy. Astron. Soc. \textbf{416}, 87 (2011)
doi:10.1111/j.1365-2966.2011.19012.x
[arXiv:1101.4039 [gr-qc]].

\bibitem{Suarez:2013iw}
A.~Su\'arez, V.H.~Robles and T.~Matos,
Astrophys. Space Sci. Proc. \textbf{38}, 107-142 (2014)
doi:10.1007/978-3-319-02063-1\_9
[arXiv:1302.0903 [astro-ph.CO]].

\bibitem{Sunyaev:1970er}
R.A.~Sunyaev and Y.B.~Zeldovich,
Astrophys. Space Sci. \textbf{7}, 20-30 (1970).

\bibitem{Szydlowski:2005nu}
M.~Szydlowski, W.~Godlowski and R.~Wojtak,
Gen. Rel. Grav. \textbf{38}, 795 (2006)
doi:10.1007/s10714-006-0265-6
[arXiv:astro-ph/0505202 [astro-ph]].

\bibitem{Synge} J.L. Synge, {\em Relativity: The General Theory}  
(North Holland, Amsterdam, 1955).

\bibitem{Tauber1967} G.E. Tauber, J. Math. Phys. {\bf 8}, 118 (1967).

\bibitem{Tolman1931} R.C. Tolman, Phys. Rev. {\bf 38}, 1758 (1931).

\bibitem{Tolman} R.C. Tolman, {\em Relativity, Thermodynamics 
and Cosmology} (Clarendon Press, Oxford, 1934), pp.~407-411.

\bibitem{Tsamparlis:2018nyo}
M.~Tsamparlis and A.~Paliathanasis,
Symmetry \textbf{10}, no.7, 233 (2018)
doi:10.3390/sym10070233
[arXiv:1806.05888 [gr-qc]].

\bibitem{Uzan01} J.-P. Uzan and R. Lehoucq, Eur. J. Phys. {\bf 22}, 371 
(2001).

\bibitem{UrenaLopez:2006ay}
L.~A.~Urena-L\'opez,
[arXiv:physics/0609181 [physics]].

\bibitem{Vajk1969} J.P. Vajk, J. Math. Phys. {\bf 10}, 1145 (1969).

\bibitem{Valiviita:2008iv}
J.~Valiviita, E.~Majerotto and R.~Maartens,
JCAP \textbf{07}, 020 (2008)
doi:10.1088/1475-7516/2008/07/020
[arXiv:0804.0232 [astro-ph]].

\bibitem{Vazquez:2012ag}
J.A.~Vazquez, S.~Hee, M.P.~Hobson, A.N.~Lasenby, M.~Ibison and 
M.~Bridges,
JCAP \textbf{07}, 062 (2018)
doi:10.1088/1475-7516/2018/07/062
[arXiv:1208.2542 [astro-ph.CO]].

\bibitem{Verma07} A.K. Verma, Astrophys. Space Sci. {\bf 321}, 73-77 
(2009).  

\bibitem{Vernov:2006dm}
S.~Y.~Vernov,
Teor. Mat. Fiz. \textbf{155}, 47-61 (2008)
doi:10.1007/s11232-008-0045-4 
[arXiv:astro-ph/0612487 [astro-ph]].

\bibitem{Vilenkin:1984rt}
A.~Vilenkin,
Phys. Rev. Lett. \textbf{53}, 1016-1018 (1984)
doi:10.1103/PhysRevLett.53.1016

\bibitem{Vincent:2008ds}
A.C.~Vincent and J.M.~Cline,
JHEP \textbf{10}, 093 (2008)
doi:10.1088/1126-6708/2008/10/093
[arXiv:0809.2982 [astro-ph]].

\bibitem{Wagoner:1970vr}
R.V.~Wagoner,
Phys. Rev. D \textbf{1}, 3209-3216 (1970)
doi:10.1103/PhysRevD.1.3209

\bibitem{WainwrightEllisBook} J. Wainwright and G.F.R. Ellis (eds.) {\em 
Dynamical Systems in Cosmology} (Cambridge University Press, Cambridge, 
UK, 1997).

\bibitem{Wald} R.M. Wald, {\em General Relativity} (Chicago University 
Press, Chicago, 1984).  

\bibitem{Wang:2009ag}
J.~Wang and S.p.~Yang,
J. Theor. Phys. \textbf{1}, 62-75 (2012)
[arXiv:0901.1441 [gr-qc]].

\bibitem{Wang:2009ae}
J.~Wang, S.w.~Cui and S.p.~Yang,
Phys. Lett. B \textbf{683}, 101-107 (2010)
doi:10.1016/j.physletb.2009.11.064
[arXiv:0901.1439 [gr-qc]].

\bibitem{Wang:2010si}
T.~Wang,
Phys. Rev. D \textbf{82}, 123515 (2010)
doi:10.1103/PhysRevD.82.123515
[arXiv:1008.3198 [astro-ph.CO]].

\bibitem{Wang:2016lxa}
B.~Wang, E.~Abdalla, F.~Atrio-Barandela and D.~Pavon,
Rept. Prog. Phys. \textbf{79}, no.9, 096901 (2016)
doi:10.1088/0034-4885/79/9/096901
[arXiv:1603.08299 [astro-ph.CO]].

\bibitem{Wei:2005nw}
H.~Wei, R.G.~Cai and D.F.~Zeng,
Class. Quant. Grav. \textbf{22}, 3189-3202 (2005)
doi:10.1088/0264-9381/22/16/005
[arXiv:hep-th/0501160 [hep-th]].

\bibitem{Wei:2007rp}
H.~Wei and S.N.~Zhang,
Phys. Rev. D \textbf{76}, 063005 (2007)
doi:10.1103/PhysRevD.76.063005
[arXiv:0705.4002 [gr-qc]].

\bibitem{Weinberg} S. Weinberg, {\em Gravitation and Cosmology} (J. Wiley 
\& Sons, New York, 1972).

\bibitem{Wetterich:2014bma}
C.~Wetterich,
Lect. Notes Phys. \textbf{892}, 57 (2015)
doi:10.1007/978-3-319-10070-8\_3
[arXiv:1402.5031 [astro-ph.CO]].

\bibitem{Willbook} C.M. Will, {\em Theory and Experiment in Gravitational 
Physics} (Cambridge University Press, Cambridge, 1993).

\bibitem{Will:2014kxa}
C.M.~Will,
Living Rev. Rel. \textbf{17}, 4 (2014)
doi:10.12942/lrr-2014-4
[arXiv:1403.7377 [gr-qc]].

\bibitem{Williamsetal06} F.L. Williams, P.G. Kevrekidis, T. 
Christodoulakis, C. Helias, G.O. 
Papadopoulos and T. Grammenos, {\em Trends in General Relativity and 
Quantum Cosmology}, edited by Ch.V. Benton (Nova Science, New York, 2006), 
p.~37.

\bibitem{Wu:2005apa}
P.x.~Wu and H.w.~Yu,
Int. J. Mod. Phys. D \textbf{14}, 1873-1882 (2005)
doi:10.1142/S0218271805007486
[arXiv:gr-qc/0509036 [gr-qc]].

\bibitem{Xia:2004rw}
J.Q.~Xia, B.~Feng and X.M.~Zhang,
Mod. Phys. Lett. A \textbf{20}, 2409-2416 (2005)
doi:10.1142/S0217732305017445
[arXiv:astro-ph/0411501 [astro-ph]].

\bibitem{Xiong:2008ic}
H.H.~Xiong, Y.F.~Cai, T.~Qiu, Y.S.~Piao and X.~Zhang,
Phys. Lett. B \textbf{666}, 212-217 (2008)
doi:10.1016/j.physletb.2008.07.053
[arXiv:0805.0413 [astro-ph]].

\bibitem{Xu:2013jma}
X.D.~Xu, B.~Wang, P.~Zhang and F.~Atrio-Barandela,
JCAP \textbf{12}, 001 (2013)
doi:10.1088/1475-7516/2013/12/001
[arXiv:1308.1475 [astro-ph.CO]].

\bibitem{Yokoyama:2007uu}
S.~Yokoyama, T.~Suyama and T.~Tanaka,
JCAP \textbf{07}, 013 (2007)
doi:10.1088/1475-7516/2007/07/013
[arXiv:0705.3178 [astro-ph]].

\bibitem{Yokoyama:2007dw}
S.~Yokoyama, T.~Suyama and T.~Tanaka,
Phys. Rev. D \textbf{77}, 083511 (2008)
doi:10.1103/PhysRevD.77.083511
[arXiv:0711.2920 [astro-ph]].

\bibitem{Yurov:2008sy}
A.V.~Yurov and V.A.~Yurov,
J. Math. Phys. \textbf{51}, 082503 (2010)
doi:10.1063/1.3460856
[arXiv:0809.1216 [hep-th]].

\bibitem{Zampeli:2015ojr}
A.~Zampeli, T.~Pailas, P.A.~Terzis and T.~Christodoulakis,
JCAP \textbf{05}, 066 (2016)
doi:10.1088/1475-7516/2016/05/066
[arXiv:1511.08382 [gr-qc]].

\bibitem{Zeldovich:1969ff}
Y.B.~Zeldovich and R.A.~Sunyaev,
Astrophys. Space Sci. \textbf{4}, 301-316 (1969)
doi:10.1007/BF00661821

\bibitem{Zhang:2005kj}
X.~Zhang,
Commun. Theor. Phys. \textbf{44}, 762-768 (2005)
doi:10.1088/6102/44/4/762

\bibitem{Zhang:2006ck}
X.F.~Zhang and T.~Qiu,
Phys. Lett. B \textbf{642}, 187-191 (2006)
doi:10.1016/j.physletb.2006.09.038
[arXiv:astro-ph/0603824 [astro-ph]].

\bibitem{Zhang:2008ac}
S.~Zhang and B.~Chen,
Phys. Lett. B \textbf{669}, 4-8 (2008)
doi:10.1016/j.physletb.2008.09.025
[arXiv:0806.4435 [hep-ph]].

\bibitem{Zhang:2009un}
X.~Zhang,
Phys. Rev. D \textbf{79}, 103509 (2009)
doi:10.1103/PhysRevD.79.103509
[arXiv:0901.2262 [astro-ph.CO]].

\bibitem{Zhao:2005vj}
G.B.~Zhao, J.Q.~Xia, M.~Li, B.~Feng and X.~Zhang,
Phys. Rev. D \textbf{72}, 123515 (2005)
doi:10.1103/PhysRevD.72.123515
[arXiv:astro-ph/0507482 [astro-ph]].

\bibitem{Zhao:2006mp}
W.~Zhao,
Phys. Rev. D \textbf{73}, 123509 (2006)
doi:10.1103/PhysRevD.73.123509
[arXiv:astro-ph/0604460 [astro-ph]].

\bibitem{Zhuravlev:1998ff}
V.M.~Zhuravlev, S.V.~Chervon and V.K.~Shchigolev,
J. Exp. Theor. Phys. \textbf{87}, 223-228 (1998)
doi:10.1134/1.558649

\bibitem{Ziaeepour:2011bq}
H.~Ziaeepour,
Phys. Rev. D \textbf{86}, 043503 (2012)
doi:10.1103/PhysRevD.86.043503
[arXiv:1112.6025 [astro-ph.CO]].

\bibitem{Zimdahl:2014jsa}
W.~Zimdahl,
Int. J. Geom. Meth. Mod. Phys. \textbf{11}, 1460014 (2014)
doi:10.1142/S0219887814600147
[arXiv:1404.7334 [astro-ph.CO]].


\end{thebibliography}
\end{document}